\documentclass{article}

\PassOptionsToPackage{numbers,compress}{natbib}

\usepackage[final]{neurips_2026}

\usepackage{etoolbox}
\makeatletter
\patchcmd{\@maketitle}{\vskip 0.3in \@minus 0.1in}{\vskip 0.2in \@minus 0.1in}{}{\errmessage{patch after-authors failed}}
\patchcmd{\@maketitle}{\rule{\z@}{24\p@}\ignorespaces}{\rule{\z@}{20\p@}\ignorespaces}{}{\errmessage{patch And failed}}
\patchcmd{\@maketitle}{\rule{\z@}{24\p@}\ignorespaces}{\rule{\z@}{20\p@}\ignorespaces}{}{\errmessage{patch AND failed}}
\patchcmd{\@maketitle}{\rule{\z@}{24\p@}\@author}{\rule{\z@}{20\p@}\@author}{}{\errmessage{patch first row failed}}
\patchcmd{\@bottomtitlebar}{\vskip 0.29in}{\vskip 0.22in}{}{\errmessage{patch bottombar failed}}
\patchcmd{\@toptitlebar}{\vskip 0.25in}{\vskip 0.20in}{}{\errmessage{patch topbar failed}}
\makeatother

\usepackage{makecell}
\usepackage[utf8]{inputenc}
\usepackage[T1]{fontenc}
\usepackage{url}
\usepackage[hypcap=true]{caption}
\usepackage{hyperref}
\usepackage{booktabs}
\usepackage{amsfonts}
\usepackage{nicefrac}
\usepackage{microtype}
\usepackage{xcolor}
\usepackage{amsmath}
\usepackage{graphicx}
\graphicspath{{./}}
\usepackage{float}
\usepackage{algorithm}
\usepackage{algpseudocode}
\usepackage{multirow}
\usepackage{tikz}
\usetikzlibrary{positioning, arrows.meta, fit, calc, shapes.geometric, decorations.pathreplacing, backgrounds}
\usepackage{bm}

\title{HarmonICA: Quasi-Time-Invariant ICA for Motor Unit Decomposition during Dynamic Contractions}

\author{%
  Alexander Kenneth Clarke \\
  Imperial College London\\
  London, UK \\
  \texttt{a.clarke18@imperial.ac.uk} \\
  \AND
  Dimitrios Halatsis \\
  Imperial College London\\
  London, UK \\
  \texttt{d.halatsis@imperial.ac.uk} \\
  \And
  Agnese Grison \\
  Imperial College London\\
  London, UK \\
  \texttt{agnese.grison16@imperial.ac.uk} \\
  \And
  Irene Mendez Guerra \\
  Imperial College London\\
  London, UK \\
  \texttt{irene.mendez17@imperial.ac.uk} \\
  \And
  Noura Ezaz-Nikpay \\
  Imperial College London\\
  London, UK \\
  \texttt{n.ezaz-nikpay24@imperial.ac.uk} \\
  \And
  Pranav Mamidanna \\
  Imperial College London\\
  London, UK \\
  \texttt{p.mamidanna22@imperial.ac.uk} \\
  \And
  Shihan Ma \\
  Imperial College London\\
  London, UK \\
  \texttt{s.ma21@imperial.ac.uk} \\
  \And
  Silvia Muceli \\
  Chalmers University of Technology\\
  Gothenburg, Sweden \\
  \texttt{muceli@chalmers.se} \\
  \And
  Dario Farina \\
  Imperial College London\\
  London, UK \\
  \texttt{d.farina@imperial.ac.uk} \\
}

\begin{document}

\maketitle

\begin{abstract}
Decomposing surface electromyography (EMG) into the spike trains of individual motor neurons is a long-standing inverse problem and a key step toward motor-neuron-driven neural interfaces such as prosthetics and exoskeletons. The standard approach, independent component analysis (ICA) of the multichannel signal, assumes that the mixing from neurons to electrodes is stationary in time. This assumption fails during movement, when volume-conductor deformation makes the mixing time-varying, and current decomposition algorithms are correspondingly restricted to isometric contractions. We introduce HarmonICA, a quasi-time-invariant ICA formulation in which a static linear separator is preceded by a learned, low-rank, time-varying invertible transformation. The separator is trained with an independence loss on the uncompensated projection, and the transformation with a stationarity loss on the recovered source. Gradients are not shared between the two, so source extraction remains a classical linear-ICA problem while non-stationary distortion is absorbed by the transformation. The closed-form inverse of the transformation enables per-spike subtraction with a time-varying template during sequential peel-off. On a public simulated benchmark of dynamic high-density EMG with ground-truth spike trains, the method outperforms five adaptive ICA baselines at every recall threshold, recovering more units at a higher accuracy.
\end{abstract}

\section{Introduction}

The discharge timings of spinal motor neurons, recovered from multichannel electromyography (EMG), provide a direct readout of motor intention and are a target signal for next-generation neural interfaces such as prosthetic limbs, exoskeletons, and gesture-based human-computer interaction~\cite{kapelner2020neuro, eden2022principles}. The recent maturation of high-density invasive and surface electrode arrays has supplied the hardware needed to record from large pools of motor units in parallel~\cite{muceli2022blind, ctrl2024generic}, but blind source separation (BSS) of the resulting multichannel signal into constituent spike trains has not kept pace~\cite{chen2025high}. Existing algorithms~\cite{negro2016multi, holobar2007multichannel, grison2024particle, grison2025unlocking} achieve reliable decomposition only on laboratory contractions that are isometric and free of electrode movement. The remaining barrier to real-world deployment is robustness to non-stationary signal.

Non-stationarity in EMG arises from a few well-understood physical sources: during movement the electrode position relative to the active muscle fibers changes, the skin-electrode impedance fluctuates, and the volume conductor itself deforms~\cite{negro2016multi, muceli2015accurate, huang2020low}; over sustained contractions, muscle fatigue also alters MUAP shape. The resulting time variation in the mixing process violates the stationarity assumption underlying linear ICA and convolutive BSS~\cite{hyvarinen2000independent, holobar2007multichannel}, and a static separator that fits one stretch of the recording cleanly fails outside that stretch. Existing remedies all reduce to processing locally stationary windows or running adaptive updates against incoming data~\cite{glaser2018motor, chen2020adaptive, kramberger2021prediction, yeung2024adaptive, guerra2024adaptive}; accuracy degrades once the mixing changes faster than the chosen segment or adaptation horizon.

An alternative is to treat the time-varying mixing as a general nonlinear one, but unconstrained nonlinear ICA is not identifiable in the unsupervised regime; recovery of the sources is determined only up to an arbitrary invertible nonlinear transformation, unless auxiliary information such as temporal indices, class labels, or contextual covariates is available~\cite{hyvarinen2018nonlinear, zheng2022identifiability, hyvarinen2023nonlinear}. That kind of auxiliary information is rarely available for EMG recorded outside controlled experimental conditions.

The physics of surface EMG suggests a more constrained model class. Electrode shift, impedance change, and volume-conductor deformation change an underlying linear convolutive mixture smoothly as the joint moves, and the resulting waveform variation empirically lives in a low-dimensional subspace~\cite{naik2016principal, huang2020low}. We therefore keep source extraction linear and carry all time variation in a separate, invertible compensation stage. The result is HarmonICA: a static linear separator preceded by a learned, low-rank, time-varying invertible transformation parameterised as a Woodbury-style perturbation of identity~\cite{lu2020woodbury}, trained with decoupled gradients so the separator remains the optimum of an unmodified linear-ICA objective. A two-component GMM stationarity loss replaces the likelihood objective of normalizing flows and the independence objective of nonlinear ICA, and neural peel-off subtracts a time-varying per-spike template via the closed-form Woodbury inverse, replacing classical peel-off's static template~\cite{chen2016progressive}. On the MUniverse simulator~\cite{mamidanna2025muniverse}, HarmonICA preserves accurate motor unit tracking in regimes where current methods degrade: it recovers more high-quality motor units than every prior adaptive ICA baseline at every recall threshold and on both array densities, including units that the FastICA seed alone does not recover at high accuracy. Unlike the windowed and online baselines, the compensator models the full angular sweep within a single optimisation, so accuracy does not degrade with the segmentation or adaptation horizon.

\section{Related Work}

\paragraph{Adaptive blind source separation for EMG.}
Linear ICA and convolutive BSS form the standard baselines for high-density EMG (HD-EMG) decomposition. Convolutive BSS~\cite{negro2016multi} and Convolution Kernel Compensation~\cite{holobar2007multichannel} both reduce the convolutive EMG model to an instantaneous linear mixture by time extension and recover sources by exploiting their sparsity, and subsequent refinements have improved their yield and convergence~\cite{grison2024particle, grison2025unlocking}. All assume approximately stationary mixing, which restricts them in practice to isometric contractions~\cite{chen2025high}. Two families have been proposed to relax this assumption: window-based methods that segment the recording into locally stationary blocks~\cite{glaser2018motor}, and online methods that update separation vectors or related parameters as new data arrives~\cite{chen2020adaptive, kramberger2021prediction, yeung2024adaptive, guerra2024adaptive}. Both improve robustness to mild non-stationarities but operate on a local segment rather than the full trace, so accuracy degrades once the mixing changes faster than the segmentation or adaptation horizon allows. HarmonICA differs in that it models non-stationarity directly within a single optimisation over the entire recording, removing the local-stationarity assumption altogether.

\paragraph{Nonlinear ICA and identifiability.}
Independent component analysis~\cite{hyvarinen2000independent} in its classical, linear form is identifiable under standard non-Gaussianity assumptions. Its nonlinear generalisation is not~\cite{khemakhem2020variational}: in the fully unsupervised setting, source recovery is determined only up to an arbitrary invertible nonlinear transformation. Identifiability can be restored under various forms of side information, including auxiliary variables and contrastive objectives~\cite{hyvarinen2018nonlinear}, sparsity assumptions on the latent sources~\cite{zheng2022identifiability}, and the broader structural conditions surveyed in~\cite{hyvarinen2023nonlinear}. These guarantees are sensitive to the modelling assumptions they invoke, and the auxiliary information they require is rarely available for neurophysiological data collected outside the laboratory. HarmonICA sidesteps the issue by keeping the source-extraction step linear and carrying all time variation in a separate compensator, whose only nonlinearity is the network that schedules it, and that the independence objective never touches.

\paragraph{Invertible neural networks and normalizing flows.}
Normalizing flows~\cite{dinh2014nice, papamakarios2021normalizing} are a family of invertible neural-network layers, typically used for generative modelling and density estimation. We borrow one such layer for use as the compensator: the Woodbury layer of~\cite{lu2020woodbury}, an invertible affine map $A = I + UV^\top$ with a closed-form inverse via the Woodbury identity. The compensator is trained against a stationarity loss on the recovered source distribution rather than a likelihood. Other invertible affine parameterisations exist, including the LU-factored $1{\times}1$ convolution of Glow~\cite{kingma2018glow}, but these are static; making them time-varying would require the conditioning network to produce $O(K^2)$ parameters per timepoint, against $O(rK)$ for the Woodbury form (Section~\ref{sec:model}). Flow architectures have also been used to obtain identifiable nonlinear ICA under structural constraints~\cite{sorrenson2020disentanglement}; here, identifiability rests on the linear separator, and the compensator only needs to contain the inverse of the drift (Section~\ref{sec:objectives}).

\section{Method}

HarmonICA recovers motor unit spike trains from a non-stationary EMG recording one source at a time. Each extraction couples a static linear separator with a learned, low-rank time-varying compensator (Section~\ref{sec:model}); the two are fit by alternating between an independence loss on the separator and a stationarity loss on the compensator (Sections~\ref{sec:objectives}--\ref{sec:altopt}, Algorithm~\ref{alg:harmonica}); once a source converges, its contribution is removed by neural peel-off so that subsequent extractions see a clean residual (Section~\ref{sec:peeloff}).

\subsection{Preliminaries}
\label{sec:method_prelim}

\paragraph{Time-varying convolutive model.}
With $t$ discrete at the sampling rate $f_s$, the $N$-channel recording is a convolutive mixture of $M$ motor unit spike trains $\mathbf{s}(t)\in\mathbb{R}^M$ through motor unit action potentials (MUAPs) of $Q$ samples whose shape changes with time,
\begin{equation}
\mathbf{x}(t) = \textstyle\sum_{k=0}^{Q-1} H(t,k)\,\mathbf{s}(t-k) + \boldsymbol{\omega}(t), \qquad \mathbf{x}(t)\in\mathbb{R}^N.
\label{eq:conv_model}
\end{equation}
As in convolutive BSS for EMG~\cite{holobar2007multichannel, negro2016multi, klotz2025revisiting}, we stack $L$ delayed copies of each channel, $\mathbf{x}_s(t) = [\mathbf{x}(t)^\top,\ldots,\mathbf{x}(t-L+1)^\top]^\top\in\mathbb{R}^{NL}$, which gives the instantaneous model $\mathbf{x}_s(t) = \mathcal{H}(t)\,\bar{\mathbf{s}}(t) + \boldsymbol{\omega}_s(t)$ in the extended source $\bar{\mathbf{s}}(t)\in\mathbb{R}^{M(L+Q-1)}$ of delayed copies $s_j(t-l)$, $0\leq l\leq L+Q-2$. $\mathcal{H}(t)$ keeps the shift structure of the extension but is block-Toeplitz only where $H(t,k)$ does not vary with $t$. Principal component analysis (PCA) whitening $W_{\mathrm{pre}}\in\mathbb{R}^{K\times NL}$ acts on the measurements, never on the sources, and keeps the top $K$ components of the extended covariance, discarding the noise-dominated subspace that destabilises ICA in high dimensions:
\begin{equation}
\tilde{\mathbf{x}}(t) = W_{\mathrm{pre}}\,\mathbf{x}_s(t) = \tilde{H}(t)\,\bar{\mathbf{s}}(t) + \tilde{\boldsymbol{\omega}}(t), \qquad \tilde{H}(t) = W_{\mathrm{pre}}\mathcal{H}(t)\in\mathbb{R}^{K\times M(L+Q-1)}.
\label{eq:source_estimate}
\end{equation}
We use $K=128$, with $L=8$ on the benchmark and $L=16$ on real data (Section~\ref{sec:real}, supplementary §\ref{supp:real}).

\paragraph{Dimensions.}
If the mixing has full column rank, $M(L+Q-1)\leq NL$, one can take $K=M(L+Q-1)$; $\tilde{H}$ is then square and invertible, and where the mixing is static, convolutive identifiability holds up to permutation and a per-unit filter~\cite{castella2009generalized}. Without full column rank, deflation still needs only one projection per unit with negligible response to the other units~\cite{delfosse1995adaptive}, a per-unit separability condition~\cite{klotz2025revisiting}. $K$ therefore budgets the effective dimension of each unit's block of delayed MUAPs, not the number of units.

\paragraph{Non-stationarity.}
Joint rotation changes fibre lengths and volume-conductor geometry, so $H(t,k)$ moves with $t$. We model this as an invertible drift of the whitened mixing relative to an anchor window over which the mixing is static,
\begin{equation}
\tilde{H}(t) = D(t)\,\tilde{H}^{\ast}, \qquad D(t)\in\mathbb{R}^{K\times K}, \quad D(t)=I \text{ on the anchor window},
\label{eq:drift}
\end{equation}
and Section~\ref{sec:model} gives $D(t)$ its low-rank structure.

\paragraph{Sources.}
Independence is required only across units. Delayed copies of one spike train are dependent by construction (refractoriness forces silence after each spike), so the extended model is multidimensional ICA~\cite{cardoso1998multidimensional}; the lag structure reduces the within-unit ambiguity to the per-unit filter above, which leaves firing times unchanged. Across units, common synaptic drive makes spike trains dependent but approximately uncorrelated, which is what whitening needs to make the columns of $\tilde{H}^{\ast}$ close to orthogonal~\cite{klotz2025revisiting}. Each spike train has a bimodal prior on the pre-mixing samples $s_j(t)$, not on PCA coordinates, with point masses at a baseline $c_{\text{null}}$ and a spike amplitude $c_{\text{spike}}$,
\begin{equation}
p(s_j) = (1-\rho_j)\,\delta(s_j - c_{\text{null}}) + \rho_j\,\delta(s_j - c_{\text{spike}}),
\label{eq:spike_model}
\end{equation}
where $\rho_j\in(0,1)$ is the firing rate and $\delta(\cdot)$ the Dirac delta. The task is to recover each $s_j(t)$ without supervision when $D(t)\neq I$.

\begin{figure}[t]
  \centering
  \includegraphics[width=\linewidth]{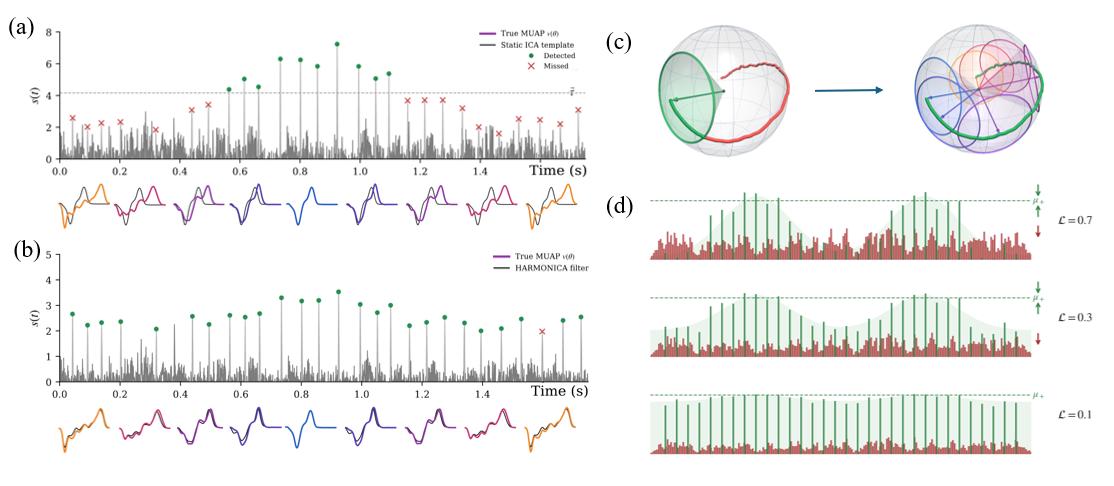}
  \caption{\textbf{(a)} A static ICA filter $\mathbf{w}_j$ recovers spikes only in the narrow local window where the mixing matches it; outside this window, the source approaches the noise floor and roughly half of the ground-truth spikes are missed. \textbf{(b)} HarmonICA's time-varying compensator $A_j(t)$ rotates the filter (equivalently, the signal space) so that the same source extracts cleanly across the full recording. \textbf{(c)} The learned rotation tracks the angle-driven MUAP variation. \textbf{(d)} The two-component GMM stationarity loss propagates this rotation reliably; classical FastICA non-Gaussianity contrasts are less reliable on this benchmark (see supplementary §\ref{supp:abl:loss}).}
  \label{fig:motivation}
\end{figure}

\subsection{Quasi-Time-Invariant Source Model}
\label{sec:model}

A static separator cannot invert a time-varying mixing (Fig.~\ref{fig:motivation}a), but because the mixing is linear at every instant, a linear time-varying separator can (Fig.~\ref{fig:motivation}b,c). We factor it into a time-varying invertible compensator $A_j(t)\in\mathbb{R}^{K\times K}$ that undoes the drift and a static separator $\mathbf{w}_j\in\mathbb{R}^{K}$ that extracts source $j$,
\begin{equation}
\hat{s}_j(t) = \mathbf{w}_j^\top\,A_j(t)\,\tilde{\mathbf{x}}(t).
\label{eq:compensation}
\end{equation}
$A_j$ is scheduled by an exogenous variable (time or joint angle) through a small network, a linear parameter-varying (LPV) construction~\cite{shamma1990analysis, shamma1991guaranteed}; we call the model \emph{quasi-time-invariant}. Keeping $\mathbf{w}_j$ static is deliberate: a fully time-varying separator can switch between sources mid-recording, the discontinuity problem of dynamic source extraction~\cite{koldovsky2021dynamic}, whereas with $\mathbf{w}_j$ fixed by the seed and $A_j$ continuous, only the scale and effective delay of the extracted unit can vary (supplementary §\ref{supp:theory:ambiguities}).

\paragraph{Low-rank compensator.} The form of $A_j$ follows from the drift of Eq.~\ref{eq:drift}. If the drift is low-rank in a fixed subspace, $D(t) = I + U(t)V^\top$ with $U(t), V\in\mathbb{R}^{K\times r}$, the Woodbury identity~\cite{hager1989updating} gives $D(t)^{-1} = I - U(t)\big(I_r + V^\top U(t)\big)^{-1}V^\top$, which has the same form. We therefore learn
\begin{equation}
A_j(t) = I + U_j(t)\,V_j^\top, \qquad U_j(t), V_j\in\mathbb{R}^{K\times r},
\label{eq:lowrank}
\end{equation}
a class that contains $D(t)^{-1}$ exactly, so that $\mathbf{w}_j$ separates at every $t$ as it would on the anchor (supplementary §\ref{supp:theory:results}, which also bounds the leakage when the drift is only approximately low-rank). We use $r=3$ (supplementary §\ref{supp:abl:rank}). The same identity gives $A_j(t)^{-1}$ in closed form at $O(rK)$ cost per timepoint, reused in neural peel-off (Section~\ref{sec:peeloff}), and the network needs only $rK$ outputs per timestep, against $O(K^2)$ for a full-rank compensator~\cite{lu2020woodbury}.

\paragraph{Conditioning.} The time-varying factor is $U_j(t) = \mathrm{NN}_j(\boldsymbol{\gamma}(t);\phi_j)$, and $V_j$ is a learned per-source matrix shared across time. By default $\boldsymbol{\gamma}(t)$ is a fixed bank of sinusoids at periods $T_1,\ldots,T_P$, a blind positional encoding in the spirit of neural radiance fields~\cite{mildenhall2021nerf},
\begin{equation}
\boldsymbol{\gamma}(t) = \big(\sin(2\pi t/T_1),\, \cos(2\pi t/T_1),\, \ldots,\, \sin(2\pi t/T_P),\, \cos(2\pi t/T_P)\big).
\label{eq:positional_encoding}
\end{equation}
When a covariate of the drift is measured, such as the joint angle, $\boldsymbol{\gamma}(t)$ can be set to it instead (``HarmonICA ($\theta$)'', Table~\ref{tab:yield_3bin}; supplementary §\ref{supp:abl:pe_vs_angle}). Setting $U_j(t)=0$ recovers standard linear ICA.

\subsection{Independence and Stationarity Losses}
\label{sec:objectives}

The separator $\mathbf{w}_j$ and the compensator $\phi_j$ are trained against two complementary losses. The separator is fit by an independence loss $L_I$ on the \emph{uncompensated} projection $\mathbf{w}_j^\top\tilde{\mathbf{x}}(t)$; the compensator is fit by a stationarity loss $L_N$ on the compensated source $\hat{s}_j(t)$ with $\mathbf{w}_j$ detached. The separator is therefore the optimum of an unmodified linear-ICA objective on $\tilde{\mathbf{x}}$, and the compensator only has to undo the drift. Under full column rank ($K=M(L+Q-1)$), an anchor window, single-delay dominance within each unit~\cite{klotz2025revisiting} and the fixed-subspace drift of Section~\ref{sec:model}, the supplement shows that a static separator and a compensator in this class exist that recover each unit up to scale and delay at every $t$, and that fitting the separator on the whole recording adds a bias of at most the order of the mean drift. These are existence results, not convergence guarantees, which is why the FastICA warm start matters (supplementary §\ref{supp:abl:headline}).

\paragraph{Independence loss $L_I$.}
$L_I$ is a projection-pursuit contrast on the uncompensated projection $u_j(t) = \mathbf{w}_j^\top\tilde{\mathbf{x}}(t)$ with $f(s) = \mathrm{sign}(s)\,|s|^e$, where larger exponents $e$ weight the largest excursions more heavily and so reward increasingly sparse sources (we use $e=3$),
\begin{equation}
L_I(\mathbf{w}_j) = -\,\mathbb{E}\!\left[\mathrm{sign}(u_j(t))\,|u_j(t)|^e\right],
\label{eq:independence_loss}
\end{equation}
where $u_j(t)$ is standardised to zero mean and unit variance before evaluating $f$, which removes the trivial scale degeneracy.

\paragraph{Stationarity loss $L_N$.}
If the compensator absorbs all non-stationarity, the empirical distribution of $\hat{s}_j(t)$ should match the bimodal prior of Eq.~\ref{eq:spike_model}: tight clusters around $c_{\text{null}}$ and $c_{\text{spike}}$. Residual non-stationarity widens these clusters. We measure the residual by relaxing the prior to a two-component Gaussian mixture model (GMM; Fig.~\ref{fig:motivation}d),
\begin{equation}
q(\hat{s}_j) = (1-\rho_j)\,\mathcal{N}\!\big(\hat{s}_j\mid\mu_{\text{null}},\sigma^2_{\text{null}}\big) + \rho_j\,\mathcal{N}\!\big(\hat{s}_j\mid\mu_{\text{spike}},\sigma^2_{\text{spike}}\big),
\label{eq:gmm}
\end{equation}
whose clusters are re-estimated from the current $\hat{s}_j(t)$ every $K_g$ compensation steps: peaks in the upper cluster of a two-means split of peak heights form the spike set, $1000$ random non-spike samples form the null set, and $\sigma^2_{\text{spike}}$, $\sigma^2_{\text{null}}$ are the sample variances of $\hat{s}_j$ over each. The compensator loss is the root-summed within-cluster spread,
\begin{equation}
L_N(\phi_j) = \big(\sigma^2_{\text{spike}} + \max(\sigma^2_{\text{null}},\,\bar{\sigma}^2_{\text{null}})\big)^{1/2},
\label{eq:compensation_loss}
\end{equation}
where $\bar{\sigma}^2_{\text{null}}$ is the same null-sample variance evaluated on the \emph{uncompensated} projection $\mathbf{w}_j^\top\tilde{\mathbf{x}}$. The $\max$ floors the null-cluster variance at its uncompensated value, preventing a degenerate solution in which the compensator collapses the baseline at the cost of inflating spike-cluster variance.

\begin{figure}[tb]
\centering
\resizebox{\linewidth}{!}{%
\begin{tikzpicture}[
  font=\small,
  >=Stealth,
  every node/.append style={align=center},
  sig/.style    ={draw, rounded corners=1.5pt, fill=blue!8,    inner sep=2.5pt, minimum height=6mm, minimum width=8mm},
  comp/.style   ={draw, rounded corners=1.5pt, fill=orange!22, inner sep=2.5pt, minimum height=6mm},
  par/.style    ={draw, rounded corners=1.5pt, fill=gray!15,   inner sep=2.5pt, minimum height=6mm, minimum width=8mm},
  upd/.style    ={draw, rounded corners=1.5pt, fill=yellow!30, inner sep=2.5pt, minimum height=6mm, minimum width=8mm, font=\small\bfseries},
  los/.style    ={draw, rounded corners=1.5pt, fill=green!22,  inner sep=2.5pt, minimum height=6mm, font=\footnotesize},
  op/.style     ={draw, circle, fill=white, inner sep=0pt, minimum size=4.2mm, font=\scriptsize},
  panel/.style  ={draw=gray!55, dashed, rounded corners=4pt, inner sep=3mm, fill=white},
  arr/.style    ={-{Stealth[length=1.6mm]}, semithick},
  garr/.style   ={-{Stealth[length=1.6mm]}, semithick, red!70!black, densely dashed},
  plabel/.style ={fill=white, inner xsep=3pt, inner ysep=1pt, anchor=west, font=\footnotesize\bfseries},
  note/.style   ={font=\scriptsize\itshape, gray!40!black}
]

\node[sig]                    (g)   {$\bm{\gamma}(t)$\\[-2pt]\scriptsize sin/cos bank};
\node[sig, below=10mm of g]   (x)   {$\tilde{\mathbf{x}}(t)$\\[-2pt]\scriptsize whitened};
\node[comp, right=5mm of g]   (nn)  {$\mathrm{NN}_j(\cdot\,;\phi_j)$};
\node[par, below=3mm of nn]   (V)   {$V_j$};
\node[comp, right=11mm of nn] (Aj)  {$A_j(t)\!=\!I{+}U_j(t)V_j^{\!\top}$};
\node[op,   right=10mm of Aj] (mA)  {$\times$};
\node[op,   right=4mm of mA]  (mW)  {$\times$};
\node[par,  below=10mm of mW] (w)   {$\mathbf{w}_j^{\top}$};
\node[sig,  right=4mm of mW, fill=blue!14] (s) {$\hat{s}_j(t)$};

\draw[arr] (g) -- (nn);
\draw[arr] (nn) -- node[above, font=\scriptsize, pos=0.45] {$U_j(t)$} (Aj);
\draw[arr] (V.east) -| (Aj.south);
\draw[arr] (Aj) -- (mA);
\draw[arr] (x.east) -| (mA.south);
\draw[arr] (mA) -- (mW);
\draw[arr] (w.north) -- (mW.south);
\draw[arr] (mW) -- (s);

\node[fit=(g)(x)(nn)(V)(Aj)(s)(w), inner sep=3mm] (panA) {};

\node[sig, below=10mm of x.south west, anchor=north west] (xb1) {$\tilde{\mathbf{x}}(t)$};
\node[font=\scriptsize\itshape, gray!35!black, left=2mm of xb1, anchor=east] (lab1) {every step\\[-2pt]($A_j\!=\!I$)};
\node[upd, right=3mm of xb1] (wb1) {$\mathbf{w}_j$};
\node[op,  right=2mm of wb1] (mB1) {$\times$};
\node[sig, right=2mm of mB1] (ub1) {$u_j(t)$};
\node[los, right=2.5mm of ub1] (LI)  {$L_I$ (Eq.~\ref*{eq:independence_loss})};
\node[font=\footnotesize, right=3mm of LI, anchor=west] (gradI) {$\Rightarrow\; \mathbf{w}_j\!\mathrel{-}\!=\eta_w\nabla_{\!\mathbf{w}_j}\!L_I$};

\draw[arr] (xb1) -- (wb1);
\draw[arr] (wb1) -- (mB1);
\draw[arr] (mB1) -- (ub1);
\draw[arr] (ub1) -- (LI);

\node[sig, below=8mm of xb1.south west, anchor=north west] (xb2) {$\tilde{\mathbf{x}}(t)$};
\node[font=\scriptsize\itshape, gray!35!black, left=2mm of xb2, anchor=east] (lab2) {after $T_w$\\[-2pt]re-split /\,$K_g$};
\node[comp, right=3mm of xb2] (Ab2) {$A_j(t)$};
\node[upd,  below=3.5mm of Ab2, font=\small\bfseries] (phi) {$\phi_j$};
\node[par,  right=2.5mm of Ab2] (wb2) {$\mathbf{w}_j$\\[-3pt]\scriptsize\itshape detach};
\node[op,   right=2.5mm of wb2] (mB2) {$\times$};
\node[sig,  right=2.5mm of mB2] (sb2) {$\hat s_j(t)$};
\node[los,  right=2.5mm of sb2, fill=green!22] (LN) {GMM\,$\to L_N$};
\node[font=\footnotesize, right=3mm of LN, anchor=west] (gradC) {$\Rightarrow\; \phi_j\!\mathrel{-}\!=\eta_\phi\nabla_{\!\phi_j}\!L_N$};

\draw[arr] (xb2) -- (Ab2);
\draw[arr] (phi.north) -- (Ab2.south);
\draw[arr] (Ab2) -- (wb2);
\draw[arr] (wb2) -- (mB2);
\draw[arr] (mB2) -- (sb2);
\draw[arr] (sb2) -- (LN);

\node[fit=(xb1)(lab1)(xb2)(lab2)(gradI)(gradC)(phi), inner sep=2.5mm] (panB) {};

\node[sig, below=8mm of panB.south west, anchor=north west, fill=blue!14] (sc) {$\hat s_j(t)$};
\node[par, right=2.5mm of sc] (thr) {GMM\\[-2pt]\scriptsize threshold};
\node[sig, right=2.5mm of thr] (tj) {$\{t_{j,i}\}$};
\node[comp, right=4mm of tj, fill=orange!22] (sta)  {$\mathrm{STA}_j^{\mathrm{comp}}(\tau)$};
\node[comp, right=14mm of sta, fill=orange!22] (Ainv) {$A_j(t_{j,i}\!+\!\tau)^{-1}$};
\node[op,   right=2.5mm of Ainv] (sub) {$-$};
\node[sig,  right=2.5mm of sub, fill=blue!8] (resid) {$\tilde{\mathbf{x}}\!\leftarrow\!\tilde{\mathbf{x}}\!-\!\hat{\mathbf{x}}_j$};
\node[font=\footnotesize, right=3mm of resid, anchor=west] (lab3) {$\Rightarrow$\;restart Alg.~\ref*{alg:harmonica}\\[-2pt]\itshape\scriptsize on residual ($j{+}1$)};

\draw[arr] (sc) -- (thr);
\draw[arr] (thr) -- (tj);
\draw[arr] (tj) -- (sta);
\draw[arr] (sta) -- node[above, font=\scriptsize, pos=0.5] (psi) {per spike $i$} (Ainv);
\draw[arr] (Ainv) -- (sub);
\draw[arr] (sub) -- (resid);

\node[fit=(sc)(thr)(tj)(sta)(Ainv)(sub)(resid)(lab3)(psi), inner sep=2.5mm] (panC) {};

\path[overlay] let
  \p1=(panA.west), \p2=(panB.west), \p3=(panC.west),
  \p4=(panA.east), \p5=(panB.east), \p6=(panC.east)
  in
  coordinate (panL) at ({min(\x1,\x2,\x3)},0)
  coordinate (panR) at ({max(\x4,\x5,\x6)},0);

\begin{pgfonlayer}{background}
  \path[draw=gray!55, dashed, rounded corners=4pt, fill=white]
    let \p1=(panA.north), \p2=(panA.south), \p3=(panL), \p4=(panR)
    in (\x3,\y1) rectangle (\x4,\y2);
  \path[draw=gray!55, dashed, rounded corners=4pt, fill=white]
    let \p1=(panB.north), \p2=(panB.south), \p3=(panL), \p4=(panR)
    in (\x3,\y1) rectangle (\x4,\y2);
  \path[draw=gray!55, dashed, rounded corners=4pt, fill=white]
    let \p1=(panC.north), \p2=(panC.south), \p3=(panL), \p4=(panR)
    in (\x3,\y1) rectangle (\x4,\y2);
\end{pgfonlayer}

\path let \p1=(panL), \p2=(panA.north) in coordinate (panA_NW) at (\x1,\y2);
\path let \p1=(panL), \p2=(panB.north) in coordinate (panB_NW) at (\x1,\y2);
\path let \p1=(panL), \p2=(panC.north) in coordinate (panC_NW) at (\x1,\y2);
\node[plabel] at (panA_NW) {(a)\;\;Compensated source $\hat s_j(t)\!=\!\mathbf{w}_j^{\top}A_j(t)\,\tilde{\mathbf{x}}(t)$};
\node[plabel] at (panB_NW) {(b)\;\;Alternating training\;\;{\rmfamily\itshape for a fixed $T_{\max}$ epochs}};
\node[plabel] at (panC_NW) {(c)\;\;Neural peel-off \& sequential extraction};

\end{tikzpicture}%
}
\caption{\textbf{HarmonICA single-source extraction.} \textbf{(a)} Forward pass: time encoding $\bm\gamma(t)$, per-source network $\mathrm{NN}_j$, low-rank invertible compensator $A_j(t)=I+U_j(t)V_j^\top$ applied to the whitened signal, and static separator $\mathbf{w}_j$. \textbf{(b)} Decoupled training: independence step on $\mathbf{w}_j$ ($L_I$) with the compensator off, and after a $T_w$-step warmup a compensation step on $\phi_j$ ($L_N$) with $\mathbf{w}_j$ detached and the spike/null sets re-split every $K_g$ steps. \textbf{(c)} Per-spike peel-off: subtract the spike-triggered average mapped back through $A_j(t_{j,i}+\tau)^{-1}$, then restart on the residual.}
\label{fig:training}
\end{figure}

\begin{algorithm}[t]
\caption{HarmonICA: extraction of a single source $j$.}
\label{alg:harmonica}
\begin{algorithmic}[1]
\Require Whitened signal $\tilde{\mathbf{x}}(t)$; time encoding $\boldsymbol{\gamma}(t)$; epoch budget $T_{\max}$; warmup $T_w$; spike/null re-split period $K_g$; learning rates $\eta_w,\eta_\phi$.
\State Initialise $\mathbf{w}_j$ from the $j$-th FastICA separation vector and $\phi_j$ such that $U_j(t)=0$, i.e., $A_j(t)=I$.
\For{$\text{step} = 1, 2, \ldots, T_{\max}$}
    \State \emph{Independence step:} compute $u_j(t) = \mathbf{w}_j^\top\tilde{\mathbf{x}}(t)$ \Comment{compensator switched off}
    \State $\mathbf{w}_j \gets \mathbf{w}_j - \eta_w\,\nabla_{\mathbf{w}_j} L_I(\mathbf{w}_j)$ \Comment{Eq.~\ref{eq:independence_loss}}
    \If{$\text{step} \geq T_w$}
        \State \emph{Compensation step:} compute $\hat{s}_j(t) = \mathbf{w}_j^\top A_j(t)\,\tilde{\mathbf{x}}(t)$
        \If{$\text{step} \bmod K_g = 0$}
            \State re-derive the spike/null sets from $\hat{s}_j(t)$
        \EndIf
        \State $\phi_j \gets \phi_j - \eta_\phi\,\nabla_{\phi_j} L_N(\phi_j)$,\; $\mathbf{w}_j$ detached \Comment{Eq.~\ref{eq:compensation_loss}}
    \EndIf
\EndFor
\State \Return $\mathbf{w}_j$, $\phi_j$, and spike times from the GMM-thresholded $\hat{s}_j(t)$.
\end{algorithmic}
\end{algorithm}

\subsection{Alternating Optimisation}
\label{sec:altopt}

$L_I$ and $L_N$ are minimised in alternation (Algorithm~\ref{alg:harmonica}, Fig.~\ref{fig:training}). The two gradients never mix, so alternation does not let $L_N$ steer $\mathbf{w}_j$; what it provides is that the spike/null sets feeding $L_N$ are re-derived from the current $\mathbf{w}_j$, so the compensator trains on labels that sharpen as the separator converges. Because the compensator sees only $L_N$, any transform that does not match the current $\mathbf{w}_j$ widens the GMM clusters instead of tightening them, so the compensator cannot recover sources the separator has not already identified. The compensator gradient is taken every step, but the spike/null sample indices that feed $L_N$ are refreshed only every $K_g$ steps so the objective is locally smooth in $\phi_j$. We delay compensation by $T_w$ warmup iterations: without this delay the compensator fits the projection before $\mathbf{w}_j$ has converged and the algorithm settles on a spurious solution. Both updates use RMSprop, and each source is trained for a fixed budget of $T_{\max}=200$ epochs (supplementary Table~\ref{tab:supp:impl}).

\subsection{Sequential Source Extraction with Neural Peel-Off}
\label{sec:peeloff}

After a source converges, spike times $\{t_{j,i}\}_{i=1}^{n_j}$ (indexed by spike $i$ over the $n_j$ accepted spikes of source $j$) are read off from $\hat{s}_j(t)$ using the GMM threshold, and its contribution is removed from the signal so that later extractions see the residual.

Classical peel-off subtracts a single static template per source~\cite{chen2016progressive}, which is faithful only under stationary mixing: when the mixing drifts, the spike waveform drifts with it, and a static template leaves spike-locked residuals that contaminate later extractions. We instead average in the compensated frame $\tilde{\mathbf{x}}^{\mathrm{comp}}(t) = A_j(t)\,\tilde{\mathbf{x}}(t)$, where the waveform is approximately stationary along the source's own projection ($L_N$ constrains only $A_j(t)^\top\mathbf{w}_j$; other directions are compensated only insofar as $V_j$ spans them), and map the average back to the whitened frame with the Woodbury inverse of $A_j$ at each spike's own time before subtracting:
\begin{align}
\mathrm{STA}_j^{\mathrm{comp}}(\tau) &= \tfrac{1}{n_j}\textstyle\sum_{i=1}^{n_j} \tilde{\mathbf{x}}^{\mathrm{comp}}(t_{j,i} + \tau), \label{equation:sta} \\
\hat{\mathbf{x}}_j^{(i)}(t_{j,i} + \tau) &= A_j(t_{j,i} + \tau)^{-1}\,\mathrm{STA}_j^{\mathrm{comp}}(\tau), \label{equation:sta_inverse} \\
\tilde{\mathbf{x}}(t_{j,i} + \tau) &\leftarrow \tilde{\mathbf{x}}(t_{j,i} + \tau) - \hat{\mathbf{x}}_j^{(i)}(t_{j,i} + \tau). \label{equation:peel_off}
\end{align}
Each Woodbury inverse evaluates analytically at $O(rK)$ per timepoint (supplementary §\ref{supp:woodbury_param}), an $O(K^2)$ reduction over the $O(K^3)$ cost of inverting a generic $K \times K$ matrix. The algorithm resumes Algorithm~\ref{alg:harmonica} on the residual; the extract--subtract cycle continues until every seed in the pool has been refined.

\section{Experiments}

\subsection{Dataset}
\label{sec:data}

We test HarmonICA on \emph{muniverse-benchmark-dynamic}, a simulated dataset built on the MUniverse stack~\cite{mamidanna2025muniverse} (NeuroMotion~\cite{ma_neuromotion_2023} volume-conductor coupled with BioMime~\cite{ma2022human} MUAP synthesis) with ground-truth spike trains for every simulated motor unit. The test split contains $21$ pools across $5$ subjects, rendered on two electrode grids ($70$- and $320$-channel) under $8$ angle/SNR conditions per grid, giving $168$ recordings per grid of $10$\,s at $f_s = 2048$\,Hz, with sinusoidal or triangular angle sweeps in $\pm 32.5^\circ$ or $\pm 65^\circ$ at $20$ or $25$\,dB SNR. Construction (within-pool crosstalk filter, force carrier, val split) is described in supplementary §\ref{supp:benchmark}.

\paragraph{Stratification.}
Decomposition difficulty scales with motor-unit density, so we stratify recordings by their number of ground-truth motor units, $n_{\text{gt}}$, into three tiers (low $n_{\text{gt}} < 40$, medium $40$--$69$, high $\geq 70$). Per-recording $n_{\text{gt}}$ spans $23$--$101$, giving $40$ / $80$ / $48$ recordings per tier per grid. Both grids decompose the same pools, so each grid contains the same $9{,}448$ ground-truth motor units, partitioned $1{,}120$ / $4{,}200$ / $4{,}128$ across tiers; the denominators for the yield counts in Table~\ref{tab:yield_3bin}.

\subsection{Metrics}
\label{sec:metrics}

The headline metric is per-source F$_1 = 2PR/(P+R)$, where $P$ and $R$ are the precision and recall of the extracted spike train against its best-matching ground-truth motor unit, with a $\pm 2$\,ms matching tolerance and a $\pm 25$\,ms shift-alignment search performed once per match. Each ground-truth unit is counted at most once, at its best-matching source, so repeated extractions of a unit do not add to yield. All numbers in Section~\ref{sec:results} aggregate over recordings and over five FastICA initialisation seeds per recording, so each row in Table~\ref{tab:yield_3bin} pools $168 \times 5 = 840$ extractions.

\subsection{Implementation}
\label{sec:impl}

\paragraph{HarmonICA pipeline.}
HarmonICA is implemented in PyTorch and runs on a single NVIDIA A40 GPU. Each recording is Toeplitz-extended, PCA-projected to $K=128$, and ZCA-whitened; FastICA seeds up to $108$ candidate sources, each refined by Algorithm~\ref{alg:harmonica} with the rank-$3$ Woodbury compensator of Eq.~\ref{eq:lowrank}, zero-initialised so $A_j(t)=I$ at training start. A recording takes ${\sim}45$\,s, against ${\sim}28$\,s for batch FastICA. Full operating-point values (sinusoidal periods, learning rates, runtime) are in supplementary §\ref{supp:impl}.

\paragraph{Baselines.}
We compare HarmonICA against a linear ICA reference and five adaptive baselines. The ``ICA / cBSS'' reference is the unrefined FastICA seed inside the HarmonICA pipeline, read off before the compensator has acted; comparing against it isolates the contribution of the rotation-and-compensation stage from the upstream extraction, since both methods share the same seed by construction. The five adaptive baselines, Chen~2020~\cite{chen2020adaptive}, Kramberger~2021~\cite{kramberger2021prediction}, Yeung~2024~\cite{yeung2024adaptive}, Glaser~2018~\cite{glaser2018motor} and Mendez Guerra~2024~\cite{guerra2024adaptive}, each obtain their initial separation vectors from the FastICA stage of our pipeline (after Toeplitz extension, PCA and ZCA whitening) run on a short calibration window rather than on HarmonICA's $10$\,s seed pool, so the upstream extraction differs from the ICA / cBSS reference. Chen 2020 and Yeung 2024 are reimplemented from the original algorithm pseudocode with two minor numerical-stability deviations from the published configurations (supplementary §\ref{supp:baseline_deviations}); Glaser 2018 and Kramberger 2021 are paper-faithful reimplementations. Mendez Guerra 2024 runs the authors' adapt\_decomp v1.0.0 engine~\cite{mendezguerra2024adaptcode} through a wrapper and harness supplied by the method's first author, with the same calibration stack as Chen 2020 and Yeung 2024 on the first $1.5$\,s and a $10$--$1000$\,Hz input band-pass. Its three online hyperparameters were tuned per grid on the disjoint validation split with an objective that uses no ground truth (supplementary §\ref{supp:baseline_deviations}). Stochastic baselines are run with the same five FastICA initialisation seeds as HarmonICA; Kramberger is deterministic given its calibration data and is reported on a single seed. Under controlled drift, HarmonICA matches linear ICA when nothing drifts and leads otherwise (supplementary §\ref{supp:sanity}).

\section{Results}
\label{sec:results}

\begin{figure}[t]
\centering
\includegraphics[width=1.\textwidth]{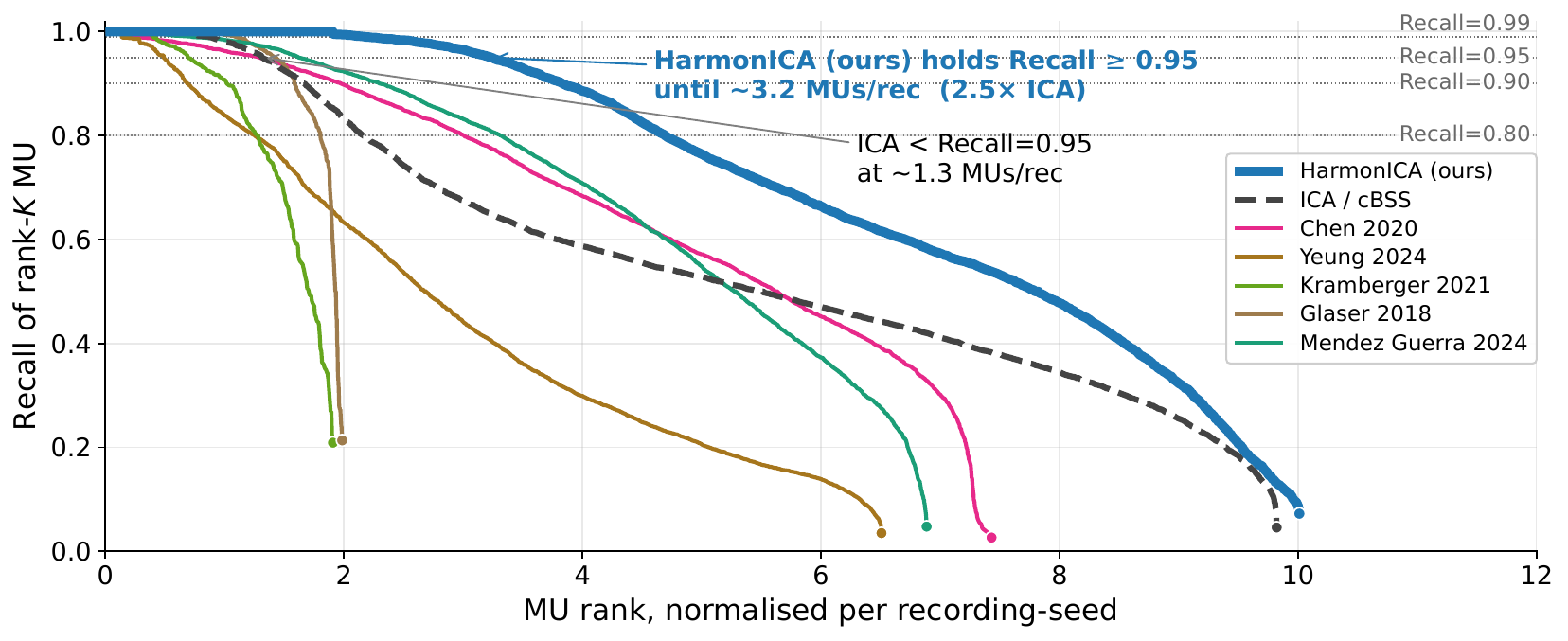}
\caption{\textbf{Yield $\times$ recall Pareto curve (ch320 benchmark, 168 recordings $\times$ 5 seeds).} $x$-axis: average MUs per recording-seed at recall $\geq R$; $y$-axis: recall threshold $R$. Higher and further-right is better. Dashed line: unrefined ICA / cBSS reference; filled circles: each method's coverage cap. Mendez Guerra 2024 is run with hyperparameters tuned on the validation split (§\ref{sec:impl}). Per-grid Pareto-crossover yields (including the 70-channel grid) are in supplementary Table~\ref{tab:supp:pe_vs_angle_pareto}.}
\label{fig:pareto_recall}
\end{figure}

\subsection{Main result}
On the \emph{muniverse-benchmark-dynamic} test split (336 recordings $\times$ 5 ICA-initialisation seeds, jointly covering the 70- and 320-channel electrode grids), HarmonICA recovers strictly more high-quality motor units than every adaptive baseline at every recall threshold and on both array densities (Fig.~\ref{fig:pareto_recall}, 320-channel grid); the same dominance holds at the tighter $F_1 \geq 0.95$ cut (Table~\ref{tab:yield_3bin}). At the standard $R \geq 0.95$ threshold the lift over the unrefined ICA / cBSS pipeline (the FastICA seed read out before the compensator has acted) is $2.5\times$ on the 320-channel grid and $2.2\times$ on the 70-channel grid. Mendez Guerra 2024, the strongest adaptive baseline at $R \geq 0.95$, exceeds the unrefined ICA seed at this threshold on the 320-channel grid ($1.62$ vs.\ $1.32$ MUs per recording-seed) but not on the 70-channel grid ($1.39$ vs.\ $1.42$), and falls below it on both grids at $R \geq 0.99$ (supplementary Table~\ref{tab:supp:pe_vs_angle_pareto}); HarmonICA stays at least $1.3$ MUs per recording-seed above Mendez Guerra 2024 at every threshold, and on the 320-channel grid holds $R \geq 0.99$ out to $2.17$ MUs per recording-seed, against $0.87$ for the ICA seed. Table~\ref{tab:yield_3bin} reports two HarmonICA configurations: ``HarmonICA (PE)'' uses the blind sinusoidal positional encoding of Eq.~\ref{eq:positional_encoding}; ``HarmonICA ($\theta$)'' substitutes the joint-angle trace for $\boldsymbol{\gamma}(t)$ when the angle is measured.

\subsection{Stratification by motor-unit density}
Table~\ref{tab:yield_3bin} reports yield at $F_1 \geq 0.95$ stratified by per-recording motor-unit count. HarmonICA (PE) recovers $476$ motor units on ch070 and $515$ on ch320, against $249$ for the unrefined ICA seed on each grid and $286 / 350$ for the strongest adaptive baseline, Mendez Guerra 2024, run with hyperparameters tuned on the validation split; it is the only adaptive baseline above the ICA seed on both grids, and the next strongest, Chen 2020, recovers $228 / 269$. The angle-conditioned HarmonICA ($\theta$) variant lifts this further to $502 / 571$, a modest gain at the cost of recording a synchronised joint-angle trace alongside the EMG. Identifiable nonlinear ICA conditioned on the same joint angle (TCL, iVAE) recovers far fewer units than linear ICA (supplementary §\ref{supp:nonlinear_ica}). The gap to ICA widens with density: the per-recording HarmonICA (PE) $-$ ICA gap grows monotonically on ch320 ($1.40 \to 1.51 \to 1.85$ MUs/rec from low to high) and peaks in the high tier on ch070, so the compensator yields its largest absolute improvement where the unrefined FastICA ceiling is tightest (supplementary Table~\ref{tab:supp:yield_per_rec}). Per-recording and seed-level breakdowns are in supplementary §\ref{supp:reproducibility}.

\begin{table}[t]
\centering
\caption{\textbf{Total motor units recovered at F$_1 \geq 0.95$,
stratified by per-recording motor-unit count $n_{\text{gt}}$.}
Cells are bin-level yield sums, mean $\pm$ s.d.\ across $5$
FastICA-initialisation seeds ($10$ for HarmonICA ($\theta$); Kramberger is deterministic and is
reported on a single seed; ``\,---'' in the s.d.\ position).
Mendez Guerra 2024 is run with hyperparameters tuned on the validation split (§\ref{sec:impl}).
Higher is better; bold marks the best method per cell.
Per-recording mean yields with across-recording dispersion are in
supplementary Table~\ref{tab:supp:yield_per_rec}; all HarmonICA $-$
baseline differences are significant at $p < 10^{-19}$ (paired
Wilcoxon signed-rank, supplementary §\ref{supp:stats}).}
\label{tab:yield_3bin}
\resizebox{\textwidth}{!}{%
\begin{tabular}{lccc|ccc}
\toprule
\multirow{2}{*}{Method} & \multicolumn{3}{c|}{\textbf{ch070} (70-channel)} & \multicolumn{3}{c}{\textbf{ch320} (320-channel)} \\
\cmidrule(lr){2-4} \cmidrule(lr){5-7}
 & $n_{\text{gt}} < 40$ & $n_{\text{gt}}\,40\!-\!69$ & $n_{\text{gt}} \geq 70$ & $n_{\text{gt}} < 40$ & $n_{\text{gt}}\,40\!-\!69$ & $n_{\text{gt}} \geq 70$ \\
\midrule
ICA / cBSS \cite{negro2016multi}                  & $36 \pm 2.5$ & $151 \pm 3.0$ & $62 \pm 1.9$ & $40 \pm 2.9$ & $142 \pm 2.9$ & $67 \pm 4.2$ \\
\midrule
\textbf{HarmonICA (PE, ours)}                     & $90 \pm 4.6$ & $252 \pm 24$ & $134 \pm 5.5$ & $96 \pm 2.2$ & $263 \pm 4.4$ & $156 \pm 5.3$ \\
\textbf{HarmonICA ($\theta$, ours)}               & $\mathbf{91 \pm 4.7}$ & $\mathbf{265 \pm 11}$ & $\mathbf{146 \pm 13}$ & $\mathbf{102 \pm 4.6}$ & $\mathbf{290 \pm 5.3}$ & $\mathbf{179 \pm 5.5}$ \\
\midrule
Chen 2020 \cite{chen2020adaptive}                 & $36 \pm 2.9$ & $133 \pm 4.3$ & $59 \pm 3.4$ & $39 \pm 4.0$ & $149 \pm 8.7$ & $81 \pm 5.4$ \\
Kramberger 2021 \cite{kramberger2021prediction}   & $20$ \,---  & $59$ \,---     & $25$ \,---     & $25$ \,---  & $73$ \,---     & $35$ \,--- \\
Yeung 2024 \cite{yeung2024adaptive}               & $22 \pm 1.5$ & $67 \pm 3.2$  & $32 \pm 2.0$  & $20 \pm 2.2$ & $62 \pm 3.8$  & $27 \pm 1.8$ \\
Glaser 2018 \cite{glaser2018motor}                & $24 \pm 1.8$ & $37 \pm 2.6$  & $13 \pm 1.5$  & $23 \pm 2.1$ & $42 \pm 3.0$  & $16 \pm 1.2$ \\
Mendez Guerra 2024 \cite{guerra2024adaptive}      & $49 \pm 1.3$ & $156 \pm 7.7$ & $81 \pm 2.1$  & $62 \pm 2.5$ & $182 \pm 8.7$ & $106 \pm 5.4$ \\
\bottomrule
\end{tabular}}
\end{table}

\subsection{Ablation}
Supplementary §\ref{supp:abl:headline} isolates four components of the headline pipeline; each row toggles one component and leaves the rest at the operating point of Section~\ref{sec:impl}. Replacing the FastICA seed with a random initialisation collapses yield to zero, and removing the per-spike peel-off step costs ${\sim}77$ / $25$ MUs per seed at $F_1 \geq 0.95$ (ch070 / ch320). Swapping the GMM bimodality criterion of Eq.~\ref{eq:gmm} for a kurtosis contrast $L_{\mathrm{kurt}} = -\mathbb{E}[\hat{s}_j(t)^{4}]$ collapses yield by $74\%$ / $81\%$ on the per-source pipeline without peel-off (supplementary §\ref{supp:abl:loss}), because kurtosis rewards a few outlying samples rather than tight spike/null clusters. Replacing the rank-$3$ Woodbury parameterisation (Eq.~\ref{eq:lowrank}) with a per-channel affine correction $A_j(t)\tilde{\mathbf{x}}(t) = \tilde{\mathbf{x}}(t) \odot \exp(\log\mathbf{w}_j(t)) + \mathbf{b}_j(t)$ drops yield by $39\%$ / $41\%$, as it cannot express cross-channel MUAP rotation and its offset lets the compensator absorb signal the separator should recover. Extended ablations are in supplementary §\ref{supp:abl:headline}--\ref{supp:ablations}.

\subsection{Real dynamic HD-sEMG}
\label{sec:real}
We also apply HarmonICA to two $65.5$\,s recordings from participant 6 of~\cite{guerra2024adaptive}: a $64$-channel forearm grid at $2048$\,Hz during isometric digit flexion at $15\%$ MVC while a motorised platform steps the wrist from $0^\circ$ to $-40^\circ$ in $10^\circ$ holds, with the measured angle as $\boldsymbol{\gamma}(t)$. MUAP correlation with the first $8$\,s falls to $0.087$ by the end of the recording, but the drift is slow and monotone (${\sim}0.6^\circ$/s). With no intramuscular reference, we score five units against a self-consistency reference built two independent ways (agreement $0.99$--$1.00$) and reproduced by four published methods. HarmonICA improves on the linear-ICA solution it refines for every unit (mean $F_1$ $0.854 \to 0.918$), has no catastrophic run (worst $0.732$, against $0.102$ for Chen 2020 and $0.079$ for Yeung 2024), and ties Chen 2020 on mean $F_1$. The tie is expected: even calibrated on a single $5.3$\,s hold at $0^\circ$, the adaptive baselines track the most-drifting unit to $-40^\circ$ (accuracy $0.94$--$1.00$), whereas HarmonICA's advantage on the benchmark comes from faster, reversing mixing changes (full results in supplementary §\ref{supp:real}).

\section{Limitations}

The benchmark is fully simulated and inherits the MUAP statistics and volume-conductor model of BioMime and NeuroMotion. Its movement spans a single degree of freedom (wrist flexion--extension) under constant drive, with no recruitment or derecruitment; multi-joint movement will likely need a higher compensator rank, and other effort levels are untested. The real-data evaluation covers one participant, two recordings and five units, scored against a reference built from the same surface data. The method is offline: each recording is decomposed with its own refit, and reuse on new data would require the measured joint angle as $\boldsymbol{\gamma}(t)$, which we do not evaluate. Recovery is bounded by the FastICA seed, and near-duplicate MUAPs can lead the compensator to amplify false positives (supplementary §\ref{supp:source_examples}).

Our identifiability results (supplementary §\ref{supp:theory}) are existence results under idealised assumptions, notably full column rank, which the benchmark does not satisfy, and independent motor units, which common synaptic input violates; they do not guarantee that the optimiser finds the separating solution. The separator is fit to the uncompensated signal and so carries a bias that grows with the drift (supplementary §\ref{supp:theory:results}, Proposition~1$'$), and $K$ limits yield from dense pools, since it must hold each unit's block of delayed MUAPs.

\section{Conclusion}

We presented HarmonICA, a quasi-time-invariant ICA method that keeps source extraction linear and carries all time variation in a low-rank invertible compensator, trained with decoupled gradients so that the separator remains the optimum of an unmodified linear-ICA objective. On a simulated dynamic HD-EMG benchmark, it sustains recall $R \geq 0.95$ out to $2.5\times$ / $2.2\times$ as many motor units per recording as the FastICA seed it refines (320- / 70-channel grids; Fig.~\ref{fig:pareto_recall}), and recovers more units at $\mathrm{F}_1 \geq 0.95$ than every adaptive ICA baseline in every density tier (Table~\ref{tab:yield_3bin}). On real forearm recordings with slow drift, it improves every scored unit over linear ICA and matches the strongest adaptive baseline.

\begin{ack}
Supported by EPSRC NISNEM (P81687) and the Imperial--Meta Wearable Neural Interface Research Centre. Code: \url{https://github.com/AlexKClarke/harmonica_neurips}.
\end{ack}

\newpage

{\small
\setlength{\bibsep}{2pt plus 0.3ex}
\bibliographystyle{unsrtnat}
\bibliography{references}
}

\newpage
\appendix
\section{Background: ICA for EMG decomposition}
\label{supp:ica_story}

This appendix sets the stage for the rest of the supplementary by
recapitulating the decomposition problem HarmonICA addresses, the
classical convolutive-BSS pipeline it inherits, the way joint
movement breaks that pipeline's stationarity assumption, and the
HarmonICA reformulation that lifts the local-window constraint.
The three figures below trace the same arc and motivate every
subsequent appendix.

\subsection{From neurons to electrodes}
\label{supp:ica:decomposition}

\begin{figure}[H]
\centering
\includegraphics[width=\linewidth]{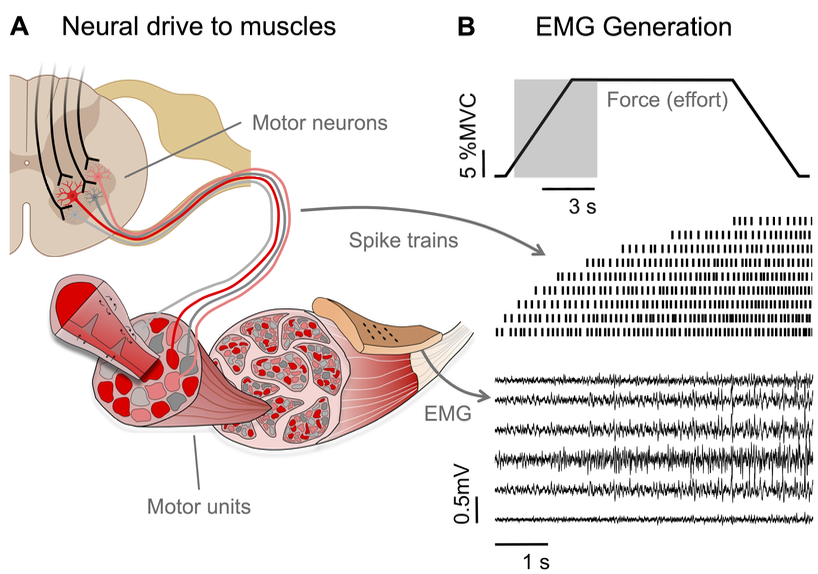}
\caption{\textbf{The EMG decomposition problem.} A motor-unit pool
sends spike trains down its motor neurons to muscle fibres; the
fibre depolarisations are picked up at the skin surface as a
multichannel signal that mixes the contributions of every active
motor unit through MUAP-shaped templates. The decomposition target
is to recover the underlying per-neuron spike trains from the
multichannel surface signal. Adapted from
\citep{mamidanna2025muniverse}.}
\label{fig:supp:ica:decomposition}
\end{figure}

A motor-unit pool is a population of motor neurons innervating a
muscle; each motor neuron drives a set of muscle fibres whose
synchronous depolarisation produces a stereotyped multichannel
signature on a surface electrode grid, the motor-unit action
potential (MUAP). EMG decomposition is the inverse problem:
recover, from the recorded multichannel surface trace, the spike
trains of the individual motor neurons that gave rise to it. For
the simulator stack used in this paper this generative chain
(neuron $\to$ muscle fibre $\to$ electrode array) is fully
accessible, with ground-truth spike trains preserved at every
stage (Fig.~\ref{fig:supp:ica:decomposition}).

\subsection{Convolutive blind source separation by FastICA}
\label{supp:ica:cbss}

The classical surface-EMG decomposition pipeline is a convolutive
blind source separation (cBSS), reduced to an instantaneous linear
mixture by Toeplitz-extending each channel with $L$ delayed copies
and PCA-whitening the result. The whitened signal
$\tilde{\mathbf{x}}(t) \in \mathbb{R}^{K}$ admits an instantaneous
linear model in the extended source $\bar{\mathbf{s}}(t)$ of delayed
spike-train copies, independent across motor units, with mixing
$\tilde{H}$ (Eq.~\ref{eq:source_estimate} of the main paper). For each source
$j$, FastICA seeks a unit-norm separator $\mathbf{w}_j$ that
maximises a non-Gaussianity contrast on the projected source
$u_j(t) = \mathbf{w}_j^\top\tilde{\mathbf{x}}(t)$. The seed pool
HarmonICA inherits is generated as follows:

\begin{enumerate}\itemsep1pt
\item \textbf{Preprocessing.} Each recording is Toeplitz-extended
  by factor $L$, projected onto
  its top-$K$ principal components, and ZCA-whitened, producing
  $\tilde{\mathbf{x}}(t) \in \mathbb{R}^{K}$.
\item \textbf{Per-attempt FastICA refinement.} For
  $n_{\text{ica}} = \min(L\,N, 108)$ random initialisations, we
  run fixed-point FastICA on $\tilde{\mathbf{x}}$ with the cubic
  contrast $g(u) = u^3$, max-iter $100$ and convergence tolerance
  $10^{-4}$, deflating successive sources with Gram--Schmidt
  orthogonalisation against previously accepted filters. After each
  attempt converges, candidate spike times are detected from the
  attempt's source $u(t)$ by k-means clustering of local peaks and
  the corresponding silhouette score is computed. This is the
  upstream FastICA spike detector used \emph{only} to score
  seed-pool candidates; the downstream HarmonICA refinement of
  Algorithm~1 of the main paper uses an entirely separate
  detector applied to the compensated source $\hat{s}_j(t)$
  (§\ref{sec:objectives} and §\ref{sec:peeloff} of the main paper).
\item \textbf{Acceptance gate.} An attempt is accepted if its
  silhouette score is at least $0.80$. Accepted filters
  $\{\mathbf{w}_j^{(0)}\}$ are added to the seed pool and used to
  deflate subsequent attempts; rejected attempts are discarded.
\item \textbf{Hand-off to HarmonICA.} Each accepted filter is passed
  to Algorithm~1 of the main paper as $\mathbf{w}_j^{(0)}$ and
  refined together with a fresh compensator $\phi_j$. No information
  beyond the filter direction is carried forward; the FastICA
  silhouette gate of $0.80$ controls only which filters \emph{enter}
  the refinement loop, while the post-refinement HarmonICA gate is
  disabled (\texttt{acceptance\_silhouette} $=0.0$) so that every
  seeded source is given a chance to be rescued by the compensator.
\end{enumerate}

The FastICA stage is identical for HarmonICA and for the
ICA / cBSS reference: the unrefined ICA seed reported throughout
§5 is read off this same seed pool before Algorithm~1 has acted on
it. The adaptive baselines (Chen 2020, Yeung 2024, Kramberger 2021,
Glaser 2018) follow the calibration recipe of their original papers
and run cBSS on a short initial calibration window
($\sim$1.5\,s), so the FastICA stage is not literally shared with
HarmonICA's $10$\,s seed pool but uses the same upstream
algorithm. Mendez Guerra 2024~\cite{guerra2024adaptive} (referred
to as ``Adapt'' in the authors' paper) is calibrated with this same
stack on the first $1.5$\,s of each recording (Toeplitz extension
$L = 8$, PCA to $K = 128$, ZCA whitening, FastICA with peel-off and
silhouette gate $0.80$) rather than with the calibration routine of
its own package; its online stage is the authors' adapt\_decomp
v1.0.0 engine~\cite{mendezguerra2024adaptcode}. See
§\ref{supp:baseline_deviations} for full deviations from each
baseline's published configuration.

\subsection{Why the stationarity assumption breaks during movement}
\label{supp:ica:nonstat}

\begin{figure}[H]
\centering
\includegraphics[width=\linewidth]{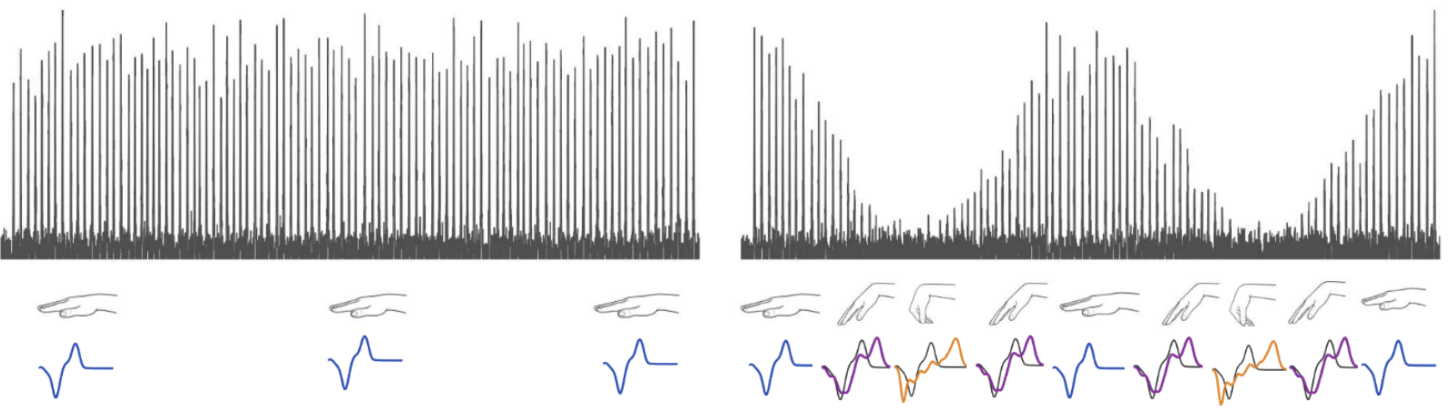}
\caption{\textbf{Stationary vs.\ non-stationary mixing.} Under
isometric conditions a single static ICA filter $\mathbf{w}_j$
extracts the full spike train of a motor unit cleanly. As joint
angle and limb position vary, each motor unit's MUAP rotates and
deforms; a fixed filter then matches the projection only inside a
local window and the recovered source decays outside it.}
\label{fig:supp:ica:stationary}
\end{figure}

Classical cBSS assumes a \emph{static} mixing, $\tilde{H}(t) =
\tilde{H}$ in Eq.~\ref{eq:source_estimate} of the main paper. During isometric contractions
this is approximately true: the muscle fibres, the volume conductor
between them and the skin, and the electrode positions all hold
still, so each motor unit projects through a fixed MUAP onto the
electrode array. During movement, joint angle, limb position and
volume-conductor deformation all change continuously, and each
motor unit's MUAP changes with them. A single static separator
$\mathbf{w}_j$ then matches the source's projection only over the
narrow time window where the mixing happens to lie close to the
filter's preferred direction, and decomposition collapses to a
\emph{local} window of the recording (or fails altogether) for two
compounding reasons: the filter is mismatched outside its training
window, and the stationary-mixture assumption ICA was derived
under no longer holds (Fig.~\ref{fig:supp:ica:stationary}).

\subsection{The HarmonICA reformulation}
\label{supp:ica:harmonica}

\begin{figure}[H]
\centering
\includegraphics[width=\linewidth]{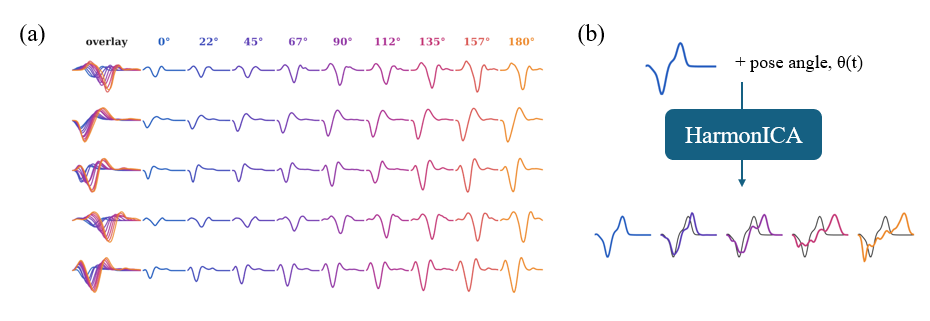}
\caption{\textbf{MUAP variation across joint angles.} The MUAP of a
single motor unit traced across the angular sweep of the simulated
joint. The waveform shape rotates and rescales smoothly with
angle; HarmonICA's compensator $A_j(t)$ targets this
low-rank, slowly varying part of the mixing.}
\label{fig:supp:ica:muap_overlay}
\end{figure}

The non-stationarity is concentrated in the MUAP, not in the spike
timing. Fig.~\ref{fig:supp:ica:muap_overlay} overlays the
angle-conditioned MUAP of a single motor unit across the joint's
range of motion: the waveform translates and rescales smoothly
with angle. HarmonICA exploits this structure. Instead of
fitting a separate static filter per local window, it learns a
soft, time-dependent modulation of a single static ICA-style filter,
parameterised as the rank-$3$ Woodbury compensator $A_j(t)$ of
Eq.~\ref{eq:lowrank} of the main paper. The compensator absorbs
the smooth angle-driven rotation of the MUAP, the static separator
$\mathbf{w}_j$ keeps the source-extraction step an unmodified
linear-ICA problem (identifiability conditions in
§\ref{supp:theory}), and the combination extends the local ICA
window to span the entire recording.

\section{Identifiability under time-varying convolutive mixing}
\label{supp:theory}

This section states and proves the results summarised in main §\ref{sec:objectives}: under stated assumptions, a static separator and a
compensator in the class of Eq.~\ref{eq:lowrank} exist that recover
each unit up to scale and delay at every $t$, and fitting the separator
on the whole recording rather than on the anchor adds a bias of at
most the order of the mean drift. All results are about existence, not about
convergence of the optimiser.

\subsection{Setup and assumptions}
\label{supp:theory:setup}

We use the notation of main §\ref{sec:method_prelim}. The whitened signal is
$\tilde{\mathbf{x}}(t) = \tilde{H}(t)\,\bar{\mathbf{s}}(t) +
\tilde{\boldsymbol{\omega}}(t)$ with $\tilde{H}(t) = D(t)\tilde{H}^\ast$
(main Eqs.~\ref{eq:source_estimate},~\ref{eq:drift}). The extended source
$\bar{\mathbf{s}}(t)\in\mathbb{R}^{\bar{M}}$, $\bar{M} = M(L+Q-1)$, is made of $M$
blocks $\bar{\mathbf{s}}_i(t) = \big(s_i(t), s_i(t-1), \ldots,
s_i(t-L-Q+2)\big)^\top$, one per unit, and $\tilde{\mathbf{h}}^\ast_{i,l}$
denotes the column of $\tilde{H}^\ast$ that multiplies $s_i(t-l)$. We
write $\Delta(t) = D(t) - I$ and $\mathbf{y}(t) =
\tilde{H}^\ast\bar{\mathbf{s}}(t)$ for the noiseless signal in the anchor
frame. Expectations are time averages over the recording, and the spike
trains are treated as stationary processes, independent of the drift
trajectory $D(t)$. The analysis is noiseless unless stated.

\paragraph{Contrast.} With the headline exponent $e=3$,
$\mathrm{sign}(u)|u|^3 = u^3$, so the independence loss of main
Eq.~\ref{eq:independence_loss} is $L_I(\mathbf{w}) = -J(\mathbf{w})$,
where
\begin{equation}
J(\mathbf{w}) = \frac{\kappa_3(u)}{\mathrm{var}(u)^{3/2}}, \qquad
u(t) = \mathbf{w}^\top\tilde{\mathbf{x}}(t),
\label{eq:supp:skew}
\end{equation}
is the skewness of the projection and $\kappa_3$ the third cumulant.
$J$ is invariant to positive rescaling of $\mathbf{w}$, so we work on the
unit sphere. $J$ is odd, so maximising it also fixes the sign of
$\mathbf{w}$: spike trains are positively skewed, and the recovered
source has its spikes pointing up.

\paragraph{Assumptions.}
\begin{itemize}
\item[(A1)] \emph{Anchor.} There is an anchor window on which $D(t)=I$.
  In the anchor frame the mixing has full column rank and $K=\bar{M}$, and
  whitening makes $\tilde{H}^\ast$ orthogonal, i.e.\ the extended
  sources have unit variance and are uncorrelated. In practice they are
  only approximately uncorrelated (main §\ref{sec:method_prelim}).
\item[(A2)] \emph{Independence across units; single-delay dominance
  within a unit.} The blocks $\bar{\mathbf{s}}_1,\ldots,\bar{\mathbf{s}}_M$
  are mutually independent. For each unit $i$, the function
  $g_i(\mathbf{v}) = \kappa_3(\mathbf{v}^\top\bar{\mathbf{s}}_i)$ on the
  unit sphere of $\mathbb{R}^{L+Q-1}$ has its positive local maxima
  exactly at the coordinate vectors $\mathbf{e}_l$ (one delay each),
  these maxima are strict, and $g_i(\mathbf{e}_l) = \kappa_3(s_i) > 0$.
\item[(A3)] \emph{Fixed-subspace low-rank drift.} $D(t) = I +
  U(t)V^\top$ with $U(t), V\in\mathbb{R}^{K\times r}$ and $V$ constant in
  time, and $D(t)$ is invertible for every $t$.
\end{itemize}
(A2) is the conditional, zero-order-dominance condition under which
\citet{klotz2025revisiting} show that local optima of the contrast are
single columns of the extended mixing. It holds exactly when the firing
of each unit is independent across samples (Lemma~1 below):
lagged copies are then independent, and $g_i(\mathbf{v}) = \kappa_3(s_i)
\sum_l v_l^3$, with $\kappa_3(s_i) = (1-2\rho_i)/\sqrt{\rho_i(1-\rho_i)}
> 0$ for a standardised spike train of firing probability $\rho_i <
1/2$. Refractoriness makes lagged copies dependent and adds cross-lag
cumulants to $g_i$; (A2) asks that they do not create new maxima.

\subsection{Results}
\label{supp:theory:results}

\paragraph{Lemma~1 (vertex lemma).}
\emph{On the unit sphere of $\mathbb{R}^n$, let $f(\mathbf{c}) =
\sum_k a_k c_k^3$ with all $a_k>0$. Every local maximum of $f$ with
$f(\mathbf{c})>0$ is a coordinate vector $\mathbf{e}_k$, and each
$\mathbf{e}_k$ is a strict local maximum.}

\noindent\emph{Proof.} Stationarity on the sphere gives $3a_kc_k^2 =
2\lambda c_k$, so every nonzero coordinate of a critical point equals
$\mu/a_k$ for a common $\mu$, and $f(\mathbf{c})>0$ forces $\mu>0$: all
nonzero coordinates are positive. If two of them, $c_k$ and $c_m$, are
nonzero, keep the others fixed and move mass between them along $c_k =
\rho\cos\theta$, $c_m = \rho\sin\theta$ with $\theta\in(0,\pi/2)$. Then
$f = \mathrm{const} + \rho^3 h(\theta)$ with $h(\theta) =
a_k\cos^3\theta + a_m\sin^3\theta$, $h'(\theta) =
3\sin\theta\cos\theta\,(a_m\sin\theta - a_k\cos\theta)$, and at the
critical point $h''(\theta) = 3\sin\theta\cos\theta\,(a_m\cos\theta +
a_k\sin\theta) > 0$: a strict local minimum along the curve, so not a
local maximum of $f$. Hence a positive local maximum has a single
nonzero coordinate, $\mathbf{e}_k$. At $\mathbf{e}_k$, any unit
perturbation $\mathbf{c} = \cos\varepsilon\,\mathbf{e}_k +
\sin\varepsilon\,\mathbf{d}$ with $\mathbf{d}\perp\mathbf{e}_k$ gives
$f(\mathbf{c}) \leq a_k(1 - \tfrac{3}{2}\varepsilon^2) +
O(\varepsilon^3) < a_k$ for small $\varepsilon\neq 0$.\hfill$\square$

\paragraph{Proposition~1 (anchor recovery).}
\emph{Assume (A1) and (A2), and let $J_0$ be the contrast
Eq.~\ref{eq:supp:skew} evaluated on the anchor-frame signal
$\mathbf{y}(t)$. (a) Every column $\tilde{\mathbf{h}}^\ast_{i,l}$ is a
strict local maximum of $J_0$, with projection $u(t) = s_i(t-l)$.
(b) Conversely, every local maximum $\mathbf{w}$ of $J_0$ with
$J_0(\mathbf{w})>0$ is a column: $\mathbf{w} =
\tilde{\mathbf{h}}^\ast_{i,l}$ for some unit $i$ and delay $l$.
The separator therefore recovers one unit up to scale and delay; which
unit is found depends on the initialisation, the permutation ambiguity
that deflation resolves.}

\noindent\emph{Proof.} Since $\tilde{H}^\ast$ is orthogonal, $\mathbf{c} =
\tilde{H}^{\ast\top}\mathbf{w}$ is a unit vector and $u = \mathbf{c}^\top
\bar{\mathbf{s}}$. Split $\mathbf{c}$ into blocks $\mathbf{c}_i =
r_i\mathbf{v}_i$ with $r_i\geq 0$ and $\|\mathbf{v}_i\|=1$. The blocks
are independent, so cumulants add, and the extended sources are
uncorrelated with unit variance, so $\mathrm{var}(u) = \sum_i r_i^2 = 1$
and
\begin{equation*}
J_0 = \textstyle\sum_i r_i^3\, g_i(\mathbf{v}_i).
\end{equation*}
(b) At a local maximum, each active block ($r_i>0$) has $\mathbf{v}_i$ a
local maximum of $g_i$, otherwise moving $\mathbf{v}_i$ alone increases
$J_0$. Write $a_i = g_i(\mathbf{v}_i)$. If two blocks $i,m$ are active,
moving mass between them is the curve of Lemma~1 with coefficients
$a_i, a_m$: an interior critical point needs $a_i,a_m$ of the same sign,
and if both are positive it is a strict minimum along the curve. If all
active $a_i\leq 0$, then $J_0\leq 0$. Hence $J_0>0$ at a local maximum
forces exactly one active block $i$, with $\mathbf{v}_i$ a local maximum
of $g_i$ at which $g_i>0$, which by (A2) is some $\mathbf{e}_l$. So $\mathbf{c} =
\mathbf{e}_{(i,l)}$ and $\mathbf{w} = \tilde{H}^\ast\mathbf{c} =
\tilde{\mathbf{h}}^\ast_{i,l}$. (a) At $\mathbf{c} = \mathbf{e}_{(i,l)}$,
a perturbation that moves mass $\sin\varepsilon$ into other blocks has
$r_i = \cos\varepsilon$, so $J_0 \leq \cos^3\!\varepsilon\;
g_i(\mathbf{v}_i) + O(\varepsilon^3) \leq \kappa_3(s_i)\big(1 -
\tfrac{3}{2}\varepsilon^2\big) + O(\varepsilon^3)$, using $g_i(\mathbf{v}_i)
\leq \kappa_3(s_i)$ near $\mathbf{e}_l$ by (A2); equality holds only for
$\varepsilon = 0$ and $\mathbf{v}_i = \mathbf{e}_l$.\hfill$\square$

\paragraph{Proposition~1$'$ (whole-recording bias).}
\emph{Assume (A1) and (A2), and let $J$ be the contrast on the whole
recording, $u(t) = \mathbf{w}^\top D(t)\mathbf{y}(t)$. Let $\mathbf{w}_0
= \tilde{\mathbf{h}}^\ast_{i,l}$ be a local maximum of $J_0$
(Proposition~1) with curvature $m>0$, i.e.\ the Hessian of $J_0$ on the
sphere at $\mathbf{w}_0$ is at most $-m$, and let $\varepsilon =
\tfrac{1}{T}\sum_t\|\Delta(t)\|_2$ be the mean drift. Then for
$\varepsilon$ small enough, $J$ has a local maximum $\mathbf{w}_T$ with
$\|\mathbf{w}_T - \mathbf{w}_0\| \leq C\varepsilon/m$, where $C$ depends
only on $\mathbb{E}\|\mathbf{y}\|^3$ and $\sup_t\|\Delta(t)\|_2$. As
$\varepsilon\to 0$ this reduces to Proposition~1.}

\noindent\emph{Proof.} $J$ and $J_0$ are the same smooth function of the
first three moments of $u$ and their derivatives in $\mathbf{w}$, with
$\mathbf{y}(t)$ replaced by $D(t)\mathbf{y}(t) = \mathbf{y}(t) +
\Delta(t)\mathbf{y}(t)$. Every difference between the two sets of moments
and of their first and second derivatives in $\mathbf{w}$ contains at
least one factor $\Delta(t)\mathbf{y}(t)$; for example,
$\big|\mathbb{E}[(\mathbf{w}^\top D\mathbf{y})^3] -
\mathbb{E}[(\mathbf{w}^\top\mathbf{y})^3]\big| \leq
3\,\tfrac{1}{T}\sum_t\|\Delta(t)\|\,(1+\|\Delta(t)\|)^2\,
\mathbb{E}\|\mathbf{y}\|^3$. Because the spike trains are stationary
and independent of $D(t)$, each such bound is a time average of
$\|\Delta(t)\|$ times a moment of $\mathbf{y}$, so $J - J_0$ and its
first two derivatives are $O(\varepsilon)$ on the sphere. For
$\varepsilon$ small the Hessian of $J$ near $\mathbf{w}_0$ stays at most
$-m/2$, so $J$ has a unique local maximum $\mathbf{w}_T$ in that
neighbourhood, and $\nabla J(\mathbf{w}_T) = 0$ gives $\|\mathbf{w}_T -
\mathbf{w}_0\| \leq 2\|\nabla(J-J_0)(\mathbf{w}_0)\|/m =
O(\varepsilon/m)$.\hfill$\square$

\noindent The bias is small when the mean drift is small relative to the
curvature of the anchor optimum, i.e.\ for modest excursions or anchors
near the centre of the trajectory. Because $L_I$ never sees the
compensator, the compensator cannot remove this bias, and coupling the
gradients instead costs yield (§\ref{supp:abl:headline}).

\paragraph{Proposition~2 (exact compensation).}
\emph{Assume (A3), and that $\mathrm{NN}_j$ can represent the factor
$U'(t) = -U(t)\big(I_r + V^\top U(t)\big)^{-1}$. Then the class of main
Eq.~\ref{eq:lowrank} contains $D(t)^{-1}$ exactly, with $V_j = V$ and
$U_j(t) = U'(t)$, and for any separator $\mathbf{w}$,
\begin{equation*}
\mathbf{w}^\top A_j(t)\,\tilde{\mathbf{x}}(t) = \mathbf{w}^\top
\mathbf{y}(t) + \mathbf{w}^\top D(t)^{-1}\tilde{\boldsymbol{\omega}}(t)
\quad\text{for every } t.
\end{equation*}
With $\mathbf{w} = \tilde{\mathbf{h}}^\ast_{i,l}$ from Proposition~1 the
output is $s_i(t-l)$ plus noise at every $t$, not only on the anchor
window; with the whole-recording separator of Proposition~1$'$ it carries
an additional $O(\varepsilon/m)$ leakage.}

\noindent\emph{Proof.} $D(t)$ is invertible if and only if $I_r +
V^\top U(t)$ is (matrix determinant lemma), and then the Woodbury
identity~\citep{hager1989updating} gives $D(t)^{-1} = I + U'(t)V^\top$.
Substituting $A_j(t) = D(t)^{-1}$ and $\tilde{\mathbf{x}}(t) =
D(t)\mathbf{y}(t) + \tilde{\boldsymbol{\omega}}(t)$ gives the
identity.\hfill$\square$

\paragraph{Approximate (A3): leakage bound.}
When the drift is only approximately low-rank in a fixed subspace, stack
the inverse drifts $E(t) = D(t)^{-1} - I$ into $\mathbf{E} = [E(t_1);
\ldots; E(t_T)]\in\mathbb{R}^{KT\times K}$, with singular values
$\sigma_1\geq\sigma_2\geq\ldots$ and top-$r$ right singular vectors
$V_r$. The compensator $A(t) = I + E(t)V_rV_r^\top$ is in the class (with
$U_j(t) = E(t)V_r$, $V_j = V_r$), and its residual $R(t) = D(t)^{-1} -
A(t) = E(t)(I - V_rV_r^\top)$ satisfies $\sum_t\|R(t)\|_F^2 =
\sum_{k>r}\sigma_k^2$, the smallest possible for any fixed rank-$r$
subspace (Eckart--Young). For a unit separator, and since
$\mathrm{cov}(\mathbf{y}) = I$ under (A1), the mean squared leakage is
\begin{equation}
\frac{1}{T}\sum_t \mathbb{E}\big[(\mathbf{w}^\top R(t)D(t)\mathbf{y}(t))^2\big]
\;\leq\; \max_t\|D(t)\|_2^2 \cdot \frac{1}{T}\sum_{k>r}\sigma_k^2,
\label{eq:supp:leakage}
\end{equation}
so recovery degrades continuously with the drift's departure from a
rank-$r$ fixed subspace. Under exact (A3), $D(t)^{-1}-I$ and $D(t)-I$
share the right factor $V$, so the two stacks have the same rank.

\paragraph{Pointwise rank is not enough.} The fixed subspace in (A3)
matters. Take $D(t)^{-1} - I = \mathbf{u}(t)\mathbf{v}(t)^\top$ with
$\mathbf{v}(t) = (\cos\omega t, \sin\omega t, 0, \ldots, 0)^\top$: the
drift has rank $1$ at every instant, but the stack has rank $2$, so no
compensator with $r=1$ inverts it, and Eq.~\ref{eq:supp:leakage} with
$r=1$ leaves the second singular value as leakage. The rank to budget
for is the rank of the stacked drift; we set $r=3$ from the rank sweep
in §\ref{supp:abl:rank}.

\subsection{Time-varying ambiguities}
\label{supp:theory:ambiguities}

A general time-varying separator can extract a different source at
different times, so that its output concatenates intervals from several
units; this is the discontinuity problem of dynamic source
extraction~\citep{koldovsky2021dynamic}. HarmonICA avoids it by
construction. Write the output as $\hat{s}_j(t) =
\mathbf{c}(t)^\top\bar{\mathbf{s}}(t)$ with $\mathbf{c}(t)^\top =
\mathbf{w}_j^\top A_j(t)\,D(t)\,\tilde{H}^\ast$. The separator
$\mathbf{w}_j$ is static and fixed at a column by the seed
(Proposition~1), the compensator starts at $A_j = I$ and is a continuous
function of a continuous input $\boldsymbol{\gamma}(t)$, and the physical
drift $D(t)$ is continuous. So $\mathbf{c}(t)$ moves continuously from
$\mathbf{e}_{(i,l)}$. Moving to another unit would require passing
through coefficient vectors with mass on two units, where the output is a
sum of two independent spike trains, which widens the spike and null
clusters that $L_N$ penalises, because the spike set is built from the
seeded unit's own spikes. Moving within unit $i$ across delays only filters
$s_i$, which leaves its firing times unchanged. What can vary with $t$ is
therefore a scale $\alpha(t)$ and an effective delay $d(t)$,
\begin{equation*}
\hat{s}_j(t) \approx \alpha(t)\,s_i\big(t - d(t)\big),
\end{equation*}
and both are harmless for a readout based on firing times. This is an
argument, not a theorem: abrupt, discontinuous changes of the mixing are
outside it, and that is the cold-start regime of §\ref{supp:abl:headline}. Its
empirical check is the spike-level score: a unit that switched sources
mid-recording would lose precision against its matched ground truth, and
every score we report is a spike-level $F_1$ (at $\pm2$\,ms tolerance on
the benchmark).

\subsection{What is not proven}
\label{supp:theory:limits}

\begin{itemize}
\item \emph{Dependent spike trains.} Proposition~1 uses independence
  across units to make cumulants additive. Real motor units share
  synaptic input and are dependent but approximately uncorrelated; their
  cross-cumulants then enter $J_0$. Degradation under such dependence is
  gradual in practice~\citep{klotz2025revisiting}, and FastICA-type
  estimators remain consistent under variance
  dependence~\citep{kawanabe2005estimating}, but an unconditional
  identifiability result for dependent spike trains is open for every
  method, including all our baselines.
\item \emph{Single-delay dominance (A2)} is proven only for firing that
  is independent across samples (Lemma~1); with refractoriness it is
  assumed, as in~\citet{klotz2025revisiting}.
\item \emph{Rank.} The propositions assume full column rank, $K=\bar{M}$.
  Without it, only a per-unit separability condition is
  available~\citep{klotz2025revisiting}, and the number of units that
  fit within $K$ is limited by the effective dimension of each unit's
  block of delayed MUAPs.
\item \emph{Existence, not convergence.} The propositions show that good
  separators and compensators exist and are local optima; they do not
  show that the optimiser finds them. From a random start it does not
  (§\ref{supp:abl:headline}, $F_1\approx0.28$), which is why the FastICA warm start
  matters.
\item \emph{Compensator training.} Proposition~2 assumes the network can
  represent $U'(t)$, and does not show that minimising $L_N$ selects
  $D(t)^{-1}$; it shows only that $D(t)^{-1}$ is in the class.
\item \emph{Noise.} $D(t)^{-1}$ amplifies noise where $D(t)$ is close to
  singular.
\end{itemize}

\section{Construction of \emph{muniverse-benchmark-dynamic}}
\label{supp:benchmark}

The MUniverse motor-unit-decomposition dataset \cite{mamidanna2025muniverse} is a
public resource that offers
(i)~high-density surface electromyography arrays,
(ii)~angle-conditioned non-stationary mixing,
(iii)~ground-truth motor-unit spike trains, and
(iv)~realistic motor-unit pool sizes ($\geq 30$ units per pool).
We adopt it as the source for our benchmark.
However, raw MUniverse cannot be used as a benchmark in unmodified form,
because the BioMime-generated motor-unit pools contain pervasive
within-pool MUAP near-duplicates that bound any decomposition method's
ceiling far below the nominal pool size. We document the empirical case
in \S\ref{supp:bench:duplicates}--\ref{supp:bench:ceiling}, then
describe our construction in \S\ref{supp:bench:construction}.

\subsection{Within-pool MUAPs form giant near-duplicate clusters}
\label{supp:bench:duplicates}

For each of 45 motor-unit pools (drawn from 10 simulated subjects and 8 forearm
muscles; $8{,}517$ raw motor units total) we computed the pairwise
cosine similarity matrix between MUAPs at the mid-pose joint angle
(BioMime angle step 65 of 130), flattened across electrodes and
extension factor. We define a near-duplicate \emph{pair} at threshold
$T$ as two motor units with similarity $\geq T$, and a \emph{cluster} as
a connected component of size $\geq 2$ in the threshold graph.
Table~\ref{tab:supp:clusters} shows the result.

\begin{table}[h]
\centering
\caption{Pair- and cluster-level structure of within-pool MUAP
similarity, aggregated across all 45 MUniverse motor-unit pools
($N_{\text{full}}=8{,}517$ motor units). At any decomposability-relevant
threshold the typical pool collapses into a single connected component
that absorbs nearly every motor unit; the pair-less / paired transition
only occurs above $T=0.95$.}
\label{tab:supp:clusters}
\begin{tabular}{cccccc}
\toprule
$T$ & \# pairs & pair-less MUs & paired MUs &
mean cluster size & max cluster size \\
\midrule
0.75 & 110\,377 & 61 (0.7\,\%) & 8\,456 (99.3\,\%) & 158.5 & \textbf{224} \\
0.85 & 47\,893 & 312 (3.7\,\%) & 8\,205 (96.3\,\%) & 105.0 & 224 \\
0.90 & 23\,132 & 988 (11.6\,\%) & 7\,529 (88.4\,\%) & 36.1 & 210 \\
0.95 & 6\,171 & 3\,428 (40.3\,\%) & 5\,089 (59.8\,\%) & 5.5 & 100 \\
0.99 & 156 & 8\,228 (96.6\,\%) & 289 (3.4\,\%) & 2.1 & 4 \\
\bottomrule
\end{tabular}
\end{table}

Two observations. First, at the standard $T=0.75$ decomposability
threshold the dataset is 99.3\,\% non-isolated: only 61 of $8{,}517$
motor units across the entire dataset have no within-pool neighbour at
similarity $\geq 0.75$. The mean cluster (connected component) size is
158.5, the maximum is 224 (an entire pool of 224 motor units forms one
near-duplicate component). Second, the ``most motor units have no
duplicate'' regime starts only above $T=0.95$ and is essentially
complete only at $T=0.99$, where 156 effective duplicate pairs are
distributed across the 45 pools.

\paragraph{Pose robustness.}
The values in Table~\ref{tab:supp:clusters} use the mid-pose for
similarity (the same definition our filter uses).
Table~\ref{tab:supp:cluster_pose} repeats the analysis at the early
(step 10) and late (step 120) poses to verify the cluster topology is
not a mid-pose artefact.

\begin{table}[h]
\centering
\caption{Cluster topology under early / mid / late joint poses.
Pose-spread is at most 3.8 percentage points at any threshold, and the
maximum cluster size at $T=0.75$ is invariant (224 across all three
poses), so the same motor units sit in the same dominant component
regardless of joint angle.}
\label{tab:supp:cluster_pose}
\begin{tabular}{lcccc}
\toprule
& \multicolumn{3}{c}{\% pair-less motor units} & spread \\
\cmidrule(lr){2-4}
$T$ & early (step 10) & mid (step 65) & late (step 120) & (pp) \\
\midrule
0.75 & 0.58 & 0.72 & 0.80 & 0.22 \\
0.85 & 3.35 & 3.66 & 3.82 & 0.47 \\
0.90 & 9.51 & 11.60 & 10.83 & 2.09 \\
0.95 & 36.48 & 40.25 & 39.10 & 3.77 \\
0.99 & 95.26 & 96.61 & 95.02 & 1.59 \\
\midrule
mean cluster size $\!@\,T\!=\!0.75$ & 169.2 & 158.5 & 167.8 & --- \\
max cluster size $\!@\,T\!=\!0.75$ & \textbf{224} & \textbf{224} & \textbf{224} & 0 \\
\bottomrule
\end{tabular}
\end{table}

\subsection{Decomposability ceiling at 14--41\,\% of nominal pool size}
\label{supp:bench:ceiling}

The cluster topology of \S\ref{supp:bench:duplicates} implies that the
maximum number of motor units a decomposition method can recover from a
raw pool is bounded above by the largest \emph{decomposable subset},
the largest set with no within-pool pair above a similarity bound $T$.
We compute this with a greedy Henneman-ordered selection: walking
through motor units in recruitment order (smallest first), keep motor
unit $i$ if its cosine similarity to every already-kept motor unit $j$ is
below the threshold, i.e.\ $\max_j \mathrm{sim}(i,j) < T$, where
$\mathrm{sim}(i,j)$ is the cosine similarity between MUAPs $i$ and $j$
(flattened across electrodes and the extension factor).
Table~\ref{tab:supp:ceiling} shows the result.

\begin{table}[h]
\centering
\caption{Decomposable-subset retention under a greedy Henneman-ordered
crosstalk filter. Even at a permissive $T=0.90$, which admits MUAP pairs
$90\,\%$ identical, only $40.8\,\%$ of nominal motor units survive;
at the standard $T=0.75$ only $13.5\,\%$ do. Reporting raw $F_1$ against
the nominal pool size therefore under-estimates every method's ceiling
by an $\sim 3$--$7\times$ factor that has nothing to do with method
quality.}
\label{tab:supp:ceiling}
\begin{tabular}{ccc}
\toprule
threshold $T$ & kept MUs / pool (mean $\pm$ std) & retention \\
\midrule
0.75 & $25.5 \pm 7.5$ & 13.5\,\% \\
0.80 & $35.0 \pm 10.6$ & 18.5\,\% \\
0.85 & $50.5 \pm 14.9$ & 26.7\,\% \\
0.90 & $77.2 \pm 20.9$ & 40.8\,\% \\
\bottomrule
\end{tabular}
\end{table}

\subsection{Construction pipeline}
\label{supp:bench:construction}

Given the empirical bounds in \S\ref{supp:bench:duplicates}--%
\ref{supp:bench:ceiling}, our benchmark is constructed by filtering
each motor-unit pool to a decomposable subset before recording
generation, then synthesising recordings on the curated pools.

\paragraph{Stage 1: BioMime motor-unit pool generation.}
For each subject--muscle pair drawn from 10 simulated subjects and 8 forearm muscles
(ECRB, ECRL, ECU, EDI, FCU\textsubscript{h}, FCU\textsubscript{u},
FDSI, PL), we generate the angle-conditioned MUAP library at the
Flexion--Extension degree of freedom on the BioMime forward model
\cite{ma2022human}, with 130 angle steps spanning $\pm 65^\circ$. Each
cache is a tensor of shape $(N_\text{MUs}, 130, 10, 32, 96)$
together with motor-neuron-pool properties (depth, angle, innervation
zone, fibre length, conduction velocity). The 45 caches contain
$8\,517$ raw motor units in total ($189 \pm 23$ per cache).

\paragraph{Stage 2: Crosstalk analysis.}
For each cache, we compute the pairwise cosine similarity matrix
between motor units at the mid-pose, flattened across the
$10 \times 32 \times 96$ electrode-time grid. This is the input to
both the analysis tables in
\S\ref{supp:bench:duplicates}--\ref{supp:bench:ceiling} and to the
filter below.

\paragraph{Stage 3: Greedy Henneman-ordered crosstalk filter.}
We apply the greedy filter from \S\ref{supp:bench:ceiling} per cache,
with the threshold chosen per pool-size tier
(Table~\ref{tab:supp:dataset}).

\begin{table}[h]
\centering
\caption{Structure of the \emph{muniverse-benchmark-dynamic} test split.
Pools are stratified by decomposable size after the tier-dependent
crosstalk filter; each pool is rendered under 8 angle/SNR conditions
$\times$ 2 electrode grids ($16$ recordings/pool). An additional
held-out val split (5 pools, 80 recordings) is included in the
public release for groups that wish to use it; all main-paper
numbers and the supplementary tables that mirror §5 are on the
$168$-per-grid test split.}
\label{tab:supp:dataset}
\begin{tabular}{lcccc}
\toprule
Tier & Pool size (MUs) & Threshold $T$ & \#~pools & \#~recordings \\
\midrule
small  & 26--45  & $0.85$           & 6  & 96  \\
medium & 47--74  & $0.85$           & 8  & 128 \\
large  & 71--113 & $0.85$ / $0.90$  & 7  & 112 \\
\midrule
\textbf{total} & 26--113 & --- & \textbf{21} & \textbf{336} \\
\bottomrule
\end{tabular}
\end{table}

At $T = 0.85$ the maximum decomposable pool size is $\sim 80$
motor units; for the large-pool tier ($\geq 80$ motor units
decomposable), reaching pool sizes representative of motor pools
at the upper end of physiological recruitment requires relaxing
to $T = 0.90$ for $6/21$ combos.

\paragraph{Stage 4: NeuroMotion recording generation.}
For every clean cache we synthesise EMG recordings under
8 conditions: $\{$sinusoidal, triangular$\}$ angle profile $\times$
$\{0.5, 1.0\}$ amplitude factor (range $\pm 32.5^\circ$ or $\pm 65^\circ$)
$\times$ $\{20, 25\}$\,dB SNR, with constant 50\,\% MVC effort,
$0.3$\,Hz carrier, $10$\,s duration at $f_s = 2048$\,Hz. Two electrode
configurations are emitted: 70-channel (a $7 \times 10$ subgrid) and
320-channel (the full $32 \times 10$ grid).

\paragraph{Stage 5: Triangular-angle correction.}
The default NeuroMotion triangular branch returns flat-zero angle
profiles when \texttt{TargetAngle = 0}. We apply a symmetric-triangle
correction, included in the code release, before generating triangular
recordings.

\paragraph{Stage 6: Test/val split.}
The benchmark is rendered with a test/val split:
$21$~pools across $5$~simulated subjects (sub-sim00 through
sub-sim04) form the test split, giving $336$ recordings ($8$
angle/SNR conditions $\times$ $2$ electrode grids per pool); a
smaller held-out val split is included in the public release for
groups that wish to use it. All numbers reported in §5 of the main
paper, and in the supplementary tables that mirror §5, are on the
$168$-per-grid test split.

\paragraph{Final benchmark.}
416 dynamic EMG recordings, with ground-truth spike trains preserved.
Pool sizes range 26--113 motor units (the per-recording $n_{\text{gt}}$ of main §\ref{sec:data}, 23--101, counts only the units active in each recording), stratified by tier across
3 buckets (small, medium, large). Decomposability ceiling matches the
per-pool clean-$N$ count by construction (no hidden un-decomposable mass).
Stratifying the main-paper results by tier, rather than aggregating
over all of MUniverse, prevents method-comparison from being
confounded by per-muscle template-collision density, which on the raw
pools varies $2.2\times$ across muscles.

\section{Detailed yield and seed reproducibility}
\label{supp:reproducibility}

This appendix expands on the main-paper headline (Fig.~\ref{fig:pareto_recall}, Table~\ref{tab:yield_3bin} of the main paper) in four
directions: (i) yield at four $F_1$ quality floors averaged over the
test split (Table~\ref{tab:supp:detailed_yield}); (ii) the same
$3$-bin headline reported as a \emph{per-recording mean} (in MUs/rec,
units identical to standard benchmark conventions) with the $\pm$
taken across recordings within each cell rather than across seeds
(Table~\ref{tab:supp:yield_per_rec}), the complementary
view to the main-paper Table~\ref{tab:yield_3bin}, which reports bin-level totals;
(iii) the same per-recording mean but with the $\pm$ taken across the
$5$~ICA-initialisation seeds rather than across recordings, isolating
seed reproducibility from benchmark heterogeneity
(Table~\ref{tab:supp:seed_std}); (iv) two visual companions to the
per-method overlap analysis (full heatmap of pairwise overlap;
bin-stratified Venn-style bars showing the discovery vs.\ denoising
split per method).

\paragraph{Methodology notes for the main-paper Table~\ref{tab:yield_3bin}.}
Each cell of Table~\ref{tab:yield_3bin} is the bin-level sum of yield over the test-split
recordings, with $\pm$ taken across the $5$
FastICA-initialisation seeds. Bin recording counts are $40$~/~$80$~/~$48$
for the low~/~medium~/~high $n_{\text{gt}}$ tiers respectively (main
§\ref{sec:data}). HarmonICA's lead over every baseline in every cell
of Table~\ref{tab:yield_3bin} holds on every individual seed. The ordering among
baselines is less stable: Chen 2020 and the ICA / cBSS reference swap
order across seeds in three cells, as do Mendez Guerra 2024 and the
ICA / cBSS reference in the ch070 $40$--$69$ bin ($156$ vs.\ $151$ on
average). Kramberger~2021 is
deterministic given its calibration data and is therefore reported on
a single seed (``\,---\,'' in the s.d.\ position); recordings on
which Kramberger's algorithm returns zero MUs contribute zero to the
bin sum and are zero-yield outcomes, not missing runs. The
per-recording views in Tables~\ref{tab:supp:detailed_yield}--\ref{tab:supp:bootstrap}
instead average Kramberger~2021 over only the $144$ (ch070) / $146$ (ch320)
recordings on which it returns $\geq 1$~MU ($39$/$71$/$34$ and $39$/$74$/$33$
per bin), which favours Kramberger; counting the remaining recordings as zeros
gives the lower means implied by Table~\ref{tab:yield_3bin}
($0.62$ / $0.79$ MUs/rec at $F_1 \geq 0.95$).

\begin{table}[h]
\centering
\caption{\textbf{Yield at four $F_1$ quality floors}.
Mean MUs per recording at the indicated $F_1$ floor, mirroring the
main-paper convention (per-recording mean over the $5$
ICA-initialisation seeds, then mean across recordings within a
density). Higher is better; bold marks the best method in each column. Kramberger 2021 is averaged over the recordings on which it returns $\geq 1$~MU only (§\ref{supp:reproducibility}).}
\label{tab:supp:detailed_yield}
\resizebox{\textwidth}{!}{%
\begin{tabular}{l|c c |c c |c c |c c}
\toprule
\multirow{2}{*}{Method}
 & \multicolumn{2}{c|}{$F_1 \geq 0.99$}
 & \multicolumn{2}{c|}{$F_1 \geq 0.95$}
 & \multicolumn{2}{c|}{$F_1 \geq 0.90$}
 & \multicolumn{2}{c}{$F_1 \geq 0.80$} \\
\cmidrule(lr){2-3} \cmidrule(lr){4-5} \cmidrule(lr){6-7} \cmidrule(lr){8-9}
 & ch070 & ch320 & ch070 & ch320 & ch070 & ch320 & ch070 & ch320 \\
\midrule
ICA / cBSS \cite{negro2016multi}                  & 0.97 & 0.93 & 1.48 & 1.49 & 1.82 & 1.84 & 2.64 & 2.81 \\
\midrule
\textbf{HarmonICA (ours)}                         & \textbf{1.94} & \textbf{2.15} & \textbf{2.83} & \textbf{3.06} & \textbf{3.44} & \textbf{3.67} & \textbf{4.59} & \textbf{5.01} \\
\midrule
Chen 2020 \cite{chen2020adaptive}                 & 0.50 & 0.50 & 1.36 & 1.60 & 1.87 & 2.30 & 2.94 & 3.41 \\
Kramberger 2021 \cite{kramberger2021prediction}   & 0.39 & 0.44 & 0.72 & 0.91 & 0.92 & 1.08 & 1.19 & 1.36 \\
Yeung 2024 \cite{yeung2024adaptive}               & 0.24 & 0.28 & 0.72 & 0.65 & 1.03 & 1.05 & 1.67 & 1.70 \\
Glaser 2018 \cite{glaser2018motor}                & 0.25 & 0.23 & 0.44 & 0.48 & 0.55 & 0.64 & 0.68 & 0.81 \\
Mendez Guerra 2024 \cite{guerra2024adaptive}                   & 0.90 & 1.07 & 1.71 & 2.08 & 2.23 & 2.87 & 3.08 & 3.87 \\
\bottomrule
\end{tabular}}
\end{table}

\paragraph{Per-recording mean yield.}
The main-paper Table~\ref{tab:yield_3bin} reports bin-level totals with the $\pm$ taken
across the $5$ ICA-initialisation seeds, which keeps standard
deviations small (a function of seed reproducibility) and gives raw
counts of motor units recovered. Table~\ref{tab:supp:yield_per_rec}
reports the complementary view: each cell is the per-recording mean
yield in MUs/rec (averaged over the $5$ seeds within each recording,
then over the recordings within the cell), with the $\pm$ taken
across recordings. The standard deviations here are large because
the test split spans seven muscles, $21$ motor-unit pools, eight
angle/SNR conditions, and per-recording pool sizes from $23$ to
$101$~MUs; for ICA in dense pools the median per-recording yield is
$0$ and the right-tail of ``easy'' recordings dominates the mean.
HarmonICA's per-recording mean is roughly $2\times$ ICA's in every
cell, mirroring the bin-totals in the main-paper table.

\begin{table}[h]
\centering
\caption{\textbf{Per-recording mean yield at F$_1 \geq 0.95$ stratified
by per-recording motor-unit count $n_{\text{gt}}$}.
Each cell is mean MUs/rec $\pm$ standard deviation \emph{across
recordings} within the cell, after each recording is collapsed to its
mean over the $5$ FastICA-initialisation seeds. Bin recording counts
are $40$ / $80$ / $48$ for the three tiers. Standard deviations are
large because the test split is heterogeneous (per-recording pool
sizes range $23$--$101$~MUs); the corresponding across-seed standard
deviation per cell is given in Table~\ref{tab:supp:seed_std}. Kramberger 2021 is averaged over the recordings on which it returns $\geq 1$~MU only (§\ref{supp:reproducibility}).}
\label{tab:supp:yield_per_rec}
\resizebox{\textwidth}{!}{%
\begin{tabular}{lccc|ccc}
\toprule
\multirow{2}{*}{Method} & \multicolumn{3}{c|}{\textbf{ch070} (70-channel)} & \multicolumn{3}{c}{\textbf{ch320} (320-channel)} \\
\cmidrule(lr){2-4} \cmidrule(lr){5-7}
 & $n_{\text{gt}} < 40$ & $n_{\text{gt}}\,40\!-\!69$ & $n_{\text{gt}} \geq 70$ & $n_{\text{gt}} < 40$ & $n_{\text{gt}}\,40\!-\!69$ & $n_{\text{gt}} \geq 70$ \\
\midrule
ICA / cBSS \cite{negro2016multi}                  & 0.90 $\pm$ 1.31 & 1.88 $\pm$ 2.39 & 1.29 $\pm$ 2.38 & 1.01 $\pm$ 1.42 & 1.78 $\pm$ 2.30 & 1.40 $\pm$ 2.19 \\
\midrule
\textbf{HarmonICA (ours)}                         & \textbf{2.25 $\pm$ 1.69} & \textbf{3.15 $\pm$ 2.45} & \textbf{2.78 $\pm$ 2.31} & \textbf{2.39 $\pm$ 1.75} & \textbf{3.28 $\pm$ 2.45} & \textbf{3.25 $\pm$ 2.64} \\
\midrule
Chen 2020 \cite{chen2020adaptive}                 & 0.91 $\pm$ 1.09 & 1.66 $\pm$ 1.66 & 1.23 $\pm$ 1.57 & 0.98 $\pm$ 1.31 & 1.87 $\pm$ 1.94 & 1.68 $\pm$ 1.91 \\
Kramberger 2021 \cite{kramberger2021prediction}   & 0.51 $\pm$ 0.88 & 0.83 $\pm$ 0.89 & 0.74 $\pm$ 0.71 & 0.64 $\pm$ 0.87 & 0.99 $\pm$ 1.03 & 1.06 $\pm$ 0.93 \\
Yeung 2024 \cite{yeung2024adaptive}               & 0.56 $\pm$ 0.46 & 0.83 $\pm$ 0.84 & 0.67 $\pm$ 0.66 & 0.50 $\pm$ 0.44 & 0.78 $\pm$ 0.89 & 0.56 $\pm$ 0.50 \\
Glaser 2018 \cite{glaser2018motor}                & 0.60 $\pm$ 0.72 & 0.46 $\pm$ 0.56 & 0.26 $\pm$ 0.57 & 0.57 $\pm$ 0.88 & 0.53 $\pm$ 0.75 & 0.33 $\pm$ 0.71 \\
Mendez Guerra 2024 \cite{guerra2024adaptive}                   & 1.22 $\pm$ 1.25 & 1.96 $\pm$ 1.73 & 1.70 $\pm$ 1.67 & 1.55 $\pm$ 1.60 & 2.28 $\pm$ 1.90 & 2.20 $\pm$ 1.91 \\
\midrule
\textit{recordings/bin} & 40 & 80 & 48 & 40 & 80 & 48 \\
\bottomrule
\end{tabular}}
\end{table}

\paragraph{Seed reproducibility (per-recording mean view).}
The main-paper Table~\ref{tab:yield_3bin} reports bin-level totals with $\pm$ taken across
the $5$ FastICA-initialisation seeds, and the per-recording mean view
in Table~\ref{tab:supp:yield_per_rec} above takes $\pm$ across recordings
within each cell. Table~\ref{tab:supp:seed_std} below provides the
across-seed counterpart of Table~\ref{tab:supp:yield_per_rec}: the same
per-recording mean yield, but with $\pm$ taken across the $5$ seeds rather
than across recordings. The seed-only spread is at most $\pm 0.30$~MUs/rec
across every method and bin (the maximum is HarmonICA on ch070 in the
medium tier; every other cell is within $\pm 0.12$), confirming that
the large bars in Table~\ref{tab:supp:yield_per_rec} are dispersion
across the heterogeneous panel of recordings, not method instability.

\begin{table}[h]
\centering
\caption{\textbf{Seed reproducibility} of the per-recording mean yield at $F_1 \geq 0.95$.
Same shape and units as Table~\ref{tab:supp:yield_per_rec} (mean MUs/rec)
but with the $\pm$ taken across the $5$ FastICA-initialisation seeds
rather than across recordings. Kramberger 2021 is averaged over the recordings on which it returns $\geq 1$~MU only (§\ref{supp:reproducibility}).}
\label{tab:supp:seed_std}
\resizebox{\textwidth}{!}{%
\begin{tabular}{lccc|ccc}
\toprule
\multirow{2}{*}{Method} & \multicolumn{3}{c|}{\textbf{ch070} (70-channel)} & \multicolumn{3}{c}{\textbf{ch320} (320-channel)} \\
\cmidrule(lr){2-4} \cmidrule(lr){5-7}
 & $n_{\text{gt}} < 40$ & $n_{\text{gt}}\,40\!-\!69$ & $n_{\text{gt}} \geq 70$ & $n_{\text{gt}} < 40$ & $n_{\text{gt}}\,40\!-\!69$ & $n_{\text{gt}} \geq 70$ \\
\midrule
ICA / cBSS \cite{negro2016multi}                  & 0.90 $\pm$ 0.06 & 1.88 $\pm$ 0.04 & 1.29 $\pm$ 0.04 & 1.01 $\pm$ 0.07 & 1.78 $\pm$ 0.04 & 1.40 $\pm$ 0.09 \\
\midrule
\textbf{HarmonICA (ours)}                         & \textbf{2.25 $\pm$ 0.11} & \textbf{3.15 $\pm$ 0.30} & \textbf{2.78 $\pm$ 0.12} & \textbf{2.39 $\pm$ 0.05} & \textbf{3.28 $\pm$ 0.06} & \textbf{3.25 $\pm$ 0.11} \\
\midrule
Chen 2020 \cite{chen2020adaptive}                 & 0.91 $\pm$ 0.07 & 1.66 $\pm$ 0.05 & 1.23 $\pm$ 0.07 & 0.98 $\pm$ 0.10 & 1.87 $\pm$ 0.11 & 1.68 $\pm$ 0.11 \\
Kramberger 2021 \cite{kramberger2021prediction}   & 0.51 $\pm$ ---  & 0.83 $\pm$ ---  & 0.74 $\pm$ ---  & 0.64 $\pm$ ---  & 0.99 $\pm$ ---  & 1.06 $\pm$ ---  \\
Yeung 2024 \cite{yeung2024adaptive}               & 0.56 $\pm$ 0.04 & 0.83 $\pm$ 0.04 & 0.67 $\pm$ 0.04 & 0.50 $\pm$ 0.06 & 0.78 $\pm$ 0.05 & 0.56 $\pm$ 0.04 \\
Glaser 2018 \cite{glaser2018motor}                & 0.60 $\pm$ 0.04 & 0.46 $\pm$ 0.03 & 0.26 $\pm$ 0.03 & 0.57 $\pm$ 0.05 & 0.53 $\pm$ 0.04 & 0.33 $\pm$ 0.03 \\
Mendez Guerra 2024 \cite{guerra2024adaptive}                   & 1.22 $\pm$ 0.03 & 1.96 $\pm$ 0.10 & 1.70 $\pm$ 0.04 & 1.55 $\pm$ 0.06 & 2.28 $\pm$ 0.11 & 2.20 $\pm$ 0.11 \\
\bottomrule
\end{tabular}}
\end{table}

\begin{figure}[h]
\centering
\includegraphics[width=\textwidth]{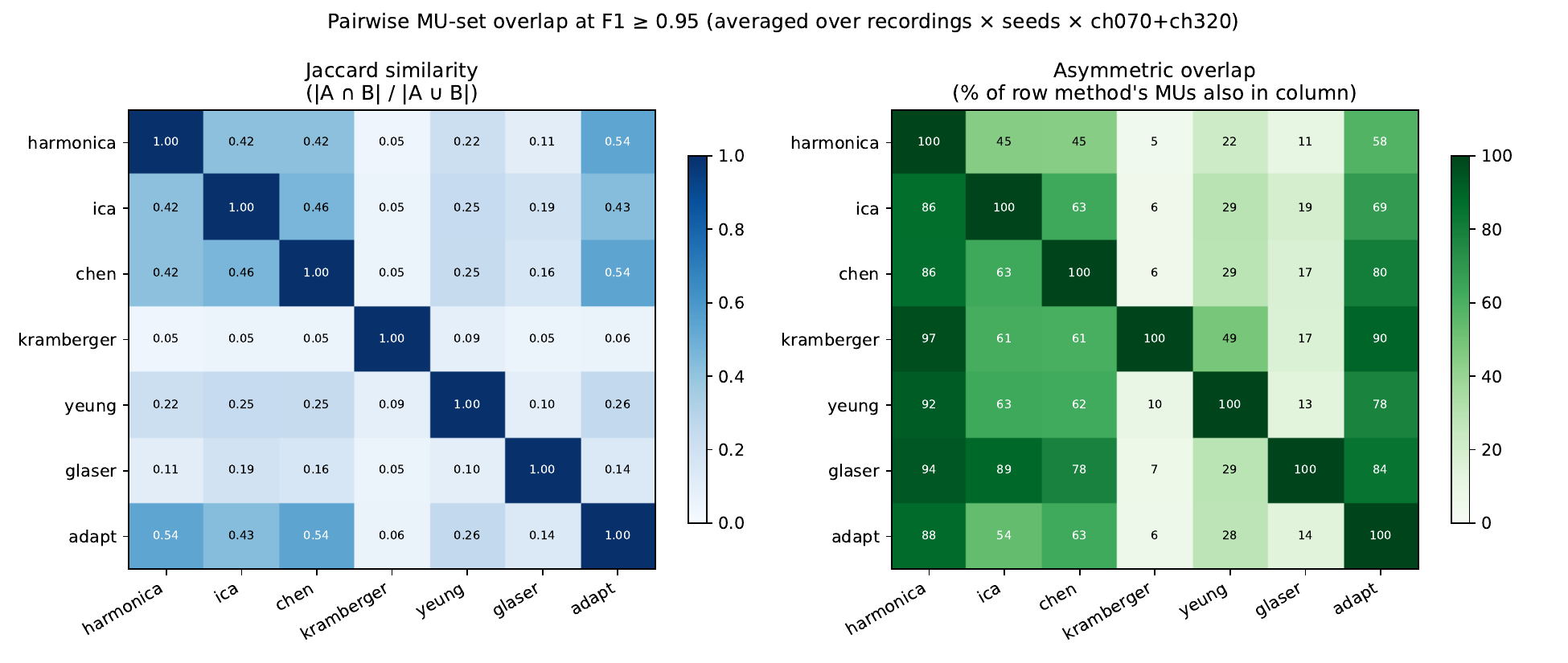}
\caption{\textbf{Pairwise MU-set overlap at $F_1 \geq 0.95$.}
Left: Jaccard similarity (symmetric). Right: \% of row method's
accepted MUs also accepted by the column method (asymmetric;
diagonal $= 100$). \emph{adapt}: Mendez Guerra 2024 with
hyperparameters tuned by the method's author on the validation split.}
\label{fig:supp:pairwise_overlap}
\end{figure}

\paragraph{Pairwise method-vs-method overlap (figure).}
Figure~\ref{fig:supp:pairwise_overlap} plots, for every pair of methods
$(A, B)$, the average overlap of their accepted-MU sets at $F_1 \geq
0.95$ across $168$ recordings $\times 5$ seeds $\times$ both densities.
The left panel shows Jaccard similarity ($|A \cap B|/|A \cup B|$),
which is symmetric. The right panel shows the asymmetric coverage
matrix: the $(i,j)$ cell is the fraction of the row method's
accepted MUs that are also in the column method's set; $86$--$97\,\%$
of every baseline's accepted MUs are also in HarmonICA's accepted set
(the \emph{harmonica} column; Mendez Guerra 2024 $88\,\%$), while at
most $58\,\%$ of HarmonICA's MUs are in any single baseline (the
\emph{harmonica} row; Mendez Guerra 2024 $58\,\%$, ICA $45\,\%$,
Chen $45\,\%$, Yeung $22\,\%$, Glaser $11\,\%$, Kramberger $5\,\%$).
Adaptive / segmented baselines cluster with each other and with ICA
(Chen $\leftrightarrow$ Mendez Guerra 2024 Jaccard $0.54$, the highest
non-HarmonICA pair; Chen $\leftrightarrow$ ICA $0.46$). Kramberger's
Jaccard is $\leq 0.09$ with every other method mainly because it is
run once (it is deterministic), so in this seed-pooled view its set is
empty on four of the five seeds; restricted to seed $0$, its Jaccard
with the other methods is $0.13$--$0.33$, and $97\,\%$ of its accepted
MUs are also in HarmonICA's set.

\begin{figure}[h]
\centering
\includegraphics[width=\textwidth]{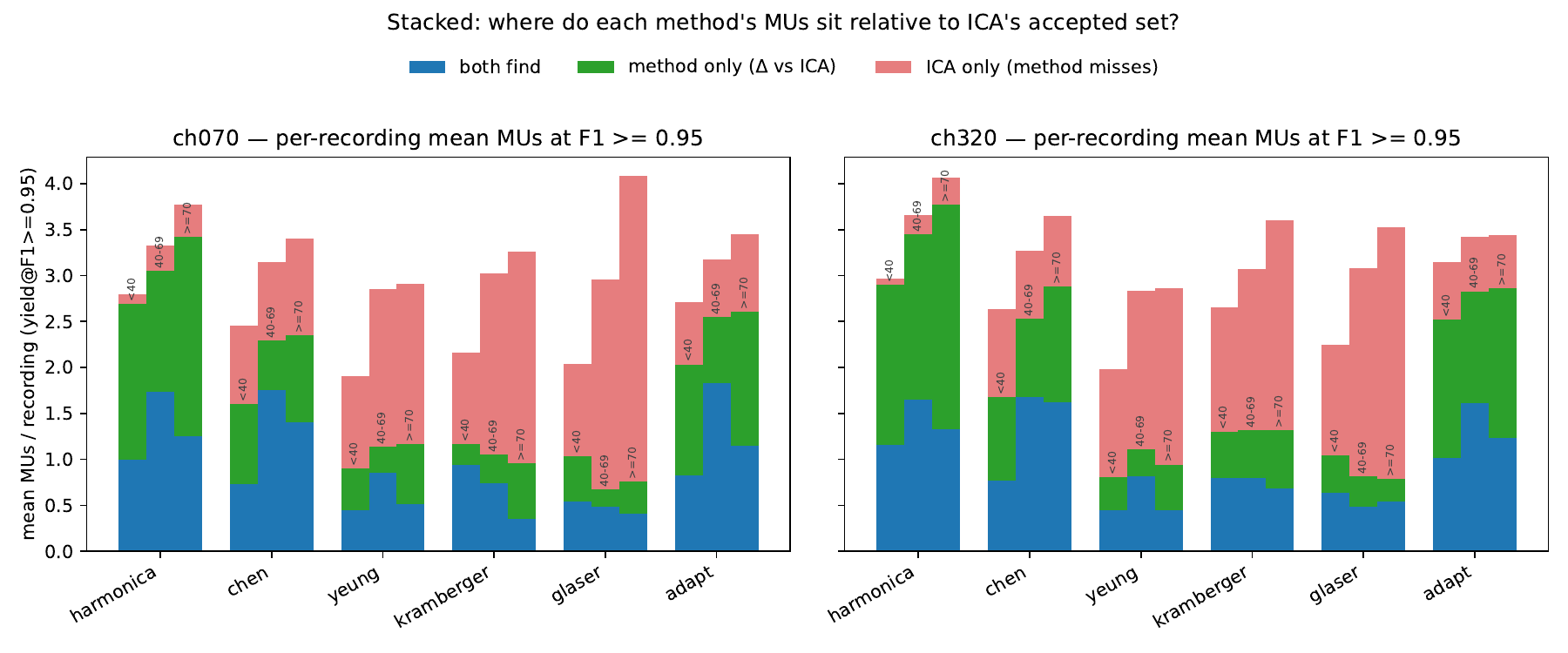}
\caption{\textbf{Per-method MU-discovery breakdown vs.\ the unrefined
FastICA seed at $F_1 \geq 0.95$.} Each cluster of bars is one method;
within each cluster the three bars correspond to GT motor-unit count
bins ($<40$, $40$--$69$, $\geq 70$). Stack components: \emph{both} =
GT MU found by both method and ICA, \emph{method only} = GT MU found
by method but not by ICA (novel), \emph{ICA only} = GT MU found by
ICA but not by method (missed). \emph{adapt}: Mendez Guerra 2024
with hyperparameters tuned by the method's author on the validation split.}
\label{fig:supp:overlap_bars}
\end{figure}

\paragraph{Per-method discovery breakdown vs.\ ICA (figure).}
Figure~\ref{fig:supp:overlap_bars} stacks, per method and per
density-bin, the mean number of MUs at $F_1 \geq 0.95$ split into three
mutually exclusive components: GT motor units found by both the method
and the unrefined FastICA seed (blue), GT MUs found by the method but
not by ICA (green; \emph{novel discoveries}), and GT MUs found by ICA
but not by the method (red; \emph{misses}). HarmonICA has the largest green stack in every density $\times$ bin
cell ($1.32$--$2.44$ MUs/rec, vs.\ at most $1.63$ for any baseline),
its green stack exceeds its blue stack in $5$ of $6$ cells, and its
green-to-total ratio peaks in dense pools (the rightmost
$n_\text{gt} \geq 70$ bin). Mendez Guerra 2024 has the second-largest
discovery component ($1.02$ / $1.39$ novel MUs/rec on ch070 / ch320,
vs.\ $1.63$ / $1.96$ for HarmonICA; green $>$ blue in $4$ of $6$ cells). Glaser, Yeung, and Kramberger sit
predominantly in the blue/red zones with small green stacks, indicating
they recover a subset of ICA's GT MUs rather than discovering new ones.

\subsection{Statistical comparisons}
\label{supp:stats}

The main claim of §5 of the main paper, that HarmonICA
outperforms every adaptive ICA baseline at every recall threshold,
is supported here by per-recording bootstrap $95\%$ confidence
intervals on yield at $F_1 \geq 0.95$
(Table~\ref{tab:supp:bootstrap}) and by paired Wilcoxon
signed-rank tests of the per-recording HarmonICA $-$ baseline
difference (Table~\ref{tab:supp:wilcoxon}). Per-recording yield is
defined as the count of extracted motor units at $F_1 \geq 0.95$,
averaged over the $5$ FastICA-initialisation seeds (a single
deterministic seed for Kramberger 2021). Pairing is at the recording
level; for Kramberger 2021, the paired test is restricted to the
$144$ (ch070) / $146$ (ch320) recordings on which Kramberger
returns $\geq 1$ MU, the same recordings over which its
per-recording means are taken (§\ref{supp:reproducibility}). On the remaining
$24$ / $22$ recordings Kramberger returns no units; pairing them as
zero-yield outcomes would only add HarmonICA wins and produce
uniformly smaller $p$-values.
Bootstrap CIs use $5\,000$ resamples drawn with replacement from
the per-recording distribution; intervals are reported at the
$95\%$ level.

\begin{table}[h]
\centering
\caption{Bootstrap $95\%$ confidence interval on per-recording mean
yield at $F_1 \geq 0.95$. The HarmonICA interval does not overlap
with any baseline interval on either grid. Kramberger 2021 is averaged over the recordings on which it returns $\geq 1$~MU only (§\ref{supp:reproducibility}).}
\label{tab:supp:bootstrap}
\small
\begin{tabular}{l c c c c c c}
\toprule
& \multicolumn{3}{c}{\textbf{ch070}} & \multicolumn{3}{c}{\textbf{ch320}} \\
\cmidrule(lr){2-4} \cmidrule(lr){5-7}
Method & mean & 95\% CI lo & 95\% CI hi & mean & 95\% CI lo & 95\% CI hi \\
\midrule
\textbf{HarmonICA (ours)}     & $\mathbf{2.83}$ & $2.51$ & $3.14$ & $\mathbf{3.06}$ & $2.71$ & $3.41$ \\
ICA / cBSS reference          & $1.48$          & $1.16$ & $1.82$ & $1.49$          & $1.18$ & $1.82$ \\
Chen 2020                     & $1.36$          & $1.13$ & $1.60$ & $1.60$          & $1.34$ & $1.88$ \\
Kramberger 2021               & $0.72$          & $0.59$ & $0.86$ & $0.91$          & $0.75$ & $1.06$ \\
Yeung 2024                    & $0.72$          & $0.61$ & $0.83$ & $0.65$          & $0.54$ & $0.76$ \\
Glaser 2018                   & $0.44$          & $0.35$ & $0.53$ & $0.48$          & $0.37$ & $0.60$ \\
Mendez Guerra 2024                         & $1.71$          & $1.47$ & $1.95$ & $2.08$          & $1.81$ & $2.36$ \\
\bottomrule
\end{tabular}
\end{table}

\begin{table}[h]
\centering
\caption{Paired Wilcoxon signed-rank tests on per-recording yield
at $F_1 \geq 0.95$. ``Wins'' is the count of recordings where
HarmonICA exceeds the baseline; ``losses'' the reverse; ``ties''
where they are equal. $\Delta$ is the mean per-recording difference
in MUs at $F_1 \geq 0.95$. All p-values are far below the
Bonferroni-corrected threshold for the $12$ tests reported here.}
\label{tab:supp:wilcoxon}
\small
\begin{tabular}{l c r r r r r}
\toprule
HarmonICA vs. & grid & $n_{\text{rec}}$ & $\Delta$ (MUs/rec) & $p$-value & wins & losses (ties) \\
\midrule
ICA / cBSS reference  & ch070 & $168$ & $+1.35$ & $2.4\!\times\!10^{-20}$  & $150$ & $\phantom{0}10$ ($\phantom{0}8$) \\
ICA / cBSS reference  & ch320 & $168$ & $+1.57$ & $2.3\!\times\!10^{-22}$  & $142$ & $\phantom{0}14$ ($12$) \\
Chen 2020             & ch070 & $168$ & $+1.47$ & $2.0\!\times\!10^{-26}$  & $148$ & $\phantom{00}6$ ($14$) \\
Chen 2020             & ch320 & $168$ & $+1.46$ & $6.0\!\times\!10^{-26}$  & $148$ & $\phantom{00}9$ ($11$) \\
Kramberger 2021       & ch070 & $144$ & $+2.23$ & $2.8\!\times\!10^{-24}$  & $136$ & $\phantom{00}2$ ($\phantom{0}6$) \\
Kramberger 2021       & ch320 & $146$ & $+2.35$ & $2.2\!\times\!10^{-25}$  & $143$ & $\phantom{00}1$ ($\phantom{0}2$) \\
Yeung 2024            & ch070 & $168$ & $+2.10$ & $2.9\!\times\!10^{-27}$  & $154$ & $\phantom{00}4$ ($10$) \\
Yeung 2024            & ch320 & $168$ & $+2.41$ & $1.1\!\times\!10^{-27}$  & $158$ & $\phantom{00}0$ ($10$) \\
Glaser 2018           & ch070 & $168$ & $+2.39$ & $1.4\!\times\!10^{-27}$  & $156$ & $\phantom{00}2$ ($10$) \\
Glaser 2018           & ch320 & $168$ & $+2.58$ & $5.1\!\times\!10^{-28}$  & $160$ & $\phantom{00}0$ ($\phantom{0}8$) \\
Mendez Guerra 2024                 & ch070 & $168$ & $+1.13$ & $4.5\!\times\!10^{-22}$  & $141$ & $\phantom{0}10$ ($17$) \\
Mendez Guerra 2024                 & ch320 & $168$ & $+0.98$ & $1.1\!\times\!10^{-23}$  & $140$ & $\phantom{0}18$ ($10$) \\
\bottomrule
\end{tabular}
\end{table}

\section{Implementation details}
\label{supp:impl}

This appendix lists the operating-point values used for the
headline benchmark on \emph{muniverse-benchmark-dynamic}. The same
settings are used on the 70- and 320-channel grids. The upstream
FastICA seed-pool generator that supplies the candidate separators
$\{\mathbf{w}_j^{(0)}\}$ to Algorithm~1 of the main paper is
described in §\ref{supp:ica:cbss}; a glossary of the symbols
introduced in §3 of the main paper is provided in
§\ref{supp:notation} at the end of this document.

\begin{table}[h]
\centering
\caption{Implementation details for HarmonICA and the baselines.}
\label{tab:supp:impl}
\small
\begin{tabular}{@{}>{\raggedright\arraybackslash}p{0.40\linewidth} >{\raggedright\arraybackslash}p{0.56\linewidth}@{}}
\toprule
\multicolumn{2}{l}{\textbf{HarmonICA pipeline}} \\
\midrule
Hardware                    & 1 $\times$ NVIDIA A40 GPU \\
Framework                   & PyTorch 2.x, CUDA 12.x \\
Sampling rate $f_s$         & $2048$ Hz \\
Toeplitz extension factor $L$ & $8$ \\
PCA dimensionality $K$      & $128$ \\
Whitening                   & ZCA \\
ICA seed pool size          & up to $108$ candidates \\
ICA silhouette gate         & $0.80$ \\
Compensation form           & low-rank Woodbury, rank $r = 3$ \\
Compensator MLP             & $3$ hidden layers, width $64$, ReLU, dropout $0.2$ \\
Output initialisation       & zero (so $U_j(t)=0$ and $A_j(t)=I$ at training start) \\
Positional-encoding periods $T_1,\ldots,T_P$ & $\{0.1, 0.2, 0.5, 1.0, 2.0\}$ s ($P=5$) \\
Independence-loss exponent $e$ & $3$ \\
Optimiser                   & RMSprop \\
Separator learning rate $\eta_w$ & $1\!\times\!10^{-5}$ \\
Compensator learning rate $\eta_\phi$ & $1\!\times\!10^{-4}$ \\
Total epochs / source       & $200$ \\
Separator-only warmup $T_w$ & $50$ epochs \\
Spike/null re-split period $K_g$ & $5$ compensation steps \\
Spike-detection threshold   & $3\sigma$ above background \\
Spike/null split            & two-means on peak heights; upper cluster = spikes \\
GMM null-mode samples       & $1000$ random background samples \\
Spike-match tolerance       & $\pm 2$ ms (with $\pm 25$ ms shift sweep) \\
Seeds per recording         & $5$ paired (FastICA, compensator) seeds: FastICA $\in \{1909, 1910, 1911, 1912, 1913\}$, compensator $\in \{42, 43, 44, 45, 46\}$ \\
Approx.\ runtime            & $44$--$47$ s / recording \\
\midrule
\multicolumn{2}{l}{\textbf{Baselines (extension factor $L = 8$ for all)}} \\
\midrule
ICA / cBSS \cite{negro2016multi}                 & inherited (FastICA seed inside HarmonICA pipeline); $\sim$$28$ s / recording \\
Mendez Guerra 2024 \cite{guerra2024adaptive}                  & v1.0.0, \cite{mendezguerra2024adaptcode}, MIT; tuned by its author on val (§\ref{supp:baseline_deviations}); $\sim$$4$ / $6$ s / recording on one CPU thread (ch070 / ch320) \\
Chen 2020 \cite{chen2020adaptive}                & reimplemented from paper; $\sim$$13$ s / recording \\
Kramberger 2021 \cite{kramberger2021prediction}  & reimplemented (deterministic, $1$ seed); $3$--$9$ s / recording \\
Yeung 2024 \cite{yeung2024adaptive}              & reimplemented from paper; $\sim$$11$ s / recording \\
Glaser 2018 \cite{glaser2018motor}               & reimplemented per Pseudocode~1 (cyclostationary CKC with PNR-gated tracking); $5$--$50$ s / recording \\
\bottomrule
\end{tabular}
\end{table}

\paragraph{Compute budget and hyperparameter selection.}
Unless stated otherwise, all experiments ran on a single NVIDIA A40
GPU. The headline
benchmark per electrode grid is $168$~recordings $\times 5$ seeds.
At HarmonICA's $\sim$$45$\,s per recording, one full grid is
$\sim$$10.5$\,GPU-hours; both grids together are
$\sim$$21$\,GPU-hours. Adding Chen 2020, Kramberger 2021, Yeung 2024
and Glaser 2018 (Table~\ref{tab:yield_3bin}) at the per-recording
runtimes given in Table~\ref{tab:supp:impl} brings that benchmark
to $\sim$$50$\,GPU-hours. Mendez Guerra 2024 runs on the CPU and adds
$\sim$$2.5$\,CPU-hours of method time over both grids; its
hyperparameter search ($100$ trials per grid on the validation split)
was run by the method's author on separate hardware.
 The five ablation rows of the headline ablation table
(\S\ref{supp:abl:headline}) and the loss-family / rank /
PE-vs-angle sweeps in this appendix add roughly another
$\sim$$200$\,GPU-hours; total compute spent on the experiments
reported in the paper is on the order of $300$ GPU-hours.

Preprocessing and
detection settings were inherited from the upstream FastICA + cBSS
pipeline~\cite{negro2016multi}; the Toeplitz extension factor was
swept over $L \in \{8, 16, 32, 64\}$ with $L = 8$ adopted;
the compensator rank, warmup length, total epochs, and
compensator architecture were validated against mean $F_1$ on val
and showed no meaningful sensitivity within the ranges tested.

\subsection{Woodbury parameterisation of the compensator}
\label{supp:woodbury_param}

Section~3.2 of the main paper parameterises the compensator at
rank $r \ll K$ as
\begin{equation}
A_j(t) = I + U_j(t)\,V_j^\top, \qquad
U_j(t) \in \mathbb{R}^{K \times r},\; V_j \in \mathbb{R}^{K \times r},
\end{equation}
where $U_j(t)$ is the output of a small per-source neural network
and $V_j$ is a learned per-source matrix shared across time. Two
design properties make this form essential rather than a
generic low-rank approximation.

\paragraph{Per-timestep network output count.}
A time-varying invertible $K \times K$ matrix expressed in any
generic factorisation (LU, QR, full-rank) requires $O(K^2)$
network outputs per timepoint. The rank-$r$ Woodbury factorisation
needs only $rK$ outputs, since $V_j$ is shared across time and
only $U_j(t)$ is emitted at every $t$. With $K = 128$ and $r = 3$
that is $384$ outputs per timepoint against $\sim 16$k for an
unconstrained form, a $43\times$ reduction in the conditioning
network's per-timestep emission load.

\paragraph{Closed-form inverse.}
The Woodbury identity~\cite{hager1989updating, lu2020woodbury} gives
\begin{equation}
A_j(t)^{-1} = I - U_j(t)\,\big(I_r + V_j^\top U_j(t)\big)^{-1}\,V_j^\top,
\label{eq:supp:woodbury}
\end{equation}
which reduces inversion to an $r \times r$ solve with no
$K \times K$ matrix materialised. At $r = 3$ the inner $r \times r$
inverse is itself analytical (closed-form determinant and adjugate),
so $A_j(t)^{-1}$ is an algebraic expression that we evaluate at
every spike time during neural peel-off (§3.5 of the main paper).
The total per-timepoint cost is $O(rK)$, against $O(K^3)$ for a
generic dense inverse of a time-varying $K \times K$ matrix.

\paragraph{Why $r = 3$ is sufficient.}
Rank $3$ is empirically sufficient on the surface-EMG benchmark
(\S\ref{supp:abl:rank}) and consistent with PCA studies of
electrode-shift-induced waveform variation, which find that the
non-stationary part of the surface-EMG mixing concentrates in a
handful of spatial directions~\cite{naik2016principal, huang2020low}.
The rank-sweep ablation in \S\ref{supp:abl:rank} confirms that
lifting $r$ from $3$ to $7$ buys roughly $2\,\%$ additional yield
at $F_1 \geq 0.95$, plateauing between $r = 5$ and $r = 7$, while
more than doubling the per-timepoint output dimensionality.

\subsection{Baseline implementation deviations}
\label{supp:baseline_deviations}

We list every place our implementations of
the published baselines depart from the configuration printed in
the original paper. Two of the four reimplemented baselines (Chen,
Yeung) carry small numerical-stability or threshold adjustments
that we adopted after observing that the literal published recipe
under-performs on \emph{muniverse-benchmark-dynamic}; both shifts
\emph{help} the baseline. Glaser and Kramberger are
paper-faithful (no deviations). Mendez Guerra 2024 runs the authors' own online engine through a
wrapper supplied by the method's first author, with three
hyperparameters tuned by the method's author on the validation split;
the last paragraph of this section describes the protocol, the
tuning, and the result with the package's untuned NeuroMotion configuration.

\paragraph{Chen 2020: EMA covariance update.}
Chen's online stage tracks the extended-signal covariance $C$
recursively. The paper writes Eq.~(5) as a cumulative sum,
$C \leftarrow C + C_\Delta$. On long recordings
($T = 10$\,s, $\sim 20480$ samples) this drives $C$ away from
the calibration centroid; we adopt an exponential moving average,
$C \leftarrow \alpha C + (1-\alpha)\,C_\Delta$ with
$\alpha = 0.85$, in the spirit of the paper's $\ell = 0.1$
learning-rate scaling on the same step but numerically stable on
long signals. The substituted form is what we report in the
headline tables.

\paragraph{Yeung 2024: spike-detection $z$-threshold.}
Yeung's published default is $z_{\text{th}} = 3.3$, tuned on
extensor carpi radialis brevis (ECRB), a sparse muscle with
$\sim 10$--$15$ MUs per recording. On our denser $23$--$101$-MU
pools (see §\ref{supp:bench:construction}, Table~\ref{tab:supp:dataset}) a $3.3$-$\sigma$ floor
rejects most low-amplitude MU spikes. We adopt
$z_{\text{th}} = 2.0$ as the density-appropriate operating point;
the paper's $3.3$ default should be restored
on ECRB-like sparse muscles.

\paragraph{Glaser 2018: paper-faithful implementation.}
Our Glaser implementation follows the full
\emph{cyclostationary CKC} algorithm of Pseudocode~1 in §II-C of
the original paper, including PNR-gated online tracking. No
deviations from the published configuration.

\paragraph{Mendez Guerra 2024: protocol and author-tuned hyperparameters.}
The row in main Table~\ref{tab:yield_3bin} runs the authors' adapt\_decomp
v1.0.0 engine~\cite{mendezguerra2024adaptcode} through a wrapper supplied by the
method's first author, starting from the package's NeuroMotion configuration. It was
produced with the method author's benchmark harness, which shares our
scoring code and band-passes every recording at $10$--$1000$\,Hz
(zero-phase) before any method runs. Calibration then uses the stack
of §\ref{supp:ica:cbss} on the first $1.5$\,s of each recording
(Toeplitz extension $L = 8$, PCA to $K = 128$, ZCA whitening, FastICA
with peel-off and silhouette gate $0.80$); the engine then updates
the whitening matrix, the separation vectors and the spike/baseline
centroids in $100$\,ms batches over the remaining $8.5$\,s. The output
spike train is the calibration spikes on the first $1.5$\,s followed by
the engine's spikes. The method's author tuned three hyperparameters
per grid (whitening learning rate, separation-vector learning rate,
centroid momentum) with the package's multi-objective search (Optuna,
$100$ trials, multivariate TPE sampler) on the $40$-recording
validation split, whose subjects are disjoint from those of the test
split, using one FastICA seed. The objective of this search uses no
ground truth: it is the package's whitening and separation adaptation
losses, summed over recordings and restricted to calibration units
whose inter-spike-interval coefficient of variation is at most $0.3$,
and the selected configuration is the Pareto-front member with the
lowest separation loss. The selected values are
$(4.6\!\cdot\!10^{-2},\, 1.9\!\cdot\!10^{-2},\, 0.3)$ on ch070 and
$(4.7\!\cdot\!10^{-2},\, 1.0\!\cdot\!10^{-2},\, 0.1)$ on ch320, against
$(7\!\cdot\!10^{-3},\, 3\!\cdot\!10^{-3},\, 0.95)$ in the package's
NeuroMotion configuration. It is the only baseline tuned by its own
authors. With the NeuroMotion
configuration, the same wrapper and the same calibration,
the method recovers $152 / 149$ motor units at $F_1 \geq 0.95$
(ch070 / ch320, mean over $5$ seeds), below the
$249 / 249$ of the ICA / cBSS reference in Table~\ref{tab:yield_3bin}; tuning raises this
to $286 / 350$. The package's MUniverse configuration was not
evaluated. The row is re-binned to the ground-truth convention of the
main paper (non-empty spike trains only; tiers $< 40$, $40$--$69$ and
$\geq 70$). As a check on the harness
difference, the two harnesses give unrounded mean ICA / cBSS totals of
$249.0$ vs.\ $248.6$ (ch070) and $258.0$ vs.\ $249.8$ (ch320) motor
units, and $1.43$ vs.\ $1.42$ (ch070) and $1.39$ vs.\ $1.32$ (ch320)
MUs per recording-seed at recall $\geq 0.95$. In the method author's
reported results for the same harness, Chen 2020 recovers
$234 / 287$, Glaser 2018 $89 / 100$ and Kramberger 2021 $109 / 132$
motor units (Table~\ref{tab:yield_3bin}: $228 / 269$, $74 / 81$ and $104 / 133$); within that
harness, too, Mendez Guerra 2024 is the strongest baseline.

\section{Side-information ablation: positional encoding vs.\ joint angle}
\label{supp:pe_vs_angle}

The compensator MLP that produces $U_j(t)$ in the rank-$3$ Woodbury layer
takes a context vector at every time step. Two natural choices are
available on \emph{muniverse-benchmark-dynamic}:

\begin{itemize}
\item \textbf{PE (positional encoding, blind).} The input is a
  $10$-channel sinusoidal positional encoding (sine and cosine) at fixed periods
  $\{0.1, 0.2, 0.5, 1.0, 2.0\}$\,s. The compensator can model arbitrary
  smooth time variation but is told nothing about the joint angle
  driving the non-stationarity. This is the headline configuration
  reported in §5 of the main paper.
\item \textbf{Angle.} The input is the scalar joint-angle trace of
  shape $(T, 1)$ used by the synthesis pipeline. The compensator is
  conditioned on the same biomechanical state variable that
  Kramberger~2021 \cite{kramberger2021prediction} requires; it is no
  longer ``blind'' in the strict sense and is more directly comparable
  to the angle-aware baseline.
\end{itemize}

Table~\ref{tab:supp:pe_vs_angle} compares the two regimes on the same
test split, with all baselines held fixed. Both versions of HarmonICA
dominate every adaptive baseline at every recall threshold and on both
electrode grids; the angle-conditioned version is uniformly stronger,
recovering between $1$ and $27$ additional motor units per
bin at $F_1 \geq 0.95$, and lifting the
Pareto-crossover yield at $R \geq 0.95$ from $3.24$ to $3.66$ MUs per
recording-seed on the $320$-channel grid (a $2.8\times$ lift over ICA
versus the $2.5\times$ obtained from PE) and from $3.16$ to $3.34$ on
the $70$-channel grid. PE is the headline configuration of the main
paper, and deliberately so: it is the more conservative claim, it
matches the side-information available to the five non-Kramberger
baselines, and it isolates the contribution of the compensator MLP
itself from any benefit conferred by direct access to the synthesis
state variable.

\begin{table}[h]
\centering
\caption{HarmonICA bin-level total motor units recovered at $F_1 \geq
0.95$ under PE (sinusoidal positional encoding, blind) versus angle
(real joint-angle trace) compensator inputs. Mean $\pm$ s.d.\ over
the $5$ FastICA initialisation seeds ($10$ for angle). Same test split as
main Table~\ref{tab:yield_3bin}. Bin recording counts
are $40$ / $80$ / $48$.}
\label{tab:supp:pe_vs_angle}
\begin{tabular}{llccc}
\toprule
Grid & Side info & $n_{\text{gt}} < 40$ & $n_{\text{gt}}\,40\!-\!69$ & $n_{\text{gt}} \geq 70$ \\
\midrule
\multirow{2}{*}{ch070} & PE      & $90 \pm 4.6$  & $252 \pm 24$  & $134 \pm 5.5$  \\
                       & angle   & $\mathbf{91 \pm 4.7}$  & $\mathbf{265 \pm 11}$  & $\mathbf{146 \pm 13}$   \\
\midrule
\multirow{2}{*}{ch320} & PE      & $96 \pm 2.2$  & $263 \pm 4.4$  & $156 \pm 5.3$  \\
                       & angle   & $\mathbf{102 \pm 4.6}$ & $\mathbf{290 \pm 5.3}$ & $\mathbf{179 \pm 5.5}$  \\
\bottomrule
\end{tabular}
\end{table}

\begin{table}[h]
\centering
\caption{Pareto-crossover yield (MUs per recording-seed at recall
$\geq R$) for HarmonICA under PE vs.\ angle, alongside the unrefined
ICA / cBSS reference (which is identical under both regimes since it
does not see the side input) and Mendez Guerra 2024, the strongest
adaptive baseline at $R \geq 0.95$.}
\label{tab:supp:pe_vs_angle_pareto}
\begin{tabular}{llcccc}
\toprule
Grid & Method & $R \geq 0.99$ & $R \geq 0.95$ & $R \geq 0.90$ & $R \geq 0.80$ \\
\midrule
\multirow{4}{*}{ch070} & HarmonICA (PE)    & $2.23$ & $3.16$ & $3.80$ & $4.71$ \\
                       & HarmonICA (angle) & $2.36$ & $3.34$ & $3.96$ & $4.85$ \\
                       & ICA / cBSS         & $0.94$ & $1.42$ & $1.67$ & $2.11$ \\
                       & Mendez Guerra 2024 & $0.57$ & $1.39$ & $1.82$ & $2.61$ \\
\midrule
\multirow{4}{*}{ch320} & HarmonICA (PE)    & $2.17$ & $3.24$ & $3.84$ & $4.70$ \\
                       & HarmonICA (angle) & $2.49$ & $3.66$ & $4.21$ & $5.02$ \\
                       & ICA / cBSS         & $0.87$ & $1.32$ & $1.67$ & $2.16$ \\
                       & Mendez Guerra 2024 & $0.72$ & $1.62$ & $2.30$ & $3.32$ \\
\bottomrule
\end{tabular}
\end{table}

\section{Headline ablations}
\label{supp:abl:headline}

This section reports the headline ablations summarised in the main
paper, adds a fifth ablation that re-couples the
separator and compensator gradients, and gives the full
explanation for each row that the main paper omits for space.
All numbers are means over $5$ FastICA-initialisation seeds
on the $168$-recording test split (same protocol as the main results).

\paragraph{Cold-start $\mathbf{w}_j$ (random init).}
Replacing the FastICA seed with a random initialisation collapses
yield at $F_1 \geq 0.95$ to zero on both grids and drops
top-$1000$ mean $F_1$ to $\sim 0.28$. The compensator cannot
recover sources the separator has not already identified, and a
random separator with no good direction at training start gives
the compensator nothing useful to refine.

\paragraph{No peel-off.}
Removing the per-spike peel-off step entirely, so that each ground-truth
unit is instead refined from its best-matching ICA source (an oracle
pairing, which favours this row), costs
$\sim 77$ MUs per seed at $F_1 \geq 0.95$ on ch070 ($-16\%$) and
$25$ on ch320 ($-5\%$); the smaller drop in top-$1000$ mean $F_1$
($0.016$ / $0.013$) reflects that peel-off acts on the residual
rather than on individual source quality.

\paragraph{Kurtosis loss instead of GMM.}
Swapping the GMM bimodality criterion of Eq.~\ref{eq:gmm} of the main paper
for the fourth-moment contrast
$L_{\mathrm{kurt}} = -\mathbb{E}[\hat{s}_j(t)^{4}]$ drops yield at
$F_1 \geq 0.95$ by $74\%$ / $81\%$ (ch070 / ch320), measured on a
filter-space variant of the per-source refiner without inter-source peel-off
(Table~\ref{tab:supp:loss}). The mechanism is the standard
caveat against kurtosis as a contrast for sparse spike sources:
maximising $\mathbb{E}[s^{4}]$ rewards the existence of a few
extreme samples rather than the existence of two well-separated
clusters, so the compensator collapses the recovered source onto
a handful of outlying spikes and the rest of the spike train,
along with the null between spikes, is pushed to similar low
amplitudes. The result is high contrast but a degenerate spike
train. The full sweep of stationarity-loss alternatives and
case-study source plots is in §\ref{supp:abl:loss}.

\paragraph{Per-channel affine transform instead of low-rank Woodbury.}
Replacing the rank-$3$ Woodbury factorisation of Eq.~\ref{eq:lowrank} of the
main paper with a per-channel affine correction
$A_j(t)\tilde{\mathbf{x}}(t) = \tilde{\mathbf{x}}(t) \odot \exp(\log\mathbf{w}_j(t)) + \mathbf{b}_j(t)$,
with $\log\mathbf{w}_j(t)$ and $\mathbf{b}_j(t)$ output by $\mathrm{NN}_j(\boldsymbol{\gamma}(t);\phi_j)$,
drops yield at $F_1 \geq 0.95$ by $39\%$ / $41\%$ and top-$1000$
mean $F_1$ by $0.043$ / $0.053$. Two failure modes compound:
the diagonal scaling acts on each channel separately and cannot
express the cross-channel MUAP rotation that the rank-$r$ term
captures; and the unconstrained offset $\mathbf{b}_j(t)$
lets the compensator absorb signal that the linear separator
should be recovering, since any offset that drives the GMM cluster
variance down is rewarded by
$L_N$ regardless of whether it preserves the
separator's identification of the source. The Woodbury form
disciplines both: cross-channel structure through a rank-$r$
perturbation of identity, and a structural cap on how much signal
the compensator can move (a multiplicative perturbation of identity
rather than an arbitrary additive vector). The compensator-rank
sweep (\S\ref{supp:abl:rank}) confirms that $r = 3$ is sufficient.

\paragraph{Coupled gradients (decoupling toggled off).}
Section~3.3 of the main paper trains the static separator $\mathbf{w}_j$
and the time-varying compensator $A_j(t)$ with
\emph{decoupled} gradients: the independence loss
$L_I$ backpropagates into the separator weights
$\mathbf{w}_j$ only, and the stationarity loss $L_N$
backpropagates into the compensator parameters $\phi_j$ only.
Letting both losses backprop into both modules means the separator
is no longer the optimum of a linear-ICA contrast, so the conditions
of §\ref{supp:theory} no longer apply to it. The pipeline still trains: raw
mean $F_1$ drops only by $0.010$ / $0.017$. The regression
concentrates in the precision-weighted yield where the design
choice matters: top-$1000$ mean $F_1$ falls by $0.025$ / $0.028$
and yield at $F_1 \geq 0.95$ drops by $97$ (ch070, $-20\%$) and
$95$ (ch320, $-19\%$) motor units per seed. The interpretation is
the one predicted by the construction: when $L_N$ is
allowed to flow back into $\mathbf{w}_j$, the compensator can
absorb signal that the linear separator should be recovering,
trading a linear-ICA identification for a joint optimisation
with no such guarantee. The decoupled construction loses none
of the expressivity (the same $A_j(t)$ family is reachable; it is
the \emph{path} the optimiser is restricted from taking that
matters) and keeps the separator within the conditions of
§\ref{supp:theory}.

\section{Extended ablations}
\label{supp:ablations}

This section extends the headline ablations of
\S\ref{supp:abl:headline} along three further axes: the stationarity loss family
(\S\ref{supp:abl:loss}), the rank of the time-varying compensator
(\S\ref{supp:abl:rank}), and the side information fed to the
compensator MLP (\S\ref{supp:abl:pe_vs_angle}). All cells are mean
$\pm$ s.d.\ over $5$ FastICA-initialisation seeds on the
$168$-recording test split.

\subsection{Stationarity loss family}
\label{supp:abl:loss}

All losses in this section are compared on a filter-space variant of
the per-source refiner without inter-source peel-off.
The compensator stationarity loss $L_N$ (Eq.~\ref{eq:compensation_loss} of the main paper)
relaxes the bimodal spike/null prior to a two-component Gaussian
mixture and minimises the within-cluster spread. Classical FastICA
contrasts target outliers in the marginal distribution of the source
rather than tightness around the spike/null modes, and they fail in
qualitatively different ways. Let $\hat{s}_j(t)$ denote the
recovered source at time $t$ and write $\hat{s}^{\,\mathrm{n}}_j =
(\hat{s}_j - \bar{\hat{s}}_j) / \mathrm{std}(\hat{s}_j)$ for its
zero-mean unit-variance standardisation; we report each loss as a
quantity to minimise.

\begin{itemize}
\item \textbf{GMM (ours).} The two-component bimodality criterion of
  Eq.~\ref{eq:compensation_loss} in the main paper, denoted $L_N$ throughout the main paper and
  written here as $L_{\mathrm{GMM}}$ to disambiguate from the
  alternative contrasts below,
  \[
    L_{\mathrm{GMM}} \;=\; L_N \;=\; \sqrt{\sigma^2_{\mathrm{spike}} +
                                            \max(\sigma^2_{\mathrm{null}},\bar{\sigma}^2_{\mathrm{null}})},
  \]
  the root-summed within-cluster spread of $\hat{s}_j$ on the
  spike/null partition of main Eq.~\ref{eq:gmm}.
\item \textbf{Negentropy $g_1$ \cite{hyvarinen2000independent}.}
  The FastICA super-Gaussian contrast,
  \[
    L_{g_1} \;=\; -\,\mathbb{E}\!\left[
        \log \cosh \hat{s}^{\,\mathrm{n}}_j(t)
    \right].
  \]
  Maximising $\mathbb{E}[\log \cosh(\cdot)]$ favours heavy-tailed
  projections.
\item \textbf{Negentropy $g_2$ \cite{hyvarinen2000independent}.}
  The FastICA robust contrast,
  \[
    L_{g_2} \;=\; \mathbb{E}\!\left[
        \exp\!\bigl(-\tfrac{1}{2}\,
                    \hat{s}^{\,\mathrm{n}}_j(t)^{\,2}\bigr)
    \right].
  \]
  Minimising $\mathbb{E}[\exp(-\hat{s}^{\,\mathrm{n}\,2}/2)]$
  pushes mass into the tails of $\hat{s}^{\,\mathrm{n}}_j$.
\item \textbf{Kurtosis.} The classical fourth-moment contrast for
  super-Gaussian sources,
  \[
    L_{\mathrm{kurt}} \;=\; -\,\mathbb{E}\!\left[
        \hat{s}^{\,\mathrm{n}}_j(t)^{\,4}
    \right].
  \]
  For unit-variance $\hat{s}^{\,\mathrm{n}}_j$ this is equivalent to
  $-(\mathrm{kurt}(\hat{s}^{\,\mathrm{n}}_j) + 3)$, so minimising
  $L_{\mathrm{kurt}}$ maximises excess kurtosis.
\item \textbf{Power-$p$.} A sign-preserving high-power contrast used
  in some EMG decomposition pipelines,
  \[
    L_{\mathrm{pow}} \;=\; -\,\mathbb{E}\!\left[
        \mathrm{sgn}(\hat{s}^{\,\mathrm{n}}_j) \cdot
        \bigl|\alpha\,\hat{s}^{\,\mathrm{n}}_j\bigr|^{\,p}
    \right],
  \]
  with $\alpha = 10$ and $p = 4$. The amplification by $\alpha$
  emphasises large-amplitude samples; the absolute value with sign
  prefactor preserves the sign of $\hat{s}^{\,\mathrm{n}}_j$.
\end{itemize}

Table~\ref{tab:supp:loss} reports the full sweep on the per-source
iso-ablation pipeline (no inter-source peel-off; the GMM row is the
within-table reference).

\begin{table}[h]
\centering
\caption{Stationarity-loss substitution on the per-source
iso-ablation. Yield is per seed, summed over $168$ recordings.
Reference: GMM, our headline choice.}
\label{tab:supp:loss}
\small
\begin{tabular}{l c c c c}
\toprule
Loss & Mean $F_1$ & $n$ at $F_1 \geq 0.85$ & $n \geq 0.90$ & $n \geq 0.95$ \\
\midrule
\multicolumn{5}{l}{\textbf{ch070}} \\
\quad GMM (ours)               & $\mathbf{0.706 \pm 0.004}$ & $\mathbf{565.8 \pm 8.1}$  & $\mathbf{491.6 \pm 5.1}$  & $\mathbf{404.6 \pm 2.5}$ \\
\quad Negentropy ($g_1$)       & $0.616 \pm 0.008$ & $354.0 \pm 14.0$ & $282.8 \pm 5.0$  & $226.8 \pm 6.8$ \\
\quad Negentropy ($g_2$)       & $0.608 \pm 0.009$ & $341.8 \pm 8.1$  & $275.0 \pm 6.0$  & $221.8 \pm 7.0$ \\
\quad Power-$p$ ($|\hat{s}|^4$) & $0.399 \pm 0.004$ & $132.6 \pm 6.1$  & $117.0 \pm 5.6$  & $102.2 \pm 6.7$ \\
\quad Kurtosis                  & $0.296 \pm 0.005$ & $159.8 \pm 6.4$  & $136.6 \pm 8.4$  & $104.6 \pm 5.4$ \\
\midrule
\multicolumn{5}{l}{\textbf{ch320}} \\
\quad GMM (ours)               & $\mathbf{0.741 \pm 0.001}$ & $\mathbf{682.8 \pm 4.6}$  & $\mathbf{599.4 \pm 5.6}$  & $\mathbf{513.6 \pm 6.7}$ \\
\quad Negentropy ($g_1$)       & $0.632 \pm 0.006$ & $382.8 \pm 10.5$ & $308.8 \pm 7.9$  & $230.0 \pm 7.6$ \\
\quad Negentropy ($g_2$)       & $0.621 \pm 0.006$ & $360.4 \pm 11.8$ & $287.2 \pm 6.3$  & $219.4 \pm 11.5$ \\
\quad Power-$p$ ($|\hat{s}|^4$) & $0.406 \pm 0.002$ & $\phantom{0}92.0 \pm 3.2$ & $\phantom{0}74.8 \pm 4.2$ & $\phantom{0}65.0 \pm 3.2$ \\
\quad Kurtosis                  & $0.310 \pm 0.009$ & $162.0 \pm 8.1$  & $136.2 \pm 8.5$  & $\phantom{0}97.0 \pm 6.3$ \\
\bottomrule
\end{tabular}
\end{table}

The two negentropy contrasts cluster together ($\Delta F_1 \approx
-0.10$ vs.\ GMM, $-45\%$ in yield at $F_1 \geq 0.95$) and represent
the milder failure mode: the compensator finds an axis along which
the projection is non-Gaussian, but the resulting source is not
sharply bimodal and a substantial fraction of detected events are
mis-aligned to either spike or null. Kurtosis and power-$p$ collapse
($\Delta F_1 \approx -0.4$, $-75\%$ yield) because the
fourth-moment objective is dominated by outliers in the projected
amplitude distribution: the compensator pushes a small number of
samples far from zero rather than tightening clusters around two
modes, and the resulting source is unusable for spike detection.
The two-component bimodality criterion is therefore not a freely
swappable choice: it determines the geometry of the
recovered source, and classical FastICA contrasts cannot drive the
compensator alone.

\begin{figure}[H]
\centering
\includegraphics[width=\linewidth]{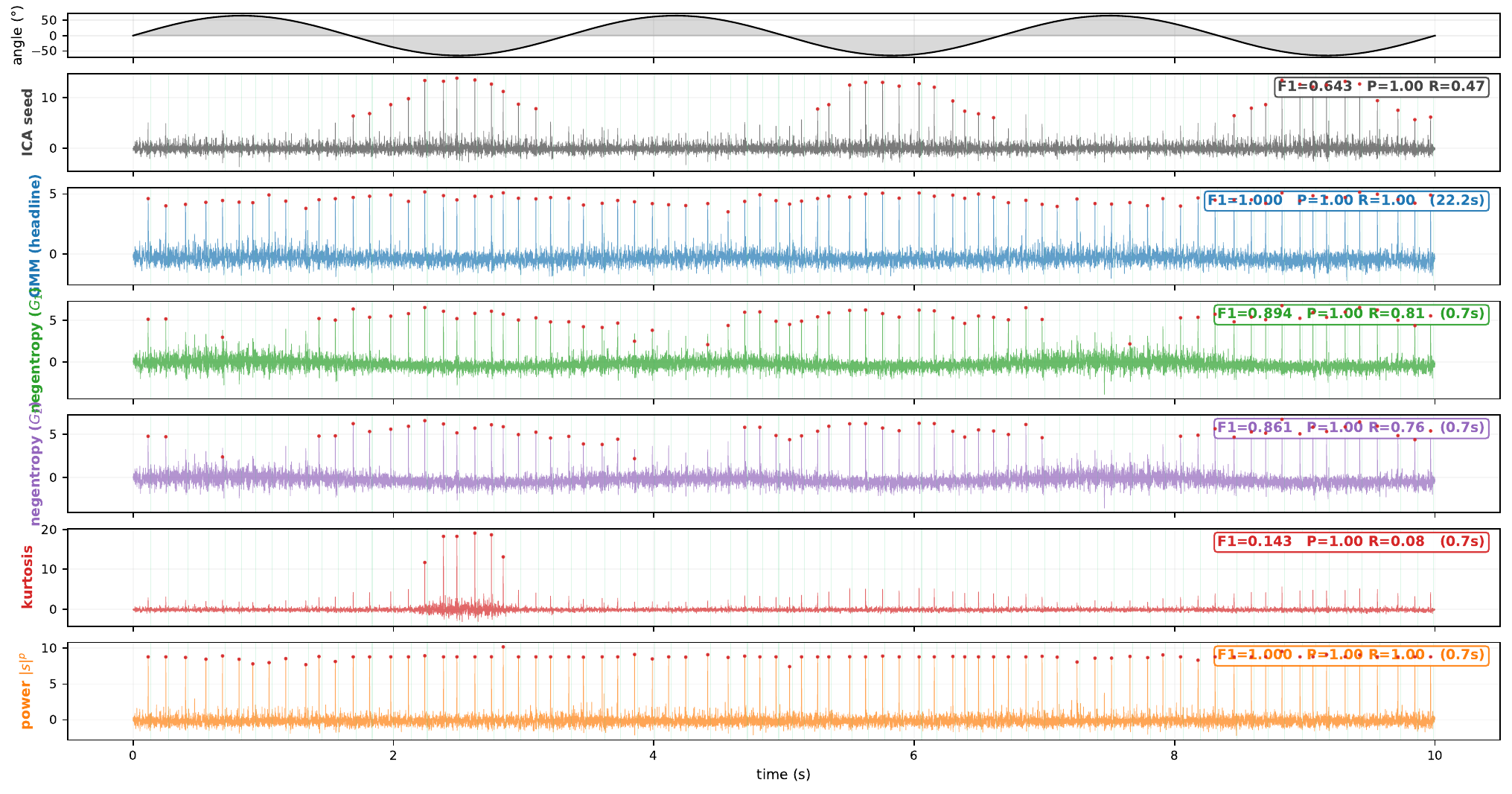}
\caption{\textbf{Rescue.} Only the GMM loss and the power-$p$ loss
work cleanly; kurtosis maximises a few spikes, the negentropy losses
fail to discover the harder spikes, and power-$p$ generally succeeds
in the easier cases.}
\label{fig:supp:loss_rescue}
\end{figure}

\begin{figure}[H]
\centering
\includegraphics[width=\linewidth]{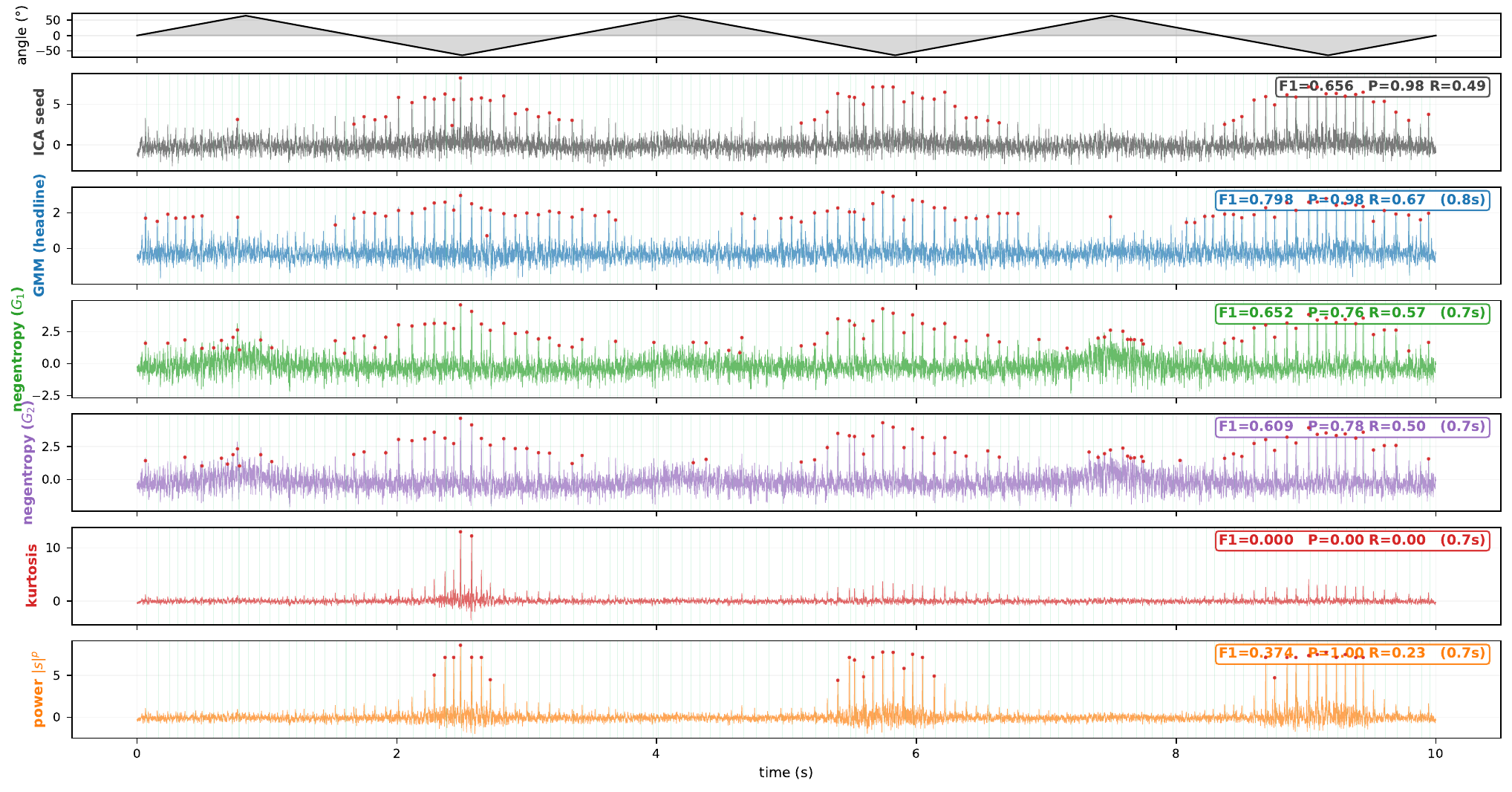}
\caption{\textbf{Partial success.} GMM yields the best result. The
negentropy losses still fail to discover a substantial fraction of
spikes; kurtosis still amplifies the central part of the source;
power-$p$ regresses.}
\label{fig:supp:loss_partial}
\end{figure}

\begin{figure}[H]
\centering
\includegraphics[width=\linewidth]{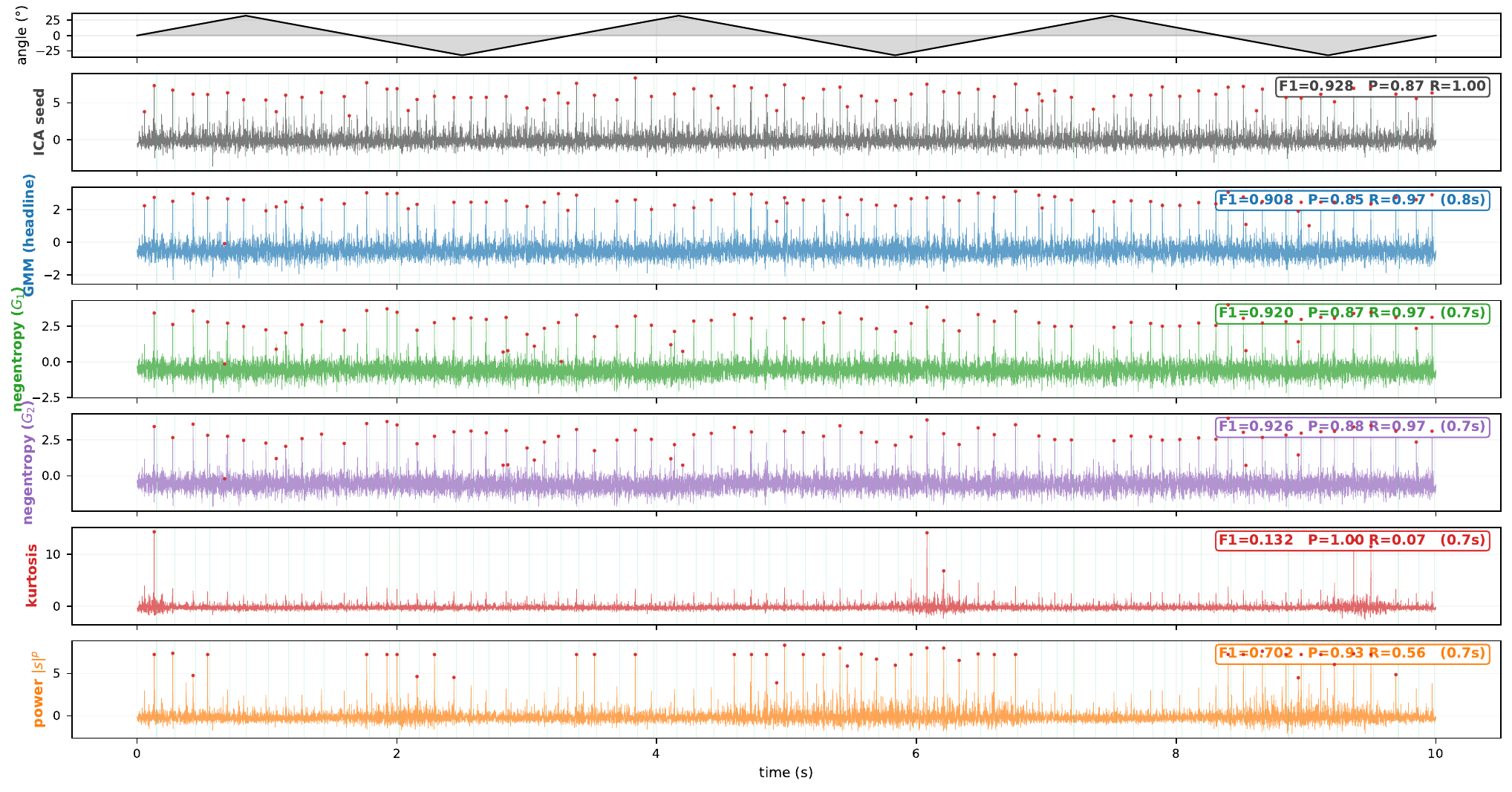}
\caption{\textbf{Regression.} GMM and the negentropy losses regress by
losing precision, while kurtosis and power-$p$ lose recall. GMM regresses slightly more than the negentropy losses
(which are more conservative in general); power-$p$ still amplifies
the bigger spikes and loses accuracy as a result; kurtosis amplifies
only a few spikes.}
\label{fig:supp:loss_regression}
\end{figure}

Figures~\ref{fig:supp:loss_rescue}--\ref{fig:supp:loss_regression}
show the same three case studies under the five loss substitutions,
making the failure modes visible directly on the recovered source.

\subsection{Compensator rank}
\label{supp:abl:rank}

The time-varying transformation $A_j(t) = I + U_j(t) V_j^\top$
(Eq.~\ref{eq:lowrank} of the main paper) is parameterised at rank $r$, with $U_j(t)
\in \mathbb{R}^{K \times r}$ and $V_j \in \mathbb{R}^{K \times r}$.
$r = 3$ is the operating point of the headline PE pipeline reported
in §5 of the main paper, and is also the operating point of the
angle-conditioned variant reported in \S\ref{supp:pe_vs_angle} and
\S\ref{supp:abl:pe_vs_angle}. Table~\ref{tab:supp:rank} reports
yield at $r \in \{3, 5, 7\}$ on the angle-conditioned pipeline; rank sensitivity carries
over to the PE pipeline by construction since the compensator
parameterisation is identical.

\begin{table}[h]
\centering
\caption{Compensator rank sweep on the angle-conditioned pipeline.
$r = 3$ is the operating point used in §5 of the main paper for
both the PE and angle variants. Yield is per seed, summed over
$168$ recordings; all three ranks come from the same sweep, run separately from Table~\ref{tab:supp:pe_vs_angle}, so the $r = 3$ row differs from it by run-to-run variation.}
\label{tab:supp:rank}
\small
\begin{tabular}{l c c c c}
\toprule
Rank & Mean $F_1$ & $n$ at $F_1 \geq 0.85$ & $n \geq 0.90$ & $n \geq 0.95$ \\
\midrule
\multicolumn{5}{l}{\textbf{ch070}} \\
\quad $r = 3$                & $0.736 \pm 0.003$ & $694.4 \pm 7.3$  & $602.0 \pm 11.3$ & $501.8 \pm 8.6$ \\
\quad $r = 5$             & $0.740 \pm 0.002$ & $710.6 \pm 10.5$ & $618.8 \pm 10.2$ & $510.0 \pm 9.4$ \\
\quad $r = 7$             & $\mathbf{0.741 \pm 0.002}$ & $\mathbf{716.6 \pm 11.0}$ & $\mathbf{624.6 \pm 11.1}$ & $\mathbf{512.6 \pm 9.9}$ \\
\midrule
\multicolumn{5}{l}{\textbf{ch320}} \\
\quad $r = 3$                & $0.759 \pm 0.003$ & $775.0 \pm 13.1$ & $677.8 \pm 10.0$ & $566.0 \pm 6.7$ \\
\quad $r = 5$             & $0.763 \pm 0.002$ & $796.6 \pm 9.6$  & $691.0 \pm 7.3$  & $\mathbf{577.8 \pm 9.1}$ \\
\quad $r = 7$             & $\mathbf{0.764 \pm 0.002}$ & $\mathbf{802.4 \pm 5.2}$  & $\mathbf{699.6 \pm 3.4}$  & $577.4 \pm 5.7$ \\
\bottomrule
\end{tabular}
\end{table}

Lifting the rank from $3$ to $7$ buys roughly $11$ additional MUs per
seed at $F_1 \geq 0.95$ on ch320 ($+2\%$) and $11$ on ch070 ($+2\%$),
with mean $F_1$ rising by $0.005$. The gain is monotonic but small,
saturating between $r = 5$ and $r = 7$. Empirically the
non-stationary part of the surface-EMG mixing concentrates in a
handful of spatial directions, consistent with PCA studies of
electrode-shift-induced waveform variation
\cite{naik2016principal, huang2020low}; rank $3$ captures most of
that subspace. The cost of higher rank is in the size of the
per-time-step compensator output ($r K$ values per source per time
step) and proportionally more parameters in the conditioning
network. We adopt $r = 3$ as the operating point of the headline
pipeline because it captures the bulk of the achievable yield at a
substantially smaller compensator cost: the additional $\sim 2\%$ at
$r = 7$ requires more than double the per-time-step output
dimensionality and a proportionally larger conditioning MLP, which
is an unfavourable trade given the diminishing returns and the
small gain relative to across-seed dispersion.

\subsection{Side information: positional encoding vs.\ joint angle}
\label{supp:abl:pe_vs_angle}

The headline runs in §5 of the main paper feed the compensator MLP a
sinusoidal positional encoding (PE) of the time index, with no
access to the joint-angle trace driving the non-stationarity. An
obvious alternative is to condition the compensator on the angle
trace directly. \S\ref{supp:pe_vs_angle} reports the bin-stratified
yield and Pareto-crossover yield for the two regimes. The angle
variant is the stronger configuration on both grids; the PE
configuration is the more conservative claim and is the headline
configuration of the main paper, for the reasons given in \S\ref{supp:pe_vs_angle}
(parity with non-Kramberger baselines, and isolation of the
compensator MLP's contribution from any benefit conferred by direct
access to the synthesis state variable).

\paragraph{Why angle helps a model that does no accumulation.}
Methods that build statistical estimates by accumulation across the
recording, such as Kramberger~2021 \cite{kramberger2021prediction},
which averages spike-triggered MUAPs into per-angle bins, benefit
naturally from explicit angle conditioning, because binning the data
by joint angle directly partitions the estimate and a separate template
is computed in each bin. HarmonICA does no such accumulation: the
compensator emits $A_j(t)$ from a side-input vector at each sample,
with no statistical estimation step or per-bin averaging. Even so,
per-angle correction is empirically somewhat stronger than per-time
PE-conditioned correction. We nonetheless make PE the headline
configuration because the joint-angle trace is not always available in
practice (recovering it requires either an instrumented setup or
post-hoc estimation, neither of which is realistic in a typical EMG
decomposition workflow), and PE keeps HarmonICA on equal footing with
the five non-Kramberger baselines.

\section{Nonlinear ICA baselines}
\label{supp:nonlinear_ica}

Identifiable nonlinear ICA restores identifiability through an
auxiliary variable~\cite{hyvarinen2016unsupervised,
hyvarinen2018nonlinear, khemakhem2020variational}. We test whether it
can stand in for HarmonICA by running time-contrastive learning
(TCL)~\cite{hyvarinen2016unsupervised} and
iVAE~\cite{khemakhem2020variational} on the same non-stationary
benchmark recordings, with the same whitened input as HarmonICA. Both
are given the auxiliary variable their theory requires: the joint
angle, and for TCL also contiguous time segments. Table~\ref{tab:supp:nlica}
reports the mean number of motor units per recording at a deliberately
lenient $F_1 > 0.5$; at the $F_1 \geq 0.95$ criterion of the main paper
neither method recovers essentially any unit.

\begin{table}[h]
\centering
\caption{Nonlinear ICA baselines on \texttt{muniverse-benchmark-dynamic}.
Mean motor units per recording at $F_1 > 0.5$. The ch070 TCL
(angle-conditioned) result varied across runs; we report the observed
range. Evaluated on $5$ recordings per electrode grid with a single
seed, not the full test split; iVAE's latent dimension was set to the
ground-truth number of units.}
\label{tab:supp:nlica}
\small
\begin{tabular}{l c c}
\toprule
Method & ch070 & ch320 \\
\midrule
Linear ICA (reference)      & $10.0$      & $8.4$ \\
TCL (angle-conditioned)     & $1$--$4$    & $2.4$ \\
TCL (time segments)         & $0.6$       & -- \\
iVAE (angle-conditioned)    & $0.6$       & $0.4$ \\
\bottomrule
\end{tabular}
\end{table}

The failure is a model mismatch rather than a tuning problem. In the
full TCL pipeline (classify time segments, take the penultimate
features, whiten, then run ICA), the downstream decomposition recovers
an order of magnitude fewer units than linear ICA
(Table~\ref{tab:supp:nlica}): the segment-classification pretext task
does not require separating the sources. Four properties of EMG explain this. First, the mixing is
convolutive, so after time extension each motor unit occupies a
subspace rather than a single direction. Second, on the whitened
extended signal a linear TCL classifier falls to chance; restoring its
accuracy requires unwhitened energy features, which amount to a
change-point detector and carry no separating information. Third, the
classifier needs only a few energy directions, so the learned features
discard the fine spatial directions in which motor units differ.
Fourth, motor-unit spike trains are sparse point processes (a duty
cycle of roughly $0.5\%$), far from the smoothly modulated
exponential-family sources that the identifiability theory assumes.

More fundamentally, these methods model non-stationarity on the
opposite axis. TCL and iVAE assume sources whose distribution is
modulated by the auxiliary variable under a fixed mixing, whereas in
dynamic EMG the mixing $H(\theta(t))$ changes with the joint angle
while the sources remain locally stationary spike trains. Conditioning
on the angle therefore does not give them the right model, which is
why a linear separator with a time-varying compensator suffices here
where a nonlinear one does not help.

\section{Baseline comparison under controlled drift}
\label{supp:sanity}

Most adaptive baselines trail the unrefined ICA seed on
\texttt{muniverse-benchmark-dynamic} (main Table~\ref{tab:yield_3bin}; Mendez Guerra 2024 exceeds it on both grids and Chen 2020 on ch320).
To check that this ordering is not an artefact of the benchmark, we
first bound what each baseline could achieve on the benchmark itself,
then run two controlled comparisons in which the amount of
non-stationarity is known.

\paragraph{An oracle upper bound on every baseline.} A baseline that
loses lock contributes a near-zero score. We remove that failure mode
from the comparison: for every ground-truth motor unit we score each
prior method at the better of its own result and plain linear ICA on
that unit, so any collapse is charged to ICA and the method inherits the
linear-ICA solution wherever it fails. HarmonICA exceeds this upper
bound for each of Chen 2020, Yeung 2024, Kramberger 2021 and Glaser 2018 on both grids
(Table~\ref{tab:supp:bestof}). Per unit, the units HarmonICA recovers and
a baseline misses far outnumber the reverse
(Table~\ref{tab:supp:coverage}); the one non-trivial complementary set
belongs to Chen 2020, the strongest of these four.

\begin{table}[h]
\centering
\caption{Oracle upper bound. Motor units per recording at
$F_1 \geq 0.8$; ``best-of'' scores each unit at
$\max(F_1^{\mathrm{ICA}}, F_1^{\mathrm{method}})$. This analysis uses its
own recording panel and extension factor, so absolute values are not
comparable with Table~\ref{tab:supp:detailed_yield}.}
\label{tab:supp:bestof}
\small
\begin{tabular}{l c c c c}
\toprule
& \multicolumn{2}{c}{ch070} & \multicolumn{2}{c}{ch320} \\
\cmidrule(lr){2-3}\cmidrule(lr){4-5}
Method & alone & best-of & alone & best-of \\
\midrule
Linear ICA                & $2.16$ & --     & $0.75$ & -- \\
Chen 2020                 & $2.96$ & $3.41$ & $3.43$ & $3.52$ \\
Yeung 2024                & $1.67$ & $2.79$ & $1.77$ & $2.06$ \\
Kramberger 2021           & $1.02$ & $2.59$ & $1.21$ & $1.84$ \\
Glaser 2018               & $0.88$ & $2.25$ & $0.83$ & $1.17$ \\
\textbf{HarmonICA (ours)} & $\mathbf{3.69}$ & $\mathbf{3.69}$ & $\mathbf{4.52}$ & $\mathbf{4.52}$ \\
\bottomrule
\end{tabular}
\end{table}

\begin{table}[h]
\centering
\caption{Per-unit coverage on ch070 (same panel as
Table~\ref{tab:supp:bestof}). Motor units per recording at
$F_1 \geq 0.7$ recovered by one method and missed by the other.}
\label{tab:supp:coverage}
\small
\begin{tabular}{l c c c}
\toprule
Method & baseline only & HarmonICA only & ratio \\
\midrule
Chen 2020       & $0.49$ & $1.53$ & $3.1\times$ \\
Yeung 2024      & $0.27$ & $3.06$ & $11\times$ \\
Glaser 2018     & $0.07$ & $3.89$ & $56\times$ \\
Kramberger 2021 & $0.07$ & $3.93$ & $56\times$ \\
\bottomrule
\end{tabular}
\end{table}

\paragraph{Drift-controlled generic simulation.} A small, generic
simulation outside MUniverse (Gaussian action potentials on an
electrode grid) has a single drift parameter that sets how fast the
mixing moves, from $0$ (static) to $0.9$. Each baseline is tuned to its
most robust configuration by grid search over its calibration and
detection settings. Table~\ref{tab:supp:drift} reports motor units per
recording at $F_1 \geq 0.7$.

\begin{table}[h]
\centering
\caption{Drift-controlled generic simulation. Motor units per recording
at $F_1 \geq 0.7$; drift $0$ is a static mixing.}
\label{tab:supp:drift}
\small
\begin{tabular}{l c c c c c c}
\toprule
Drift & ICA & HarmonICA & Chen & Yeung & Kramberger & Glaser \\
\midrule
$0.0$ (static) & $29.7$ & $29.0$ & $17.7$ & $15.3$ & $25.0$ & $18.7$ \\
$0.3$          & $22.3$ & $24.7$ & $16.7$ & $14.0$ & $14.3$ & $15.7$ \\
$0.6$          & $8.3$  & $13.3$ & $9.3$  & $8.3$  & $8.7$  & $7.0$ \\
$0.9$          & $3.7$  & $8.0$  & $4.0$  & $2.3$  & $4.0$  & $2.3$ \\
\bottomrule
\end{tabular}
\end{table}

HarmonICA is within one unit of linear ICA when the mixing is static
and ahead of every method at every nonzero drift, with the margin
over ICA widening from static to moderate drift.

\paragraph{Isometric MUniverse.} On one representative simulated
isometric MUniverse recording (constant joint angle, $104$-unit pool),
there is no drift to exploit. Linear ICA recovers $8$ units at
$F_1 \geq 0.7$, HarmonICA $7$, Kramberger 2021 $5$, Glaser 2018 $4$,
and Chen 2020 and Yeung 2024 $1$ each. The pool is dense, so even
linear ICA recovers only a small fraction of it.

\paragraph{Why the baselines trail ICA when nothing drifts.} The
adaptive baselines are real-time, short-calibration variants of
convolutive BSS, and pay three costs that a batch decomposition of the
whole recording does not: the calibration window caps how many units
can be found; their online spike detectors are weaker than a batch
detector; and adaptation injects noise when there is nothing to track.
HarmonICA refines the batch ICA solution rather than replacing it, so
it stays near ICA when the mixing is static and exceeds it once the
mixing moves.

\section{Real dynamic HD-sEMG}
\label{supp:real}

This section gives the full protocol and results summarised in main
§\ref{sec:real}.

\paragraph{Recordings.} Two $65.5$\,s recordings from participant 6 of
Mendez Guerra et al.~\cite{guerra2024adaptive}: a $64$-channel forearm
grid at $2048$\,Hz during isometric digit flexion at $15\%$ MVC while a
motorised platform steps the wrist from $0^\circ$ to $-40^\circ$ in
$10^\circ$ holds. The measured wrist angle is the compensator input
$\boldsymbol{\gamma}(t)$, and only $39\%$ of each recording is a hold.
MUAP correlation with the first $8$\,s window falls to $0.087$ by the end
of the recording, but the drift is slow and monotone (${\sim}0.6^\circ$/s).
We use $L=16$ ($NL=1024$) and $K=128$. The data are reused from the
original study, which collected them under its own ethics approval with
informed consent; no new data were collected for this work.

\paragraph{Reference.} No intramuscular reference exists, so the
reference is built from the same surface data as a self-consistency
proxy, in two independent ways: a global-anchored splice against the
whole-recording ICA backbone, and a local chaining construction that
decomposes each $15$\,s window on its own data alone, without seeing the
whole-recording result. The two agree at a rate of agreement of
$0.99$--$1.00$; because the local construction never sees the
whole-recording decomposition, this is also a held-out consistency check
on the discovered units. Four external published methods independently
recover the same units, four of the five at $F_1$ $0.93$--$0.99$ and the
hardest at $0.77$. The five units that pass these checks are scored
against the fixed reference, so all spread in
Table~\ref{tab:supp:real} is method variance.

\begin{table}[h]
\centering
\caption{Real dynamic HD-sEMG: $F_1$ per validated unit against the
self-consistency reference, mean over $5$ seeds, for $6$ methods. ``Worst
run'' is the lowest single-run $F_1$ over all $25$ unit--seed runs.}
\label{tab:supp:real}
\small
\begin{tabular}{l c c c c c c c}
\toprule
Method & MU3 & MU4 & MU1+MU14 & MU11 & MU15 & Mean & Worst run \\
\midrule
Chen 2020 \cite{chen2020adaptive}               & $0.986$ & $0.979$ & $0.993$ & $0.932$ & $0.700$ & $0.918$ & $0.102$ \\
\textbf{HarmonICA (ours)}                       & $0.999$ & $0.969$ & $0.989$ & $0.874$ & $0.757$ & $0.918$ & $0.732$ \\
Kramberger 2021 \cite{kramberger2021prediction} & $0.979$ & $0.979$ & $0.989$ & $0.809$ & $0.773$ & $0.906$ & $0.773$ \\
Yeung 2024 \cite{yeung2024adaptive}             & $0.980$ & $0.980$ & $0.989$ & $0.910$ & $0.574$ & $0.887$ & $0.079$ \\
Linear ICA (reference)                          & $0.986$ & $0.832$ & $0.886$ & $0.837$ & $0.727$ & $0.854$ & $0.706$ \\
Glaser 2018 \cite{glaser2018motor}              & $0.956$ & $0.988$ & $0.861$ & $0.597$ & $0.622$ & $0.805$ & $0.438$ \\
\bottomrule
\end{tabular}
\end{table}

\begin{figure}[h]
\centering
\includegraphics[width=\linewidth]{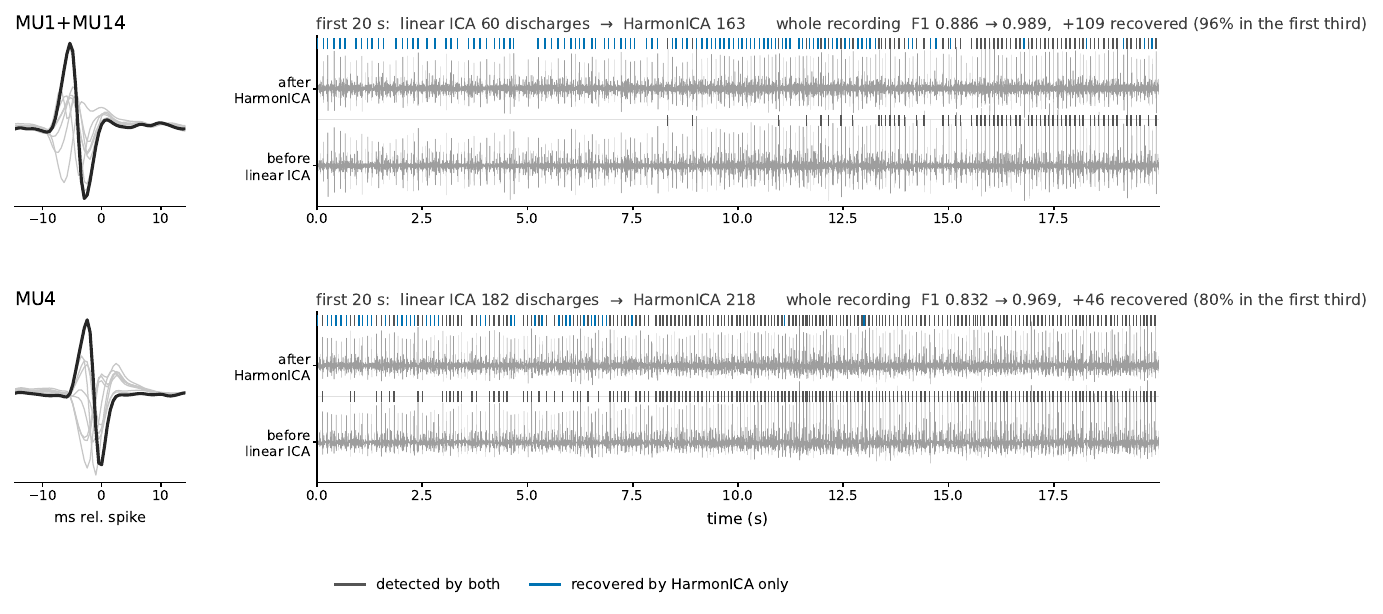}
\caption{\textbf{Where linear ICA fails on real dynamic HD-sEMG.} The two
units whose accuracy moves materially under compensation; the other three
validated units shift by $\leq 0.04$. \emph{Left:} the motor-unit action
potential on its best channel (black) and the eight next-best (grey).
\emph{Right:} the first $20$\,s of the recording through the static
linear-ICA filter (lower) and through HarmonICA's compensated filter
(upper), each normalised by its own baseline, with ticks marking detected
discharges. On MU1$+$MU14 linear ICA detects nothing for the first
$8$\,s and $60$ discharges over the $20$\,s shown, against HarmonICA's
$163$. Across the whole recording, $96\%$ of the $109$ discharges
HarmonICA recovers on MU1$+$MU14, and $80\%$ of the $46$ on MU4, fall in
the first third. HarmonICA seed $1909$.}
\label{fig:supp:real_exemplars}
\end{figure}

HarmonICA improves on the linear-ICA solution it refines for every unit
(mean $F_1$ $0.854 \to 0.918$) and has no catastrophic run across the
$25$ runs (worst $0.732$, against $0.102$ for Chen 2020 and $0.079$ for
Yeung 2024). It ties Chen 2020 on mean $F_1$, and the ordering among the
top four is within seed noise. Kramberger 2021 is deterministic and has
no seed parameter. Fig.~\ref{fig:supp:real_exemplars} shows the two units
whose accuracy moves materially under compensation.

\paragraph{Why the baselines keep up here.} To test whether this
recording flatters the adaptive baselines, we cut each one's calibration
from $20$\,s to a single $5.3$\,s hold at $0^\circ$ and required it to
track the most-drifting unit across the full sweep
(Table~\ref{tab:supp:real_calib}). Accuracy does not degrade with
distance from calibration: the drift is slow enough that filter
adaptation never loses lock, so the tie in
Table~\ref{tab:supp:real} is expected. The benchmark's advantage for
HarmonICA comes from faster, reversing mixing changes (§\ref{supp:sanity}),
which this recording does not contain. Separating the methods on real
data needs faster, non-monotone mixing together with recruitment or
derecruitment within the trial, and we know of no public dataset that
combines these with ground-truth labels.

\begin{table}[h]
\centering
\caption{Tracking accuracy of the most-drifting unit with calibration
cut to a single $5.3$\,s hold at $0^\circ$.}
\label{tab:supp:real_calib}
\small
\begin{tabular}{l c c c c c}
\toprule
Method & $0^\circ$ & $-10^\circ$ & $-20^\circ$ & $-30^\circ$ & $-40^\circ$ \\
\midrule
Chen 2020       & $1.000$ & $0.980$ & $0.979$ & $1.000$ & $0.972$ \\
Yeung 2024      & $1.000$ & $0.980$ & $1.000$ & $1.000$ & $0.982$ \\
Kramberger 2021 & $1.000$ & $1.000$ & $1.000$ & $1.000$ & $0.943$ \\
\bottomrule
\end{tabular}
\end{table}

\paragraph{A unit split by linear ICA.} Whole-recording ICA reports MU1
and MU14 as two units. They are one unit whose waveform changed enough
during the rotation to split under linear ICA ($28\times$ above-chance
coincidence, MUAP correlation $0.948$). A $28$-spike seed from a $4$\,s
window, $4.6\%$ of the unit's $611$ spikes, recovers the merged unit on
its own, raising $F_1$ from $0.44$ to $0.92$.

\paragraph{Recovery from weakened seeds.} Across $93$ weakened starting
filters over the five validated units, mean $F_1$ improves from $0.72$ to
$0.88$, with $53$ of $93$ reaching $F_1 \geq 0.9$, including $0.96$ from
five spikes in half a second. The gain is in recall ($+0.26$) at a small
cost in precision ($-0.03$), and firing becomes more regular, not less
(unfiltered ISI CoV $1.27 \to 0.38$ for MU4), consistent with recovering
missed spikes rather than adding false ones.

\paragraph{Failure case and scope.} Refinement does not always help: a
thumb--middle unit that fires only in the last $20$\,s degrades from
$F_1$ $0.973$ to $0.690$, stable over $10$ seeds. The evaluation covers
one participant, two recordings and five units, and the reference is
built from the same surface data it is scored against. Spike trains for
all five units and $25$ runs, both reference constructions, MUAP
templates, and a script that regenerates every table in this section
are included in the code release.

\paragraph{Concurrent intramuscular and surface data.} We also attempted
the dataset of Bradford et al.~\cite{bradford2023high}, which pairs
high-density surface and intramuscular EMG from the tibialis anterior
during dynamic contractions. It ships raw signals only, so a reference
decomposition of its fine-wire channels during movement would have to be
produced in-house, and automatic intramuscular decomposition during
movement is itself an open problem. The
recordings also combine fast mixing changes with high noise,
recruitment and derecruitment, and movement artefacts, well outside the
regime validated here. We could not build a reference we would trust, so
we report no results on it and leave it to future work.

\section{Source extraction examples}
\label{supp:source_examples}

This section shows representative single-source extractions on
\emph{muniverse-benchmark-dynamic} test recordings. Each panel plots
the recovered source signal $\hat{s}_j(t)$ \emph{before} the GMM
spike-detection threshold of \S\ref{sec:peeloff} is applied, for both
the unrefined ICA seed and HarmonICA, alongside the ground-truth
spike train. The first three examples (\S\ref{supp:src:shared})
illustrate behaviours shared with the underlying FastICA stage; the
last two (\S\ref{supp:src:method}) illustrate failure modes specific
to the HarmonICA refinement step.

\subsection{Behaviours inherited from FastICA}
\label{supp:src:shared}

\begin{figure}[H]
\centering
\includegraphics[width=\linewidth]{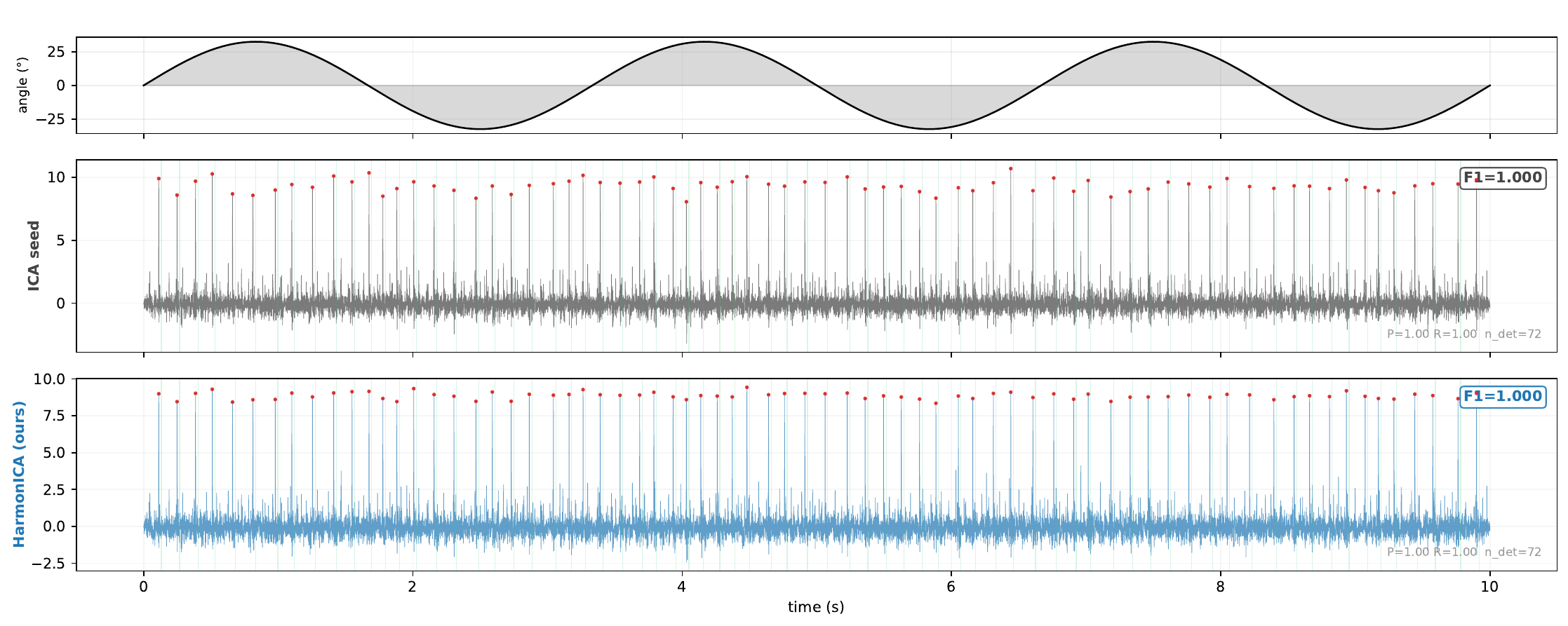}
\caption{\textbf{Easy success.} ICA cleanly extracts the spikes with
a single filter; the non-stationarity is trivial. These are the
straightforward ICA successes within the benchmark: the angle
modulation is small and the filter changes too little to matter.}
\end{figure}

\begin{figure}[H]
\centering
\includegraphics[width=\linewidth]{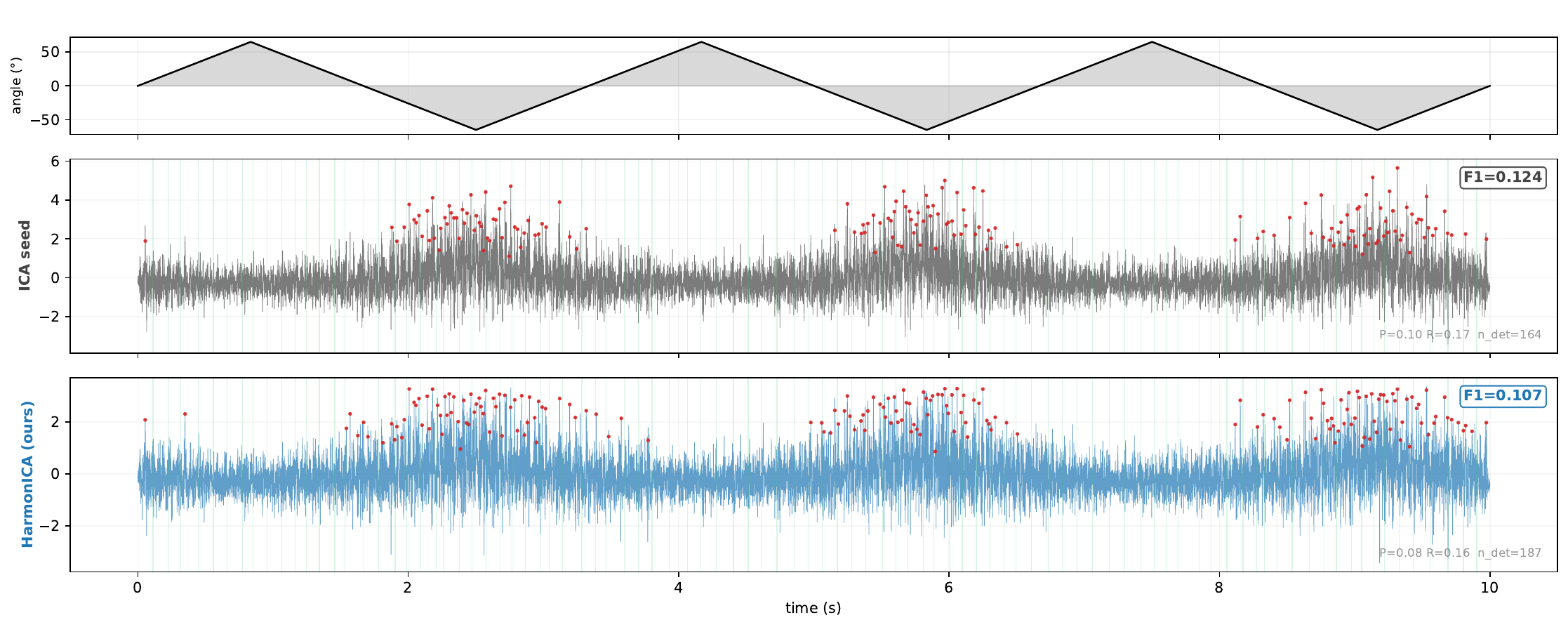}
\caption{\textbf{ICA complete failure.} A scenario where ICA would
fail even in the absence of non-stationarity; HarmonICA cannot
refine pure noise, so both methods fail completely. This accounts
for most of the unrecovered motor units in the benchmark, with the
failure driven by some combination of amplitude, crosstalk, and
non-stationarity.}
\end{figure}

\begin{figure}[H]
\centering
\includegraphics[width=\linewidth]{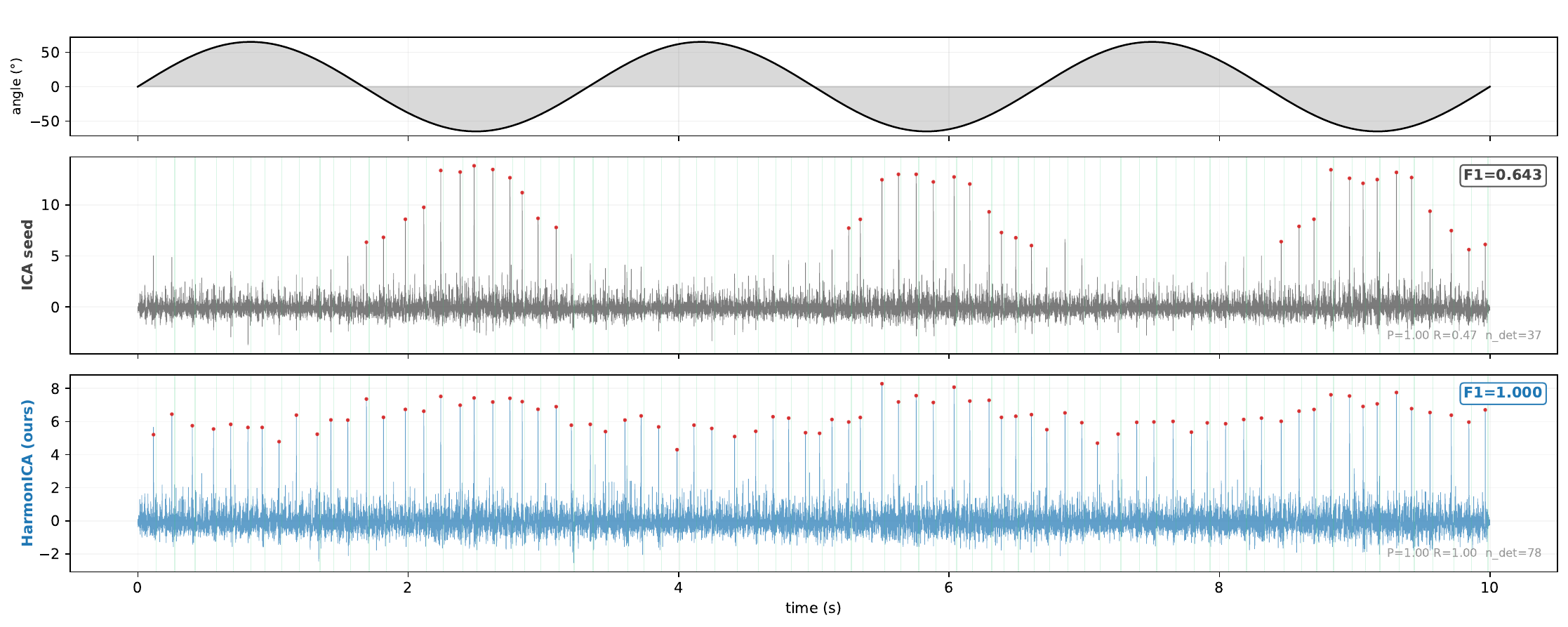}
\caption{\textbf{Stuck source rescued.} ICA yields spikes around a
slow-moving non-stationarity; the spikes are clean but weaken as
$t$ moves away from the correct filter window. HarmonICA
successfully rotates the filter to discover the missing spikes,
lifting recall from $\sim\!0.5$ to $1$.}
\end{figure}

\subsection{Method-specific failure modes}
\label{supp:src:method}

\begin{figure}[H]
\centering
\includegraphics[width=\linewidth]{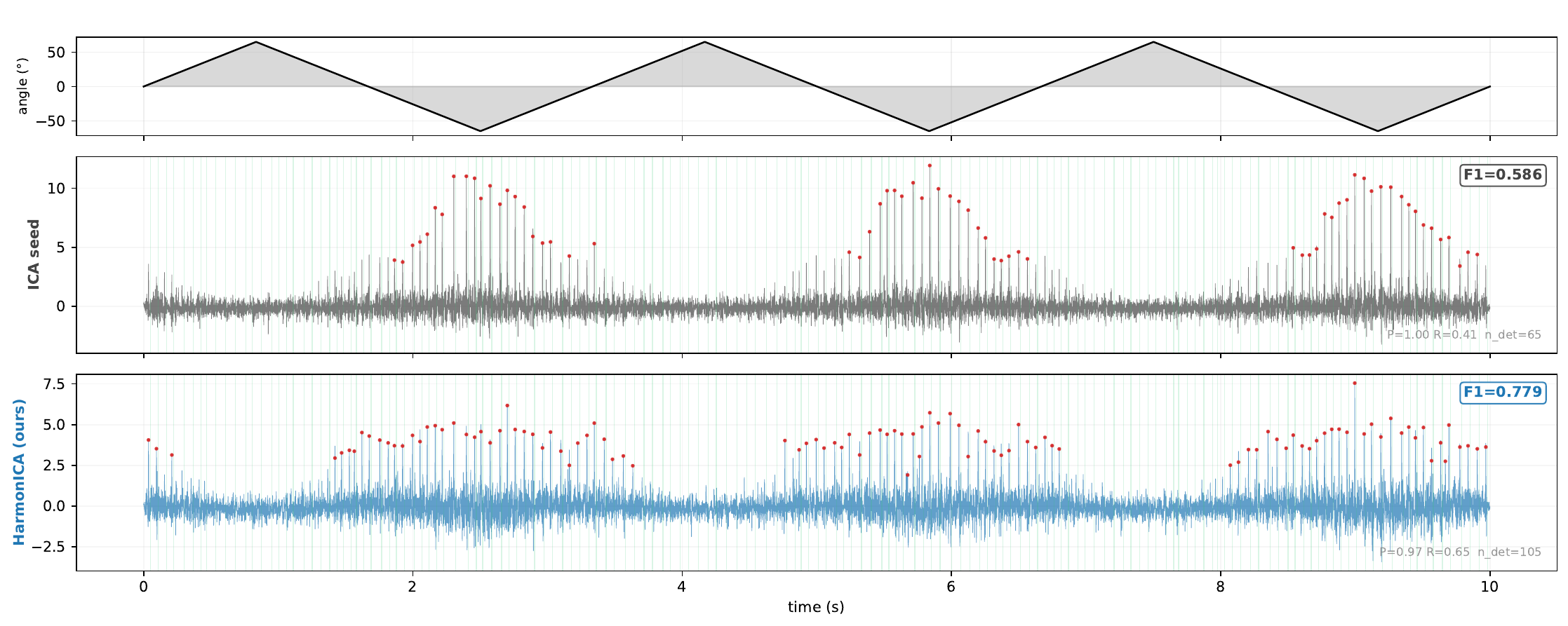}
\caption{\textbf{Partial success.} HarmonICA discovers part of the
spike train but cannot extend coverage to the entire signal. This
happens when the filter rotates too much or the local degradation
of the source is too abrupt for the compensator to track gracefully.}
\end{figure}

\begin{figure}[H]
\centering
\includegraphics[width=\linewidth]{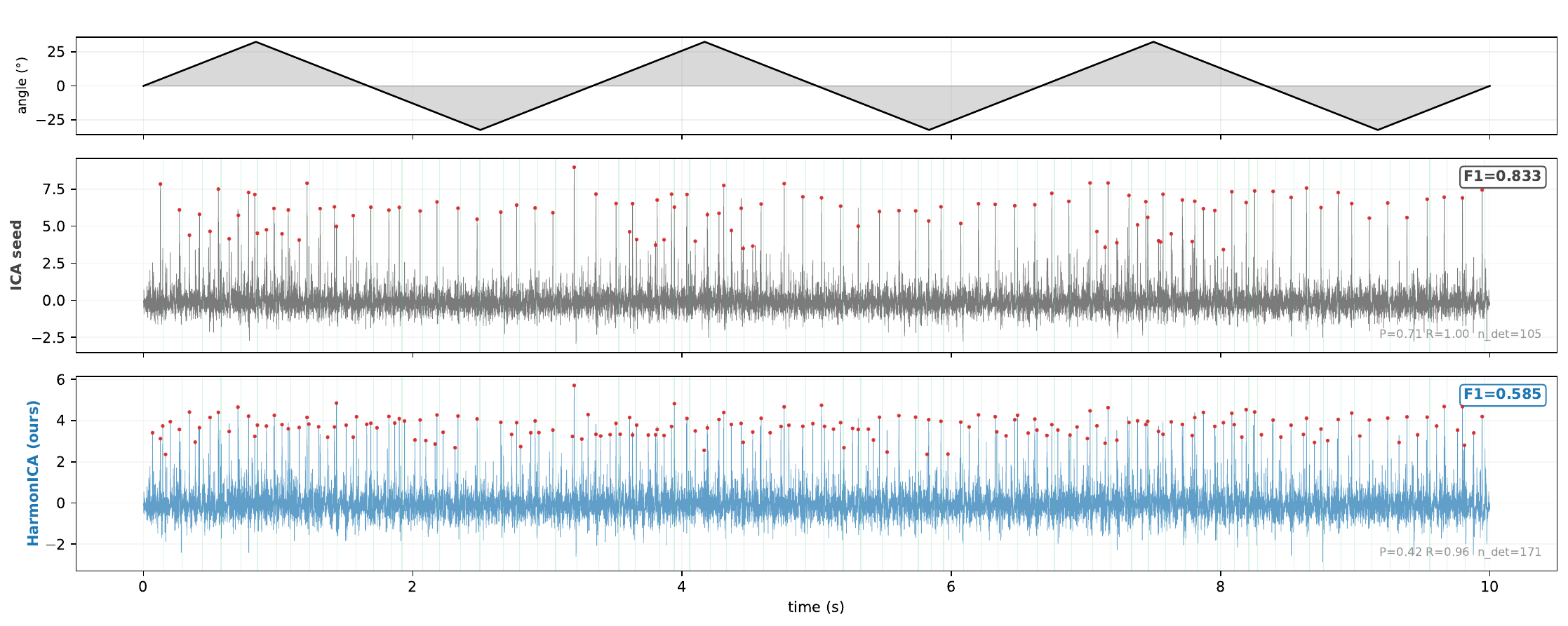}
\caption{\textbf{Noise amplification.} ICA yields a low-precision
source; HarmonICA has no way of distinguishing correct spikes from
incorrect ones in this regime, since no auxiliary information
breaks the tie. The compensator ends
up amplifying false-positive spikes and the source regresses.}
\end{figure}

\begin{figure}[H]
\centering
\includegraphics[width=\linewidth]{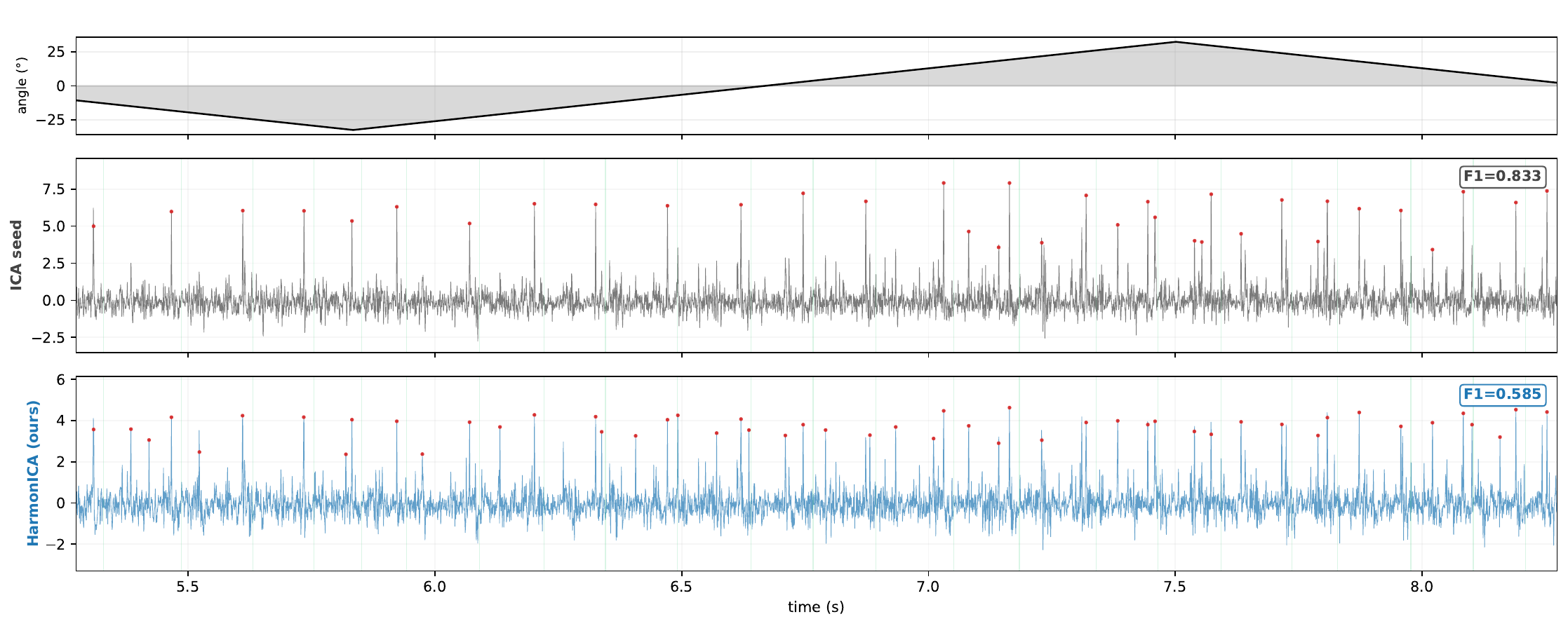}
\caption{Zoomed-in view of the noise-amplification failure mode of
the previous figure.}
\end{figure}

\section{Notation}
\label{supp:notation}

Symbols used in §3 of the main paper, in the identifiability
analysis (§\ref{supp:theory}) and in the supplementary
implementation appendix (§\ref{supp:impl}), with definitions and
operating-point values used in the headline benchmark.

\begin{table}[H]
\centering
\caption{Symbols used in §3 of the main paper and in
supplementary §\ref{supp:impl}.}
\label{tab:supp:notation}
\small
\begin{tabular}{l p{0.52\linewidth} l}
\toprule
Symbol & Meaning & Value / shape \\
\midrule
\multicolumn{3}{l}{\textbf{Signal and preprocessing}} \\
$f_s$ & sampling rate & $2048$ Hz \\
$N$ & number of recording channels & $70$ or $320$ \\
$L$ & Toeplitz extension factor (delayed copies / channel) & $8$ \\
$Q$ & MUAP length & samples \\
$K$ & PCA-whitened dimensionality & $128$ \\
$M$ & number of motor units & dataset-dependent \\
$\mathbf{x}(t)$ & raw multichannel EMG & $\mathbb{R}^{N}$ \\
$H(t,k)$ & time-varying MUAP mixing at lag $k$ (main Eq.~\ref{eq:conv_model}) & $\mathbb{R}^{N\times M}$ \\
$\mathbf{x}_s(t)$ & $L$-fold stacked (extended) signal & $\mathbb{R}^{NL}$ \\
$\bar{\mathbf{s}}(t)$ & extended source: delayed copies $s_j(t-l)$ & $\mathbb{R}^{\bar{M}}$, $\bar{M}=M(L+Q-1)$ \\
$W_{\mathrm{pre}}$ & PCA whitening of the extended signal & $\mathbb{R}^{K\times NL}$ \\
$\tilde{\mathbf{x}}(t)$ & extended, PCA-projected, ZCA-whitened signal & $\mathbb{R}^{K}$ \\
$\tilde{H}(t)$ & whitened extended mixing, $W_{\mathrm{pre}}\mathcal{H}(t)$ & $\mathbb{R}^{K\times\bar{M}}$ \\
$\tilde{H}^{\ast}$ & whitened mixing on the anchor window & $\mathbb{R}^{K\times\bar{M}}$ \\
$D(t)$ & drift relative to the anchor, $\tilde{H}(t)=D(t)\tilde{H}^{\ast}$ & $\mathbb{R}^{K\times K}$ \\
$\tilde{\boldsymbol{\omega}}(t)$ & residual Gaussian noise & $\mathbb{R}^{K}$ \\
\midrule
\multicolumn{3}{l}{\textbf{Source model}} \\
$s_j(t)$ & latent spike train of source $j$ &  \\
$\hat{s}_j(t)$ & recovered source signal & scalar \\
$c_{\text{null}}, c_{\text{spike}}$ & spike/null point-mass amplitudes & data-driven \\
$\rho_j \in (0,1)$ & firing-rate / mixture weight of spike component &  \\
$\delta(\cdot)$ & Dirac delta &  \\
\midrule
\multicolumn{3}{l}{\textbf{Quasi-time-invariant separator $+$ compensator}} \\
$\mathbf{w}_j$ & static linear separator for source $j$ & $\mathbb{R}^{K}$ \\
$A_j(t)$ & time-varying compensator at time $t$ & $\mathbb{R}^{K\times K}$ \\
$r$ & rank of the Woodbury parameterisation of $A_j(t)$ & $3$ \\
$U_j(t)$ & time-varying Woodbury factor & $\mathbb{R}^{K\times r}$ \\
$V_j$ & per-source static Woodbury factor & $\mathbb{R}^{K\times r}$ \\
$\phi_j$ & weights of the conditioning network $\mathrm{NN}_j$ &  \\
$\boldsymbol{\gamma}(t)$ & sinusoidal positional encoding of $t$ & $\mathbb{R}^{2P}$ \\
$T_1,\ldots,T_P$ & sinusoidal periods used in $\boldsymbol{\gamma}(t)$ & $\{0.1, 0.2, 0.5, 1.0, 2.0\}$ s \\
\midrule
\multicolumn{3}{l}{\textbf{Losses and optimisation}} \\
$u_j(t)$ & uncompensated projection $\mathbf{w}_j^\top\tilde{\mathbf{x}}(t)$ & scalar \\
$L_I$ & projection-pursuit independence loss on $u_j$ &  \\
$L_N$ & two-component-GMM stationarity loss on $\hat{s}_j$ (also written $L_{\mathrm{GMM}}$ in supp.\ \S\ref{supp:abl:loss}) &  \\
$e$ & exponent of the independence-loss contrast $\mathrm{sgn}(u)|u|^e$ & $3$ \\
$\sigma^2_{\text{null}}, \sigma^2_{\text{spike}}$ & sample variances of $\hat{s}_j(t)$ over the null / spike sets &  \\
$\bar{\sigma}^2_{\text{null}}$ & null-cluster variance evaluated on $u_j(t)$ &  \\
$T_w$ & separator-only warmup iterations & $50$ \\
$K_g$ & spike/null re-split period (compensation steps) & $5$ \\
$\eta_w, \eta_\phi$ & RMSprop learning rates ($\mathbf{w}_j$, $\phi_j$) & $10^{-5}$, $10^{-4}$ \\
\midrule
\multicolumn{3}{l}{\textbf{Sequential extraction and peel-off}} \\
$n_j$ & accepted spike count for source $j$ &  \\
$\{t_{j,i}\}_{i=1}^{n_j}$ & accepted spike times of source $j$ &  \\
$\tau$ & within-template lag offset relative to spike time $t_{j,i}$ &  \\
$\mathrm{STA}_j^{\mathrm{comp}}$ & spike-triggered average in compensated frame &  \\
$\hat{\mathbf{x}}_j^{(i)}$ & per-spike contribution mapped back to whitened frame &  \\
\bottomrule
\end{tabular}
\end{table}

\newpage
\section*{NeurIPS Paper Checklist}

\begin{enumerate}

\item {\bf Claims}
    \item[] Question: Do the main claims made in the abstract and introduction accurately reflect the paper's contributions and scope?
    \item[] Answer: \answerYes{}
    \item[] Justification: The headline claims in the abstract and Section~\ref{sec:results} (2.5$\times$ yield over the underlying linear-ICA pipeline at recall $\geq 0.95$ on the 320-channel grid; dominance over five adaptive baselines at every recall threshold) are evidenced by Figure~\ref{fig:pareto_recall} and Table~\ref{tab:yield_3bin}, which aggregate over 168 recordings $\times$ 5 FastICA seeds per electrode density. The headline claims are on simulated dynamic surface EMG (Section~\ref{sec:data}); a small real-data evaluation (Section~\ref{sec:real}, supplementary §\ref{supp:real}) is reported separately, with its scope stated in the Limitations section.

\item {\bf Limitations}
    \item[] Question: Does the paper discuss the limitations of the work performed by the authors?
    \item[] Answer: \answerYes{}
    \item[] Justification: Section~\ref{sec:results} is followed by a dedicated Limitations section covering the simulated benchmark (single degree of freedom, constant drive, no recruitment or derecruitment), the limited real-data evaluation (one participant, two recordings, five units, a self-consistency reference, and a unit where refinement hurts), the offline, per-recording construction, the idealised assumptions of the identifiability results, the bias from fitting the separator to the uncompensated signal, and the limit $K$ places on dense pools. Two failure modes are flagged: recovery bounded by the upstream FastICA seed, and high within-pool MUAP crosstalk.

\item {\bf Theory, Assumptions and Proofs}
    \item[] Question: For each theoretical result, does the paper provide the full set of assumptions and a complete (and correct) proof?
    \item[] Answer: \answerYes{}
    \item[] Justification: Supplementary §\ref{supp:theory} states assumptions (A1)--(A3) and gives full proofs of Lemma~1 and Propositions~1, 1$'$ and 2, summarised in main Section~\ref{sec:objectives}; it also lists what is not proven. The results are existence results, not convergence guarantees.

    \item {\bf Experimental Result Reproducibility}
    \item[] Question: Does the paper fully disclose all the information needed to reproduce the main experimental results of the paper to the extent that it affects the main claims and/or conclusions of the paper (regardless of whether the code and data are provided or not)?
    \item[] Answer: \answerYes{}
    \item[] Justification: Section~\ref{sec:model} specifies the Woodbury compensator and time encoding, Section~\ref{sec:objectives} defines the independence and stationarity losses, Algorithm~\ref{alg:harmonica} (with the surrounding alternating-optimisation prose naming the RMSprop updates) gives the full optimisation procedure, and Section~\ref{sec:peeloff} details the per-spike peel-off. Implementation Section~\ref{sec:impl} reports preprocessing, $K=128$, rank $r=3$, hardware, and the FastICA seed protocol; the full operating-point hyperparameters (sinusoidal periods, learning rates, runtime) are listed in supplementary §\ref{supp:impl}. Code is released (link in the Acknowledgments).

\item {\bf Open access to data and code}
    \item[] Question: Does the paper provide open access to the data and code, with sufficient instructions to faithfully reproduce the main experimental results, as described in supplemental material?
    \item[] Answer: \answerYes{}
    \item[] Justification: The code release is linked from the Acknowledgments, and the \emph{muniverse-benchmark-dynamic} dataset construction is described in supplementary §\ref{supp:benchmark} (referenced from Section~\ref{sec:data}). Additional data are available on reasonable request.

\item {\bf Experimental Setting/Details}
    \item[] Question: Does the paper specify all the training and test details (e.g., data splits, hyperparameters, how they were chosen, type of optimizer, etc.) necessary to understand the results?
    \item[] Answer: \answerYes{}
    \item[] Justification: Section~\ref{sec:data} reports the test split at the simulated-subject level (21 pools across 5 subjects, 168 recordings per electrode grid) along with array and angle/SNR conditions; the held-out val split (5 pools, 80 recordings, 40 per grid) is described in supplementary §\ref{supp:benchmark}. Section~\ref{sec:impl} states the preprocessing pipeline, compensator rank, FastICA seed protocol, and hardware; the optimiser (RMSprop) is named in the alternating-optimisation subsection alongside Algorithm~\ref{alg:harmonica}. Full operating-point values (sinusoidal periods, learning rates, runtime) are in supplementary §\ref{supp:impl}; the three online hyperparameters of the Mendez Guerra 2024 baseline were tuned by the method's author on the disjoint validation split (subjects 5--9, 40 recordings per grid) with an objective that uses no ground truth (supplementary §\ref{supp:baseline_deviations}).

\item {\bf Experiment Statistical Significance}
    \item[] Question: Does the paper report error bars suitably and correctly defined or other appropriate information about the statistical significance of the experiments?
    \item[] Answer: \answerYes{}
    \item[] Justification: Each cell of Table~\ref{tab:yield_3bin} reports mean $\pm$ 1 standard deviation over the $5$ FastICA initialisation seeds (the per-recording yields are first summed within each bin, then averaged and dispersed across seeds). The complementary view that aggregates per-recording mean yields and reports the across-recording dispersion is in supplementary Table~\ref{tab:supp:yield_per_rec}, which also discusses the seed-vs-recording decomposition of the table standard deviations.

\item {\bf Experiments Compute Resources}
    \item[] Question: For each experiment, does the paper provide sufficient information on the computer resources (type of compute workers, memory, time of execution) needed to reproduce the experiments?
    \item[] Answer: \answerYes{}
    \item[] Justification: Section~\ref{sec:impl} reports the hardware (single NVIDIA A40 GPU; the Mendez Guerra 2024 baseline runs on the CPU, and its hyperparameter search was run by the method's author on separate hardware). Supplementary §\ref{supp:impl} gives the per-recording runtime ($\sim$45\,s), per-method per-recording runtimes for HarmonICA and the six baselines, the total compute budget ($\sim$$300$ GPU-hours across the headline benchmark and all reported ablations), and the hyperparameter exploration for HarmonICA (an extension-factor sweep $L \in \{8, 16, 32, 64\}$; no grid search). Baselines in the drift-controlled simulation of supplementary §\ref{supp:sanity} were configured by grid search.
    
\item {\bf Code Of Ethics}
    \item[] Question: Does the research conducted in the paper conform, in every respect, with the NeurIPS Code of Ethics \url{https://neurips.cc/public/EthicsGuidelines}?
    \item[] Answer: \answerYes{}
    \item[] Justification: The headline experiments use simulated EMG from the publicly released MUniverse benchmark. The real-data evaluation reuses existing recordings from a published study~\cite{guerra2024adaptive}, collected under that study's ethics approval with informed consent; no new human-subject data were collected. The work involves no scraped or sensitive datasets and no deployable system with foreseeable misuse pathways.

\item {\bf Broader Impacts}
    \item[] Question: If appropriate for the scope and focus of your paper, did you discuss potential negative societal impacts of your work?
    \item[] Answer: \answerNA{}
    \item[] Justification: The contribution is a foundational source-separation method evaluated mainly on simulated EMG; downstream applications (prosthetics, exoskeletons, gesture interfaces) carry the same risk profile as existing motor-unit decomposition pipelines and are not advanced into deployment by this work.
    
\item {\bf Safeguards}
    \item[] Question: Does the paper describe safeguards that have been put in place for responsible release of data or models that have a high risk for misuse (e.g., pretrained language models, image generators, or scraped datasets)?
    \item[] Answer: \answerNA{}
    \item[] Justification: The released artefacts are an unsupervised source-separation algorithm and a simulated EMG benchmark; neither poses a high risk for misuse.

\item {\bf Licenses for existing assets}
    \item[] Question: Are the creators or original owners of assets (e.g., code, data, models), used in the paper, properly credited and are the license and terms of use explicitly mentioned and properly respected?
    \item[] Answer: \answerYes{}
    \item[] Justification: The MUniverse simulator and BioMime/NeuroMotion forward models are cited at point of use (Section~\ref{sec:data}); the real HD-sEMG recordings are from the published study of Mendez Guerra et al.~\cite{guerra2024adaptive}, cited at point of use (Section~\ref{sec:real}); the Mendez Guerra 2024 reference implementation (adapt\_decomp v1.0.0, MIT licence~\cite{mendezguerra2024adaptcode}), used as a baseline in Table~\ref{tab:yield_3bin}, is cited in Section~\ref{sec:impl} and supplementary §\ref{supp:impl}; all reused assets are properly credited and used within their stated terms.

\item {\bf New Assets}
    \item[] Question: Are new assets introduced in the paper well documented and is the documentation provided alongside the assets?
    \item[] Answer: \answerYes{}
    \item[] Justification: The HarmonICA codebase is released (link in the Acknowledgments) with documented training and inference scripts, and the \emph{muniverse-benchmark-dynamic} dataset is documented with the construction details given in supplementary §\ref{supp:benchmark}.

\item {\bf Crowdsourcing and Research with Human Subjects}
    \item[] Question: For crowdsourcing experiments and research with human subjects, does the paper include the full text of instructions given to participants and screenshots, if applicable, as well as details about compensation (if any)?
    \item[] Answer: \answerNA{}
    \item[] Justification: The paper involves no crowdsourcing and no new research with human subjects. The real-data evaluation reuses existing recordings from a published study~\cite{guerra2024adaptive}; all other experiments are on simulated EMG.

\item {\bf Institutional Review Board (IRB) Approvals or Equivalent for Research with Human Subjects}
    \item[] Question: Does the paper describe potential risks incurred by study participants, whether such risks were disclosed to the subjects, and whether Institutional Review Board (IRB) approvals (or an equivalent approval/review based on the requirements of your country or institution) were obtained?
    \item[] Answer: \answerNA{}
    \item[] Justification: No new human-subject data were collected. The real recordings are reused from a published study~\cite{guerra2024adaptive}, which obtained ethics approval and informed consent for their collection; all other data are simulated.

\item {\bf Declaration of LLM usage}
    \item[] Question: Does the paper describe the usage of LLMs if it is an important, original, or non-standard component of the core methods in this research? Note that if the LLM is used only for writing, editing, or formatting purposes and does not impact the core methodology, scientific rigorousness, or originality of the research, declaration is not required.
    \item[] Answer: \answerNA{}
    \item[] Justification: LLMs were not used as a component of the core methodology; any use was limited to writing and editing assistance, which does not require declaration.
    \item[] Guidelines:
    \begin{itemize}
        \item The answer NA means that the core method development in this research does not involve LLMs as any important, original, or non-standard components.
        \item Please refer to our MainTrackHandbook for what should or should not be described.
    \end{itemize}

\end{enumerate}

\end{document}